UNIVERSIDAD DEL PAÍS VASCO - EUSKAL HERRIKO UNIBERTSITATEA

ESCUELA SUPERIOR DE INGENIERÍA DE BILBAO

DEPARTAMENTO DE INGENIERÍA DE COMUNICACIONES

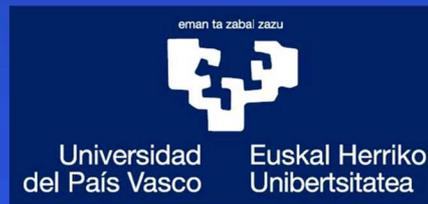

eman ta zabal zazu

Universidad
del País Vasco | Euskal Herriko
Unibertsitatea

# MODELO DE EVALUACIÓN TÉCNICO-ECONÓMICA DE TECNOLOGÍAS DE ACCESO

# MODEL FOR TECHNO-ECONOMIC ASSESSMENT OF ACCESS TECHNOLOGIES

## DOCTORAL DISSERTATION

## TESIS DOCTORAL

*Carlos Bendicho Julián*

iD https://orcid.org/0000-0002-8538-0043

**DIRECTORES:**

Prof. Dr. Juan José Unzilla Galán

Profª. Dra. Nerea Toledo Gandarias

December, 2015

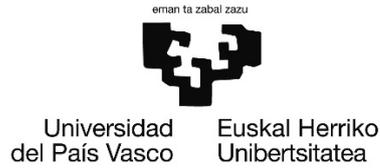

# DOCTORAL DISSERTATION

## MODEL FOR TECHNO-ECONOMIC ASSESSMENT OF ACCESS TECHNOLOGIES

**Author:**        Carlos Bendicho Julián, M.Sc.

**Supervisors:**   Prof. Dr. Juan José Unzilla Galán

Prof. Dra. Nerea Toledo Gandarias

Bilbao, December, 2015

Page intentionally left blank

This document is an English Translation of the original PhD Thesis in Spanish

titled "Modelo de Evaluación Técnico-Económica de Tecnologías de Acceso" by

Dr. Carlos Bendicho, PhD | MBA

-------------------------------------------------------------------------------------------------

Chapters 3 and 4 of this Doctoral Dissertation are protected as they are confidential.

The developed Model and Methodology are available under specific license of use.

For more information, please contact the author via e-mail in:
carlos.bendicho@coit.es

# General Index

Page intentionally left blank

# Index of Illustrations

Page intentionally left blank

# Tables Index

Page intentionally left blank

Page intentionally left blank

# Chapter 1

# INTRODUCTION

## 1.1 Context

Private data networks evolved from the interconnection of nodes through dedicated point-to-point lines, to the first packet-switched networks that allowed greater flexibility in management, to the Virtual Private IP Networks, which allowed us to offer access solutions and connectivity much more robust and cheaper than predecessor technologies. The revolution came hand in hand with IP Virtual Private Network services that allowed not only the use of dedicated point-to-point accesses from the customer's home to the access point, but also to incorporate new xDSL broadband access technologies, thereby reducing the cost of the whole. of the technical solution by opting for more economical access solutions.

These IP Virtual Private Networks allowed and also allow access to remote users through tunnels to guarantee a minimum level of security in access to information, such as IPSec tunnels. These remote users could access from fixed locations or could be on the move. The different tunneling technologies even made it possible to create virtual private networks on an operator's own IP network without contracting any virtual private network service. The use of this technology continues to this day.

In turn, the different agents in the telecommunications sector always have technical needs such as bandwidth and availability, among others, and economic ones: price, income, CapEx, OpEx, etc. related to the access network, which must be satisfied by choosing between the different, increasingly greater alternatives of access technologies, including alternatives with virtualized accesses.

The access network, due to being the part of the network that requires the greatest capillarity in order to serve all end users, is the part of the network that consumes the highest volume of CapEx investment and OpEx operating costs, constituting the called the "last mile" problem.

The growing need for interconnection of devices of different nature and in very diverse environments coined in the Internet of Things, the Industrial Internet, connected vehicles or the Internet of Everything, as well as the improvement of video resolution in broadcast or streaming of Ultra High Definition 4K / 8K, causes those technical needs for bandwidth and availability, among others, to evolve, as well as economic needs, impacting the choice of access technologies and the design of access networks.

There are also different end-user profiles with different needs, also dynamic, in the Residential area, SMEs, Large Companies, Public Administrations, conditioned by the activity sector in the case of companies, the geographical area, the socio-demographic profile, etc. Likewise, there is a great variety of agents that intervene in the telecommunications market, each one with their particular needs, whether they are clients, end users, operators, consultancies, analysis firms, Regulatory Bodies, Public Administrations, Service Providers of Contents, OTTs, etc.

Therefore, access networks are in continuous evolution, in a dynamic environment, in such a way that it is necessary to develop technical-economic models of universal application to facilitate the choice of the most appropriate technical solution in each scenario, which meet those changing technical and economic needs.

Traditionally, "technical-economic models are a method used to evaluate the economic viability of complex technical systems", according to Smura's doctoral thesis [99], which obviates any evaluation of technical viability in said definition.

In line with this definition, current decision-making processes regarding the choice of access technologies are based almost exclusively on economic criteria, which carries the risk of committing serious technical errors that may compromise the expected economic viability.

In this regard, within the framework of the European BONE project and prior to Smura's doctoral thesis, the article entitled "General Framework for Techno-Economic Analysis of Access Networks" [68] was published in 2010, which suggests, although not develops, the additional need for evaluation of technical feasibility. This aspect was also addressed by the author of this research work in the article entitled "What about 'first mile' availability?" in 2004 [25], in which it was demonstrated that the redundancy of ADSL accesses allowed reaching 'carrier-grade' availabilities (99.9999%), in addition to improving the set of technical benefits of equivalent access.

Therefore, a broader extension of the concept of technical-economic analysis is required, which emphasizes the evaluation of technical feasibility, and is supported by technical-economic models that develop it.

Therefore, the concept of technical-economic model of this doctoral thesis goes beyond the traditional scope mentioned, since the author also includes and underlines the inescapable need to carry out the evaluation of technical feasibility. From this perspective, the new definition that the author proposes is as follows:

*"The technical-economic models are a method that allows the evaluation of the technical and economic viability of complex technical systems."*

This new definition emphasizes both the technical and economic aspects of modeling, considering the technical feasibility and the satisfaction of specific technical requirements and needs.

On the other hand, the technical-economic models in the literature are eminently oriented towards the dynamics of the deployment of access networks promoted by manufacturers and operators, ignoring the perspective of end users, for which technical-economic models are required. that can reflect and respond to both perspectives, in order to contribute to a market equilibrium.

This doctoral thesis arises from a detailed analysis of the literature on technical-economic modeling in the field of access networks, which addresses the problem of shortages in the last mile, corroborated by public financing of projects in this field, and by the professional experience of the author in his professional practice as a telecommunications engineer, in order to make decisions and design innovative technical solutions in the area of access technologies. All this, in order to satisfy the needs and interests of the different agents of the telecommunications market: operators, clients, end users, regulatory bodies, content service providers, standardization bodies, etc.

## 1.2 Motivation and objectives

Within the framework of this research, universal and generalizable technical-economic models are called, those that allow choosing the most appropriate technical solution that satisfies the required needs, adapted to a very dynamic sector, in continuous evolution, with different agents that intervene in the sector, different end-user profiles, with changing technical and economic needs, a growing demand for bandwidth, robust and, at the same time, economic solutions, a great and growing variety of devices, and a great diversity of access technologies current and future.

As has already been advanced in the previous section, the need arises to extend and redefine the traditional concept of the technical-economic model of the literature enunciated by Smura [99], in such a way as to underline the evaluation of technical feasibility, compared to current decision-making processes based on purely economic criteria. As already mentioned in section 1.1, [68] suggests the evaluation of technical feasibility, but ultimately does not develop it. The author intends to develop it in this research work, proposing a universal and generalizable technical-economic model for the evaluation of access technologies. A universal technical-economic model that allows taking into account the variability of access technologies, users and agents of the sector, and generalizable in the future,

Therefore, the characteristics of a universal, scalable, flexible and generalizable theoretical technical-economic model will be defined, in a coherent way with the aforementioned redefinition, that allows to contemplate all the contextual aspects mentioned, and satisfy all the needs, allowing to integrate and serve all prospects.

A study of the technical-economic models in the literature will be carried out, classifying them based on the characteristics of the theoretical technical-economic model, identifying their degree of compliance and the existing improvement path.

From this improvement path, a new proposal for a technical-economic model will be made that presents a higher degree of compliance than the models in the literature.

Consequently, the main dual objective of this research is the following:

1) Define a technical-economic model of universal, scalable, flexible and generalizable application that allows the evaluation and comparison of multiple access technologies in different scenarios.

2) Develop a methodology for applying the technical-economic model to facilitate its use by different market agents, providing guidelines for the design of scenarios, the application of the model and the proper interpretation of the results obtained.

The secondary objectives of this doctoral thesis derived from the previous double objective are the following:

- Redefine the concept of technical-economic analysis in such a way that it emphasizes the evaluation of technical feasibility in the face of current decision-making processes based on purely economic criteria.
- Define the characteristics of a universal theoretical technical-economic model.
- Prepare a classification of the technical-economic models in the literature based on these characteristics.
- To enable the technical and economic characterization of any access technology in any configuration or combination of serial or parallel elements.
- Define specific metrics of technical and / or economic performance and efficiency of access technologies that allow their evaluation and technical-economic comparison.
- Identify the degree of compliance with the technical and / or economic customer requirements established by the user of the model for each technology.
- Identify the minimum number of redundant accesses that allows a given access technology to meet technical and / or economic customer requirements
- Allow the technical-economic evaluation of redundant accesses and parallel combinations of the same or different access technology
- Predict behaviors and trends in access technologies, given the dynamism of the market and technology.
- Allow the technical-economic evaluation of access technologies in 'top-down' approaches (from the deployment perspective) and 'bottom-up' (from the customer or end-user perspective).
- Allow comparison between any access technologies.

Motivated by the identified and exposed research niche, the research question or problem of this doctoral thesis is stated as follows:

*"Is it possible to define universally applicable, scalable, flexible and generalizable technical-economic models that make it possible to compare any access technologies in order to help the different market agents make decisions?"*

This doctoral thesis proposes, defines and develops a new model called UTEM (Universal Techno-Economic Model: Universal Technical-Economic Model) and a methodology for its application, in order to answer the research question posed and comply with the proposed objectives.

## 1.3 Structure of the thesis

Figure 1.1 shows the structure of this doctoral thesis.

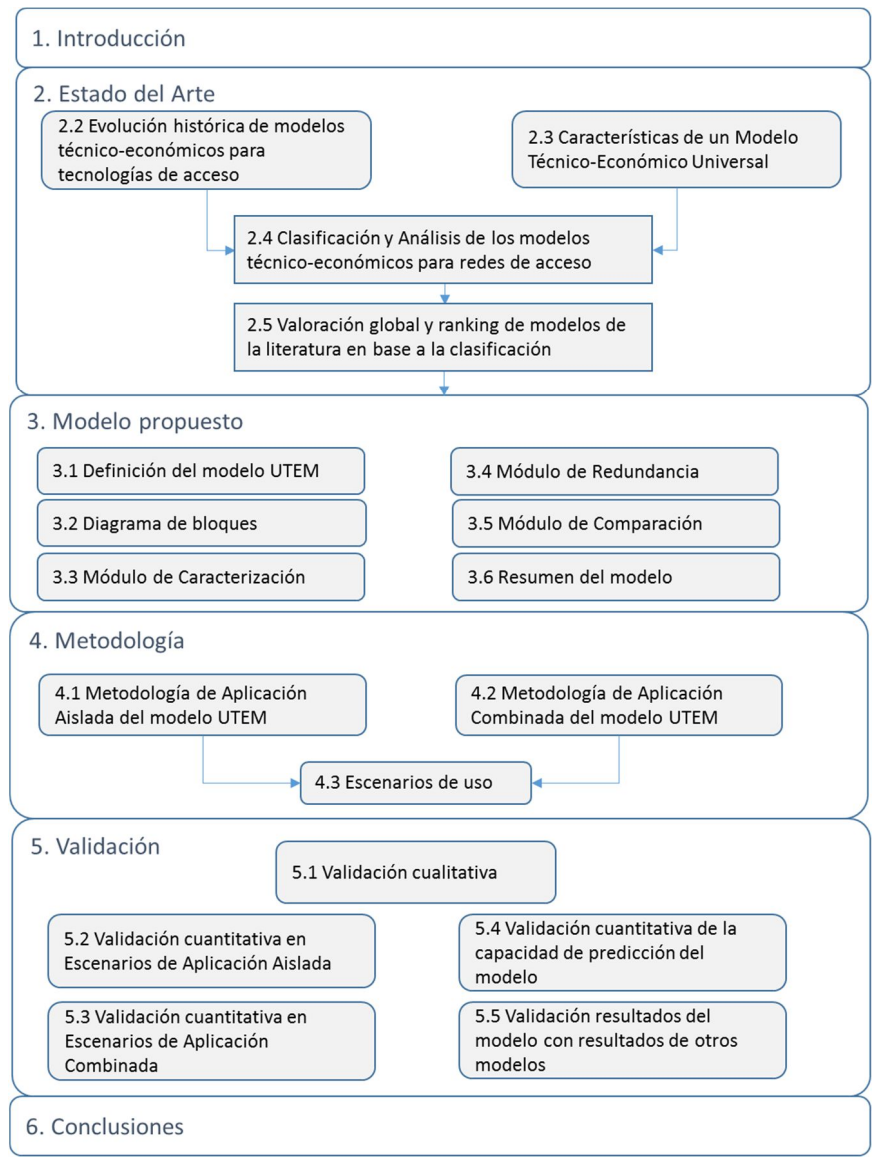

*Figure 1.1: Thesis structure*

The development of this doctoral thesis is structured after the present Chapter 1 of Introduction, in Chapter 2 of the State of the Art in which a historical evolution of technical-economic models of access technologies in the literature is presented, after which they are established and define the Characteristics of a theoretical Universal Technical-Economic Model for access technologies. It continues with a Classification and Analysis of technical-economic models of the literature, based on these characteristics. Finally, the Global Assessment and the ranking of models in the literature based on this classification are shown.

Next, Chapter 3 is presented, in which the proposed Model is defined and developed (called UTEM Model: Universal Techno-Economic Model) with its definition, block diagram, description of the modules that compose it, and a summary of the model.

In Chapter 4, the application methodology of the model is shown in its two aspects: Isolated Application and Combined or integrated application with other models, as well as the use scenarios.

In Chapter 5 of Validation, the qualitative validation of the proposed model is presented based on the Characteristics of a theoretical Universal Technical-Economic Model established in the State of the Art, the quantitative validation in isolated application scenarios and in combined application scenarios, ending with the quantitative validation of the predictive capacity of the model with the results of an analysis firm.

Finally, the Conclusions of the present doctoral thesis are exposed in Chapter 6.

## 1.4 Conclusions

As a result of this research work, the following contributions to the literature are expected:

- Redefine the concept of the technical-economic model beyond the traditional scope, including and underlining the need to carry out the evaluation of technical feasibility.
- Define the characteristics of a universal, flexible, generalizable and scalable technical-economic model for access technologies.
- Present a classification of the technical-economic models in the literature based on the characteristics of a theoretical technical-economic model.
- Define and develop a universal, generalizable, scalable and flexible model for the technical-economic evaluation of access technologies that presents a higher degree of compliance than the literature models with respect to the characteristics of the theoretical model, and whose global evaluation capacity technical and economic is consistent with the aforementioned redefinition.
- It is expected that the proposed UTEM model can be applied in combination with other models, to which it contributes the technical characterization of the access technology under study, the minimum number of redundant accesses to meet customer requirements, as well as the possibility to choose the most appropriate technology based on performance indicators and technical-economic efficiency.
- It is expected that the global technical and economic evaluation capacity of the UTEM model will allow to validate the redefinition of the concept of technical-economic analysis beyond the mere economic evaluation of complex technical systems, adding and emphasizing technical evaluation. That allows evaluating any access technologies, not only from the point of view of deployment by an operator, but also from other angles, allowing a customer to make decisions, or for a regulatory body to define policies in the field of access network, etc. It is therefore expected that the UTEM model will allow us to contemplate the perspectives of any client, end user or agent in the sector, now and in the future.

Page intentionally left blank

# Chapter 2

# STATE OF THE ART: Classification and Analysis of techno-economic models for access networks

## 2.1 Introduction

In the literature, there are different techno-economic models fundamentally oriented towards decision-making regarding the deployment of access networks by telecommunications operators. Given that technology evolves with agility and the type of agents involved in the telecommunications market is varied, it is worth investigating the state of the art and making contributions in order to propose a generalizable technical-economic model that provides greater universality in its application .

In this chapter of State of the Art, based on the main objective of the thesis, it is presented:

- in section 2.2 the historical evolution of technical-economic models for access networks in the literature and related publicly funded projects.
- Section 2.3 establishes the characteristics of a universal, scalable, flexible and generalizable technical-economic model for access technologies.
- In section 2.4 a classification and analysis of technical-economic models from the literature is elaborated, based on the characteristics of the theoretical technical-economic model exposed in section 2.2.
- In section 2.5 a global assessment is made and a ranking of technical-economic models in the literature is presented based on this classification.
- In section 2.6 it is concluded that it makes sense to deepen and investigate the development of models for the technical-economic evaluation of access technologies, which achieve greater compliance than the models in the literature, thus approaching the theoretical technical-economic model whose characteristics They are explained in section 2.3.

Both the definition of the characteristics of a universal technical-economic model for access technologies, as well as the aforementioned classification and analysis, as well as the global assessment and ranking that are presented, are the result of a detailed review and analysis of the literature, supported by the author's professional experience as a telecommunications engineer in the design of innovative solutions in the field of access networks for different agents in the telecommunications sector.

## 2.2 Historical evolution of techno-economic models for access technologies

The technical-economic models in the literature are based on the traditional definition of the technical-economic model as a "method for evaluating the economic viability of complex technical systems", according to Smura's doctoral thesis [99], as has already been advanced in the introductory chapter.

Regarding the origin of the technical-economic modeling, Smura writes the following [99]:

*"The nature of technical-economic modeling and analysis is usually future-oriented, uses and combines a number of methods from the field of Future-oriented Technology Analysis (FTA). These include cost-benefit analysis, scenario analysis, trend analysis, expert panels, and quantitative modeling (for an exhaustive list of other FTA families and methods, see TFAMWG, 2004, and [Scapolo & Porter, 2008, p 152]). Although these methods and their combinations have been widely used by both academics and professionals, academic work under the term "techno-economic" (eg: modeling, analysis, evaluation, valuation) has been published mainly related to energy (eg : [Zoulias & Lymberopoulos 2006]),*

*In the field of telecommunications, the term 'technical-economic' was introduced during the European research program (RACE) (Research into Advanced Communications for Europe: Research in Advanced Communications for Europe) in the period 1985-1995. The first techno-economic modeling work was carried out in the RACE 1014 ATMOSPHERIC [6] [4] [5] project and in the RACE 1044 [9] project in which different scenarios and evolution alternatives towards broadband systems were analyzed. Subsequently, the RACE 2087 TITAN project (Tool for Introduction scenarios and Techno-economic studies for the Access Network: Tool for Introduction scenarios and techno-economic studies for the Access Network) developed a methodology and a tool for the techno-economic evaluation of new narrowband services, broadband and access networks (see [13] [14]) . Since the end of the 90s, many European research projects have used and extended the methodologies and tools created in these first projects ".*

As advanced in the introductory chapter, the following exception found in the literature to the traditional definition of the technical-economic model indicated by Smura in 2012. In 2010, [68] states: "All business modeling should be accompanied by a technical-economic evaluation in order to provide the reader with information on the financial perspective and technical feasibility of an investment project in telecommunications ", after introducing the need to carry out a performance analysis of the access network, in which the relationship between the cost and the reliability of the network is considered, limiting itself to relating both aspects: cost and reliability, in a specific indicator or figure of merit. [68] suggests the evaluation of technical feasibility, but ultimately does not develop it.

After a detailed review and analysis of the models in the literature, it is found that they are imbued by the traditional definition of the technical-economic model indicated by Smura [99], and are eminently oriented to the deployment of access technologies from the perspective of users. operators, manufacturers and standardization bodies. It also adds that only some models have the capacity to evaluate different access technologies, and a few of combination of technologies. Only in some model has a slight hint of orientation to the end user or to other agents different from those mentioned been detected. In addition, it is added that all its output parameters are cheap, except for a very exceptional wink.

Therefore, the review and analysis of the literature, the existence of a great variety of access scenarios and technologies, the high cost of investments and maintenance of the access network, the significant volume of scientific production in this regard, the interest of the EU institutions demonstrated by the public financing of projects that promote and use technical-economic modeling, and all the contextual aspects raised in the introductory chapter, lead to the conclusion that it is interesting to deepen and lay the foundations regarding the characteristics that should have a universal and generalizable theoretical model for the technical-economic evaluation of access technologies, in order to develop a specific classification and detect with greater precision the areas for improvement of the technical-economic models in the literature.

Next, we proceed to show the review and analysis of the literature on techno-economic modeling of access technologies, as well as the chronology of projects related to public funding, given the interest in this area of research demonstrated by the institutions of the Union European (EU).

## 2.2.1 Review and analysis of the literature

The literature on technical-economic evaluation models for access technologies has been reviewed and analyzed, finding that it is based on the aforementioned traditional concept of technical-economic model enunciated by Smura [99], with the exceptional suggestion already indicated, and not developed [68].

A representative sample of the most relevant articles in the literature has been selected, which allow us to provide a vision on the State of the Art of technical-economic models for access technologies. These articles are listed below in Table 2.1.

| Author, year | Qualification | Techno-economic model | Access technologies | Investigation programme | Source |
|---|---|---|---|---|---|
| Reed & Sirbu (1989) [3] | 'An Optimal Investment Strategy Model for Fiber to the Home' | Dynamic scheduling | FTTH | BELL (scholarship) | USA |
| Lu et al. (1990) [7] | 'System and Cost Analyzes of Broad-band Fiber Loop Architectures' | Cost Modeling | B-ISDN, 4 alternative fiber loop architectures (ADS, PPL, HPPL, PON) | BELLCORE | USA |

| Graff et al. (1990) [6] | 'Techno-Economic Evaluation of the Transition to Broadband Networks' | STEM | Evolution from STM to ATM | RACE I | Europe |
|---|---|---|---|---|---|
| Ims et al. (1996) [12] | 'Multiservice Access Nework Upgrading in Europe: A Techno-Economic Analysis' | TITAN | xDSL, FTTx, HFC, FTTH (PON) | EURESCOM | Europe |
| Olsen et al. (1996) [14] | 'Techno-Economic Evaluation of Narrowband and Broadband Access Network Alternatives and Evolution Scenario Assessment' | TITAN | ADSL, PON, CATV, ISDN, FTTx, HFC | RACE II | Europe |
| Ims et al. (1997) [13] | 'Risk Analysis of Residential Broadband upgrade in a Competitive and Changing Market' | TITAN | xDSL, HFC, ATM PON | RACE II | Europe |
| Stordahl et al. (1998) [15] | 'Risk Analysis of Residential Broadband upgrade based on Market Evolution and Competition' | OPTIMUM (based on TITAN) | FTTN, FTTB, HFC | ACTS | Europe |
| Jankovic et al. (2000) [19] | 'A Techno-Economic Study of Broadband Access Network Implementation Models' | P614 | ISDN, xDSL, HFC, FTTx, WLL, Satellite | EURESCOM | Europe |
| Katsianis et al. (2001) [20] | 'The Financial Perspective of the Mobile Networks in Europe' | TERA | GPRS, UMTS | ACTS | Europe |
| Welling et al. (2003) [24] | 'Techno-Economic Evaluation of 3G & WLAN Business Case Feasibility Under Varying Conditions' | TONIC | UMTS, WLAN | EU FP5 | Europe |
| Smura (2005) [31] | 'Competitive Potential of WiMAX in the Broadband Access Market: A Techno-Economic Analysis' | based on ECOSYS / TONIC | WiMAX | EUREKA / CELTIC | Europe |
| Monath et al. (2005) [29] | 'MUSE- Techno-economics for fixed access network evolution scenarios - DA3.2p' | MUSE | FTTx, ADSL, SHDSL, VDSL, xDSL over Optics | EU FP6 | Europe |
| Sananes et al. (2005) [30] | 'Techno-Economic Comparison of Optical Access Networks' | e-Photon / One | FTTH | EU FP6 | Europe |
| Lahteenoja et al. (2006) [33] | 'ECOSYS "techno-ECOnomics of integrated communication SYStems and services". Deliverable 16: "Report on techno-economic metholology" ' | ECOSYS | ISDN, B-ISDN (FITL), xDSL, HFC, FTTx, WLL, Satellite, WiMAX | CELTIC | Europe |
| Olsen et al. (2006) [34] | 'Technoeconomic Evaluation of the Major Telecommunication Investment Options for European Players' | ECOSYS / TONIC | HFC, ADSL, VDSL, LMDS, Satellite, 3G, WLAN, FTTC, FTTH, | EUREKA / CELTIC, IST | Europe |

| Pereira (2007) [42] | 'A Cost Model for Broadband Access Networks: FTTx versus WiMAX' | Owner (BATET) | FTTx, WiMAX | | Portugal |
|---|---|---|---|---|---|
| Chowdhury et al. (2008) [46] | 'Comparative Cost Study of Broadband Access Technologies' | Owner | xDSL, Cable Modem, FTTx, WiFi, Hybrid FTTx + WiFi, Hybrid FTTx + WiMAX (WOBAN) | | USA |
| Pereira & Ferreira (2009) [58] | 'Access Networks for Mobility: A Techno-Economic Model for Broadband Access Technologies' | Owner (BATET) | Static Layer: FTTH (PON), xDSL, HFC, PLC; Nomadic Layer (mobile users): WiMAX | | Portugal |
| Van der Merwe et al. (2009) [56] | 'A Model-based Techno-Economic Comparison of Optical Access Technologies' | Owner | FTTH optical networks: GPON, AON / Active Ethernet (AE), P2P | | Germany |
| Ödling et al. (2009) [57] | 'The Fourth Generation Broadband Concept' | ECOSYS | FTTdp (G.fast) | CELTIC-4GBB | Europe |
| Ghazisaidi & Maier (2009) [53] | 'Fiber-Wireless (FiWi) Networks: A Comparative Techno-Economic Analysis of EPON and WiMAX' | Owner | FTTH + WiMAX | | Canada |
| Verbrugge et al. (2009) [59] | 'White Paper: Practical Steps in Techno-Economic Evaluation of Network Deployment Planning' | OASE | FTTH | EU FP7 | Europe |
| Casier et al. (2010) [63] | '"Overview of Methods and Tools" Deliverable 5.1. OASE ' | OASE | FTTH | EU FP7 | Europe |
| Zagar & Krizanovic (2010) [76] | 'Analyzes and Comparisons of Technologies for Rural Broadband Implementation' | Owner (Rural Broadband in Croatia) | ADSL, WiMAX | Government of Croatia | Croatia |
| Vergara et al. (2010) [74] | 'COSTA: A Model to Analyze Next Generation Broadband Access Platform Competition' | COSTA (based on BREAD & TONIC & MUSE) | FTTH / GPON, FTTN / VDSL, FTTH / P2P, HFC / Docsis, WiMAX, LTE | | Spain |
| Chatzi et al. (2010) [64] | 'Techno-economic Comparison of Current and Next generation Long Reach Optical Access Networks' | BONE | FTTH with duplicated fibers for reliability and FTTH with WDM / TDM PON ring (SARDANA architecture) | STREP-SARDANAICT-BONE (EU FP7) | Europe |

| Rokkas et al. (2010) [73] | 'Techno-economic Evaluation of FTTC / VDSL and FTTH Roll-Out Scenarios: Discounted Cash Flows and Real Option Valuation' | ECOSYS | FTTC / VDSL and FTTH | Government of Greece | Greece |
|---|---|---|---|---|---|
| Casier et al. (2011) [77] | 'Techno-economic Study of Optical Networks' | OASE | FTTH | EU FP7 | Europe |
| Feijóo et al. (2011) [79] | 'An Analysis of Next Generation Access Networks Deployment in Rural Areas' | Owner (Cost Model) | FTTH (GPON), FTTC / FTTB / VDSL, HFC DOCSIS 3.0, LTE (4G) | | Spain |
| Martín et al. (2011) [85] | 'Which could be the role of Hybrid Fiber Coax in Next Generation Access Networks?' | Owner (Cost Model) | FTTH (GPON), HFC DOCSIS 3.0 | | Spain |
| Machuca et al. (2012) [96] | 'Cost-based assessment of NGOA architectures and its impact on the business model' | OASE | Wavelength-routed WDM PON, Ultra Dense WDM, PON, AON with WDM | EU FP7 | Europe |
| Van der Wee et al. (2012) [100] | 'A modular and hierarchically structured techno-economic model for FTTH deployments' | OASE | FTTH (PON), FTTH (AON) | EU FP7 | Europe |
| Walcyk & Gravey (2012) [101] | 'Techno-Economic Comparison of Next-Generation Access Networks for the French Market' | BONE | xDSL, FTTH (GPON), FTTH (LROA-SARDANA) | | Europe |
| Pecur (2013) [112] | 'Techno-Economic Analysis of Long Tailed Hybrid Fixed-Wireless Access' | Owner | FiWi (Fixed-Wireless); Fixed: xDSL, FTTx, FSO; Wireless: WiFi, WiMAX, LTE (4G) | | Saudi Arabia |
| Bock et al. (2014) [119] | 'Techno-Economics and Performance of Convergent Radio and Fiber Architectures' | TITAN cost analysis | FTTH Active Remote Node combining PON + Radio Base Station (SODALES architecture) | EU FP7 | Europe |
| Moreira & Zucchi (2014) [122] | 'Techno-economic evaluation of wireless access technologies for campi network environments' | TONIC & ECOSYS | WiFi, WiMAX, LTE | | Brazil |
| Ruffini et al. (2014) [123] | 'DISCUS: An End-to-End Solution for Ubiquitous Broadband Optical Access' | OASE | FTTP | EU FP7 | Europe |
| Katsianis & Smura (2015) [132] | 'A cost model for radio access data networks' | Owner | LTE | | Finland |

| Forzati et al. (2015) [129] | 'Next-Generation Optical Access Seamless Evolution: Concluding Results of the European FP7 Project OASE' | OASE | FTTH | EU FP7 | Europe |
|---|---|---|---|---|---|
| Van der Wee et al. (2015) [138] | 'Techno-economic Evaluation of Open Access on FTTH Networks' | OASE | FTTH | EU-FP7 | Europe |

*Table 2.1 Articles that develop or use techno-economic evaluation models*

A review of the literature shows that there is an American germ in the field of techno-economic modeling in the field of access networks, in the late 1980s and early 1990s. Specifically, in 1989 , predictions have already been published regarding the most appropriate time to massively invest in the deployment of FTTH (Fiber To The Home: Fiber To The Home) access technology, using Dynamic Programming [3], identifying possible investment paths from a network of pure copper access to an FTTH network through hybrid networks, concluding that the optimal time to start a massive deployment would not be before 2010, considering the prediction of costs, income and interest rates. In the 90s and starting from [7] and [6], the studies began to focus, as Smura well commented in his doctoral thesis [99], in the detailed analysis of costs, starting from the components, with a 'bottom-up' approach, but ignoring the perspective of the end user, and always oriented to the deployment of access networks , in order to be able to compare the economic viability of the different technical alternatives and identify the parts of the access network that present the greatest contribution to costs, considering different scenarios of evolution of the access network, as well as of the evolution patterns Of demand. It is striking that in the US they continued to contemplate the FTTH scenarios [7] starting from the Narrowband ISDN (Integrated Services Digital Network or N-ISDN) towards the Broadband ISDN (B-ISDN: Broadband ISDN),

In the 90s, the techno-economic modeling for access networks in Europe also germinated, with the first European projects with public funding from the EU (European Union), also focused on the evaluation of costs and oriented towards the evaluation of alternatives deployment techniques and network evolution. The STEM model [6] stands out in this European germination as the precursor of a more complete model, TITAN [12] [13], which includes a cost prediction model based on the so-called extended learning curve [14], the which provides greater predictive precision to successive models such as OPTIMUM [15] and lays the foundations for more complete models such as TONIC [24] [23],

The stage of development of more complete technical-economic models, initiated and inspired by TITAN, begins its consolidation with the ECOSYS [31] [28] model, which incorporates the traditional techno-economic modeling based on the calculation of economic indicators such as the NPV (Net Present Value) or VAN (Net Present Value), with DCF analysis (Discounted Cash Flows: Depreciated Cash Flows considering the interest rate), an ROA (Real Options Analysis) analysis inspired by financial options or futures, in order to increase the precision of the economic output parameters, and allows the technical-economic evaluation of fixed, wireless, and mixed or hybrid technologies, in different scenarios and geographical areas to be covered [33] [34] [38]

As a result of the aforementioned consolidation with the ECOSYS model, there is an effect of disclosure or dissemination beyond the projects with public funding from the EU, which is detected by identifying new proprietary models such as the [27] model for PLC technology (Power Line Communications ), [32] for optical networks, [41] for 3G-LTE, the BATET model [43] [42], and its subsequent evolution distinguishing between fixed and nomadic layers, the latter for mobile users [58], and identifying in addition, general input parameters in which there is a hint of orientation to the end user, incorporating requirements for transmission and reception bandwidth. The proprietary COSTA model [74] for modeling the costs of the access network and based on UXO, also arises outside of EU public funding, TONIC extension for the entire access and aggregation network [29], which is developed in parallel to the ECOSYS project. More proprietary models appear such as [56] oriented to the comparison of optical access technologies, [46], [53] oriented to hybrid networks that combine FTTx and WiFi or WiMAX.

Consolidation and dissemination continues, multiple works are published with orientation to specific scenarios that are supported by the ECOSYS model as [57] oriented to FTTdp (Fiber To The Distribution Point: Fiber To the Distribution Point) within the framework of the 4GBB initiative leading to the current G.fast standard [73].

The BONE project arises, aimed at a future European optical network, which incorporates cost modeling for optical networks in the access / metro area, seeking long-range optical network architectures that provide high reliability [64] [65] [89] [ 101].Within the framework of the BONE project, the article mentioned at the beginning this State of the Art Chapter [68] is published, in which the evaluation of the technical feasibility is suggested, after introducing the need to carry out a performance analysis of the access network / meter, in which the relationship between cost and network reliability is considered. [68] finally does not develop said evaluation of technical feasibility, and only limits itself to relating cost and reliability, in a specific indicator or figure of merit, in order to evaluate the different technical alternatives of long-range optical networks with different mechanisms for increase reliability (example: fiber duplication, fiber duplication and OLTs, fiber duplication-OLTs-ONUs).

Hand in hand with public funding, the OASE [59] project emerges, which proposes a methodology based on the Plan-Do-Check-Act (PDCA) cycle of Shewhart / Deming (Plan, Execute, Verify, Act), and the adapts as Scope-Model-Evaluate-Refine (Visualize, Model, Evaluate and Refine), as well as a modular design of techno-economic modeling that integrates auxiliary models and methods [63] around TONIC as a framework tool . OASE broadens the vision from ECOSYS, conceiving a modular techno-economic modeling framework with the aforementioned methodology, enabling top-down and bottom-up approaches in the technical-economic evaluation of optical access networks, becoming a model of relevance in said field [52] [77] [80] [81] [84] [96] [100] [91] [106] [124] [127] [129] [125] [138].

As a result of the aforementioned dissemination effect, more proprietary models emerge such as [55], [98], [92], [116] and [115] for optical networks, [76], [75], [69], [82] , [79], [94] and [95] for Broadband deployment in rural areas, [85] that compares FTTH and HFC DOCSIS 3.0, [112] for FiWi (Fixed-Wireless) hybrid networks, which, as a particularity, distinguishes between investors and lenders in terms of financial agents. [132] for deployment of LTE networks in Finland, models the energy consumption of radio access data networks as a function of data traffic, [107] for deployment of wireless access points. Likewise, studies based on the TONIC and ECOSYS models are published for specific scenarios (example: wireless networks on campus) [122],

According to the above, as has been advanced when introducing this section, the models in the literature are based on the traditional definition of the technical-economic model indicated by Smura [99], and are fundamentally oriented to the deployment of access technologies from the perspectives of operators, manufacturers and standardization bodies.

Given the limitation of characteristics that is detected in the literature in terms of the ability to compare multi-access, combination of technologies, targeting the end user or the rest of the agents in the telecommunications sector, the dynamism of the market, and the aforementioned context In the introductory chapter, it is concluded that it is interesting to delve into the characteristics that a theoretical, universal and generalizable technical-economic model should have for access technologies, which will be explained in section 2.2 of this Chapter.

## 2.2.2 Chronology of projects with public funding

At the European level, from the public institutions of the European Union, different projects have been promoted and financed in the last two decades, oriented to the development of models of technical-economic evaluation of access technologies from the pioneers RACE 1014 ATMOSPHERIC, RACE 1028 REVOLVE, RACE 1044 IBC, RACE 2087 TITAN, AC226 OPTIMUM, AC364 TERA, IST-25172 TONIC, through the projects EURESCOM, MUSE, BREAD, ECOSYS, OASE, etc. [6] [4] [8] [33] [ 37] [39] [99].

The aforementioned publicly funded projects give rise to a large part of the literature, as can be seen in Table 2.1, since many of the technical-economic models bear the name of the project that defines them. There are also other projects with public funding and different objectives, for example, related to aggregation networks or even backbones, which use and rely on technical-economic models already developed by previous or parallel projects.[105] [123] [119].

The technical-economic evaluation scenarios of the aforementioned projects are closely related to the evolution of access technologies. Figure 2.1 shows the historical evolution of access technologies, together with the chronology of projects that develop or use technical-economic models, the object of literature on the matter, in relevant publications and conferences.

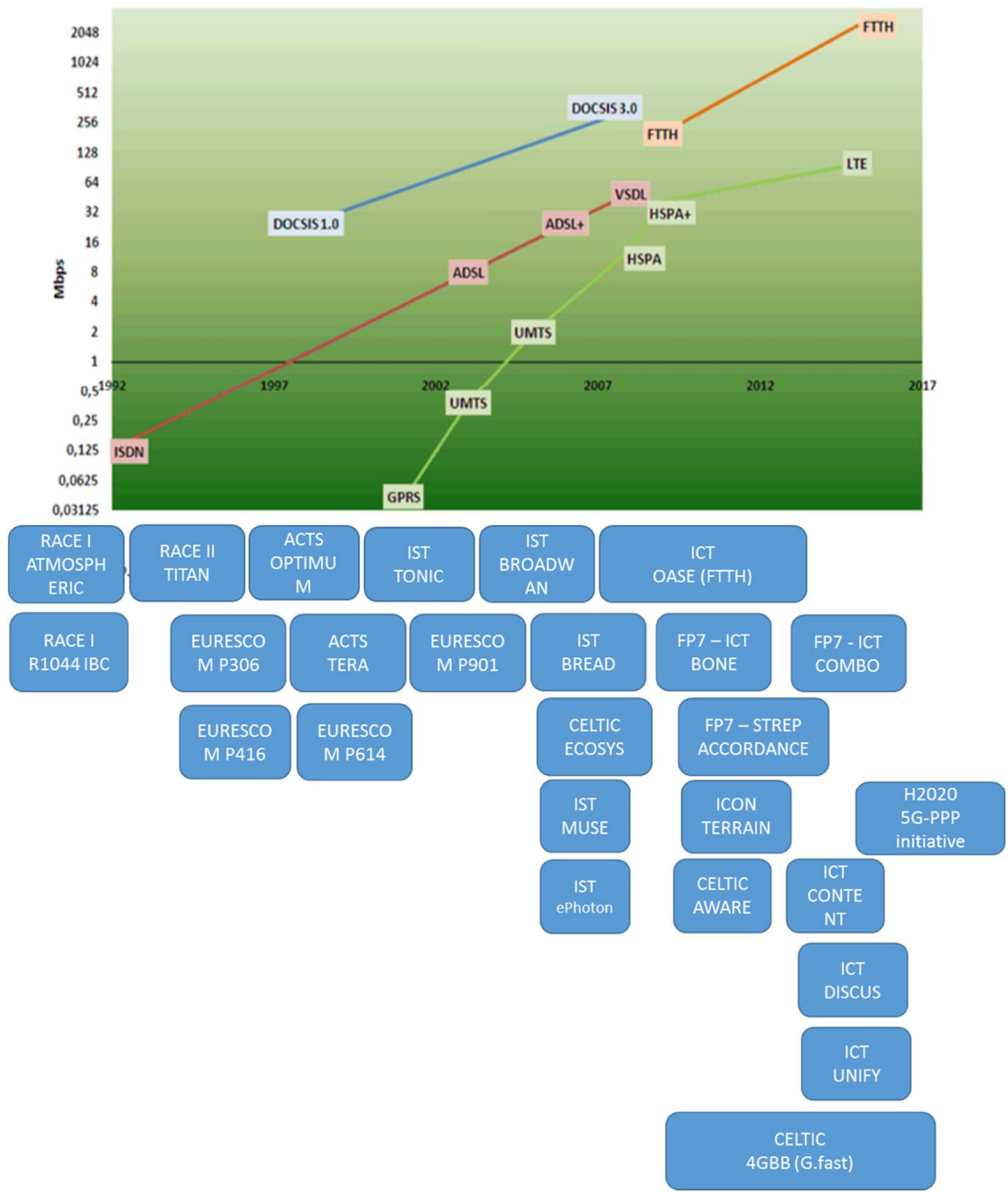

*Figure 2.1: Historical evolution of projects that develop or use technical-economic evaluation models for access networks, together with the historical evolution of access technologies. Source: ICP-ANACOM, European Commission (ec.europa.eu) and websites of each project.*

A brief description of the aforementioned projects is presented below, indicating the technical-economic model or models they use, and including a brief description of the most relevant models:

- (1988-1991) Within the European program RACE (Research into Advanced Communications-Technologies for Europe: Research in Advanced Communication Technologies for Europe), the pioneering project RACE 1044 IBC developed a techno-economic model called SYNTHESIS, which included a geometric model Simple for access networks to calculate cable and conduit lengths, an investment calculation model (CapEx), and a cost model that took into account depreciation to calculate its Present Value, and it did not consider income [2] [OLS09]. The RACE 1014 ATMOSPHERIC and RACE 1028 REVOLVE projects also relied on the first framework tool for technical-economic analysis in telecommunications that was developed within the framework of the RACE I program called Analysis STEM [6]. Analysis STEM was a framework tool, within which models such as the SYNTHESIS geometric model could be built and used to compare the implementation of different technical alternatives over time [6]. The user entered parameters of expected demand, interest rate, equipment investment and operation costs, equipment depreciation, pricing and provision policy, and provided annualized economic indicators of Income, Investments (CapEx), Operating Costs (OpEx) and Cash Flow. Figure 2.2 shows the basic annual calculation cycle of the STEM tool, in which customer demand is met by the installation of equipment, the costs of which feed into the calculation of the rates that in turn influence demand [ GRA90]. within which models such as the SYNTHESIS geometric model could be built and used to compare the implementation of different technical alternatives over time [6]. The user entered parameters of expected demand, interest rate, equipment investment and operation costs, equipment depreciation, pricing and provision policy, and provided annualized economic indicators of Income, Investments (CapEx), Operating Costs (OpEx) and Cash Flow. Figure 2.2 shows the basic annual calculation cycle of the STEM tool, in which customer demand is met by the installation of equipment, the costs of which feed into the calculation of the rates that in turn influence demand [ GRA90]. within which models such as the SYNTHESIS geometric model could be built and used to compare the implementation of different technical alternatives over time [6]. The user entered parameters of expected demand, interest rate, equipment investment and operation costs, equipment depreciation, pricing and provision policy, and provided annualized economic indicators of Income, Investments (CapEx), Operating Costs (OpEx) and Cash Flow. Figure 2.2 shows the basic annual calculation cycle of the STEM tool, in which customer demand is met by the installation of equipment, the costs of which feed into the calculation of the rates that in turn influence demand [ GRA90]. the SYNTHESIS geometric model, to compare the implementation of different technical alternatives over time [6]. The user entered parameters of expected demand, interest rate, equipment investment and operation costs, equipment depreciation, pricing and provision policy, and provided annualized economic indicators of Income, Investments (CapEx), Operating Costs (OpEx) and Cash Flow. Figure 2.2 shows the basic annual calculation cycle of the STEM tool, in which customer demand is met by the installation of equipment, the costs of which feed into the calculation of the rates that in turn influence demand [ GRA90]. the SYNTHESIS geometric model, to compare the implementation of different technical alternatives over time [6]. The user entered parameters of expected demand, interest rate, equipment investment and operation costs, equipment depreciation, pricing and provision policy, and provided annualized economic indicators of Income, Investments (CapEx), Operating Costs (OpEx) and Cash Flow. Figure 2.2 shows the basic annual calculation cycle of the STEM tool, in which customer demand is met by the installation of equipment, the costs of which feed into the calculation of the rates that in turn influence demand [ GRA90]. The user entered parameters of expected demand, interest rate, equipment investment and operation costs, equipment depreciation, pricing and provision policy, and provided annualized economic indicators of Income, Investments (CapEx), Operating Costs (OpEx) and Cash Flow. Figure 2.2 shows the basic annual calculation cycle of the STEM tool, in which customer demand is met by the installation of equipment, the costs of which feed into the calculation of the rates that in turn influence demand [ GRA90]. The user entered parameters of expected demand, interest rate, equipment investment and operation costs, equipment depreciation, pricing and provision policy, and provided annualized economic indicators of Income, Investments (CapEx), Operating Costs (OpEx) and Cash Flow. Figure 2.2 shows the basic annual calculation cycle of the STEM tool, in which customer demand is met by the installation of equipment, the costs of which feed into the calculation of the rates that in turn influence demand [ GRA90].

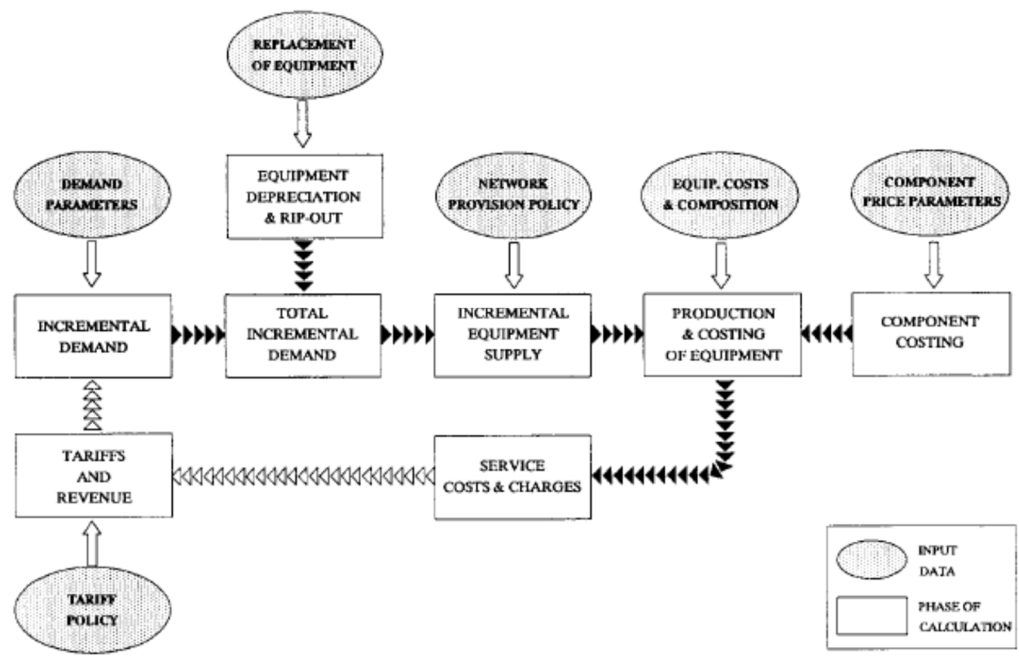

*Figure 2.2: Annual basic calculation cycle of the STEM model [6].*

- (1990-1994) RACE 2087 TITAN [14], [33]: (Tool for Introduction scenarios and Techno-economic evaluation of Access Network: Tool for Introduction and Techno-economic Evaluation scenarios of the Access Network). The main objective of the TITAN project was to develop a methodology and a tool for the technical-economic evaluation of alternatives for new narrowband and broadband services in the residential and small business market. It was framed in the RACE II (Research in Advanced Communications in Europe) program, whose main objective was the introduction of integrated broadband communications, representing the effort of the European Commission to support pre-competitive technological R&D in the telecommunications area. during the Third Framework Program for scientific research and development (FP3: Framework Program 3). Figure 2.3 shows the general structure of the TITAN methodology and tool, initially oriented to the evaluation of any type of access network architecture (eg: star, bus, ring, or combinations), as well as the incorporation of predictions.

regarding the demand or penetration of services, in line with the STEM model [6]. The flexibility regarding the study period allowed the user to consider the evolution of the network and services. Always with a 'top-down' approach (from the operator's deployment perspective), the customer density of the area to be covered was initially defined, from which one or more geometric models were applied, for example: [GAR89 ], in order to calculate the wiring length. The costs were obtained from a cost database developed in the project, coming from various European sources, and using extended learning curves [14]. Adding the OAM costs (Operation, Administration and Maintenance), the global costs were obtained. Revenues were obtained as a result of the estimates resulting from the Delphi survey or panel of experts method, together with the evolution of service rates. TITAN provided annualized economic indicators of Income, Investments (CapEx), Operating Costs (OpEx) and Cash Flow, as well as the amortization period (Payback Period) [14] [16]. the overall costs were obtained. Revenues were obtained as a result of the estimates resulting from the Delphi survey or panel of experts method, together with the evolution of service rates. TITAN provided annualized economic indicators of Income, Investments (CapEx), Operating Costs (OpEx) and Cash Flow, as well as the amortization period (Payback Period) [14] [16]. the overall costs were obtained. Revenues were obtained as a result of the estimates resulting from the Delphi survey or panel of experts method, together with the evolution of service rates. TITAN provided annualized economic indicators of Income, Investments (CapEx), Operating Costs (OpEx) and Cash Flow, as well as the amortization period (Payback Period) [14] [16].

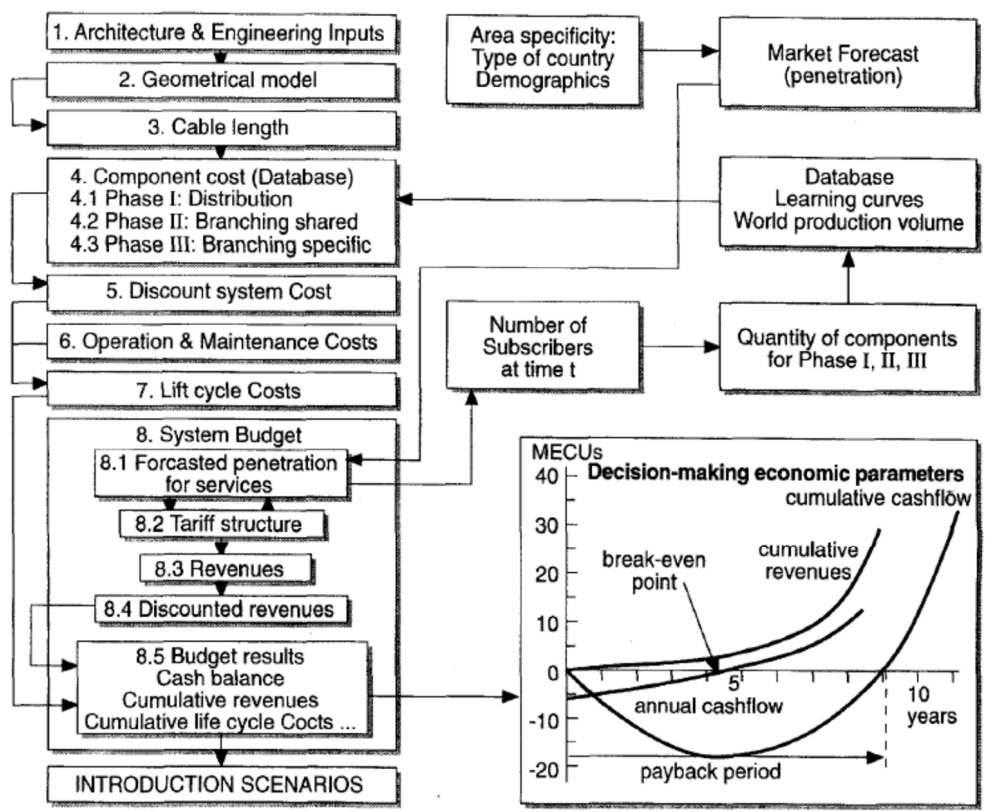

*Figure 2.3: General structure of the TITAN methodology and tool [14].*

- (1993-1996) EURESCOM P306 (Access network evolution and preparation for implementation: Evolution of the access network and preparation for its implementation). The EURESCOM P306 project integrated by European telecommunication operators arises from the consensus among operators that fiber optics will provide the future solution of the access network in terms of service capacity, response to the demand for new services, network availability and cost reduction. In the medium term, the participating operators agree that other technologies will be used, such as advanced transmission over copper or radio pairs. This project evaluates the different technologies and strategies, in order to address the viability of advanced transmission technologies over copper pair, strategic and operational recommendations for FITL systems (Fiber In The Loop: Fiber over the loop), as well as the planning and evolution of radio access. This project uses TITAN as a technical-economic assessment tool [10] [12].

- (1994-1996) EURESCOM P416 (Optical Networking: Optical Networks), as a continuation of the EURESCOM P306 project, and focused on the development of optical networks. This project is based on the consensus of European telecommunications operators regarding fiber optics as the access network technology of the future, and hand in hand with TITAN as a technical-economic evaluation tool [99].

- (1996-1998) EURESCOM P614 (Implementation strategies for advanced access networks: Implementation strategies for advanced access networks). Project continuation of the previous EURESCOM projects, which develops models of implementation of access networks, oriented towards the provision of multimedia services, and carries out the technical-economic evaluation of the same based on TITAN and OPTIMUM [99] [19]. When using OPTIMUM, the economic indicators of previous models are complemented with the NPV (Net Present Value) or VAN (Net Present Value).

- (1994-1998) AC226 OPTIMUM [33]. Project within the ACTS Program (Advanced Communications Technologies and Services). Its objective was to establish guidelines for the introduction of advanced multimedia communications networks and stimulate the increase in the use of their services, analyzing the technical-economic consequences in different case studies using the TITAN methodology and tool that was enriched to manage multimedia networks and services. , including not only the access network, but also the transport and switching part. Figure 2.1 shows the OPTIMUM techno-economic evaluation model. As can be seen, in the OPTIMUM model the economic output indicators are the NPV (Net Present Value), the IRR (Internal Rate of Return or TIR: Internal Rate of Return) and the Payback (Amortization Period), which will be common to later models.

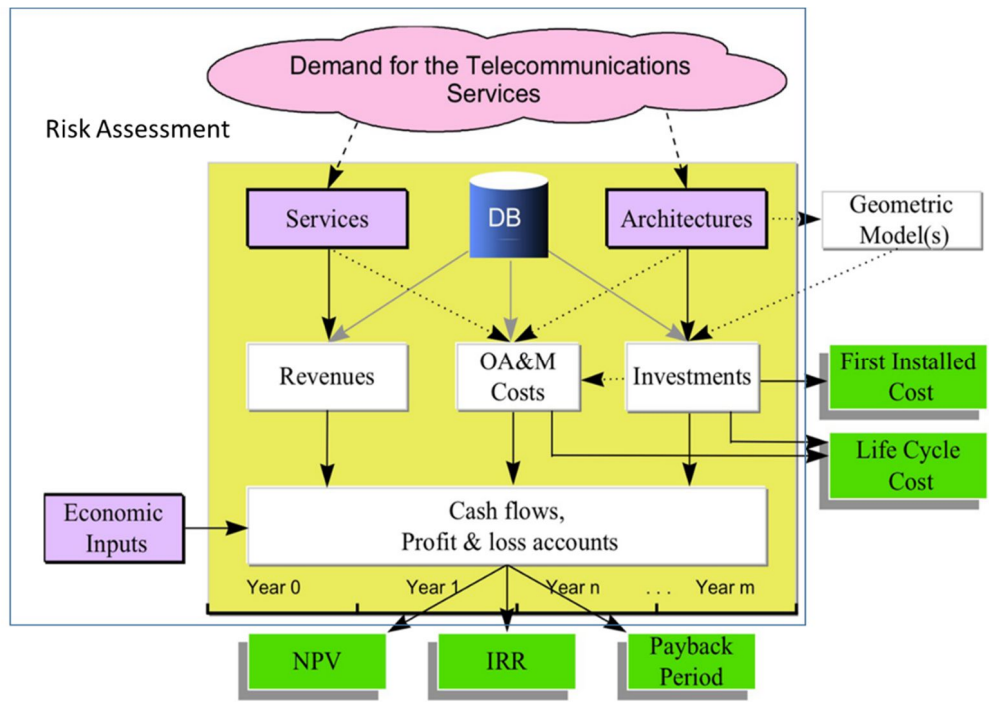

*Figure 2.4: OPTIMUM techno-economic evaluation model [17].*

- (1998-1999) AC364 TERA [33]. Its objective was to support the consolidation of development lines for the introduction of advanced communications services and networks, carrying out technical-economic evaluations of the results of ACTS program projects and field tests. He applied the technical-economic methodology and the tool developed in TITAN and OPTIMUM. The results were oriented to the ACTS community, network and service providers, equipment providers, public authorities and regulators [18].

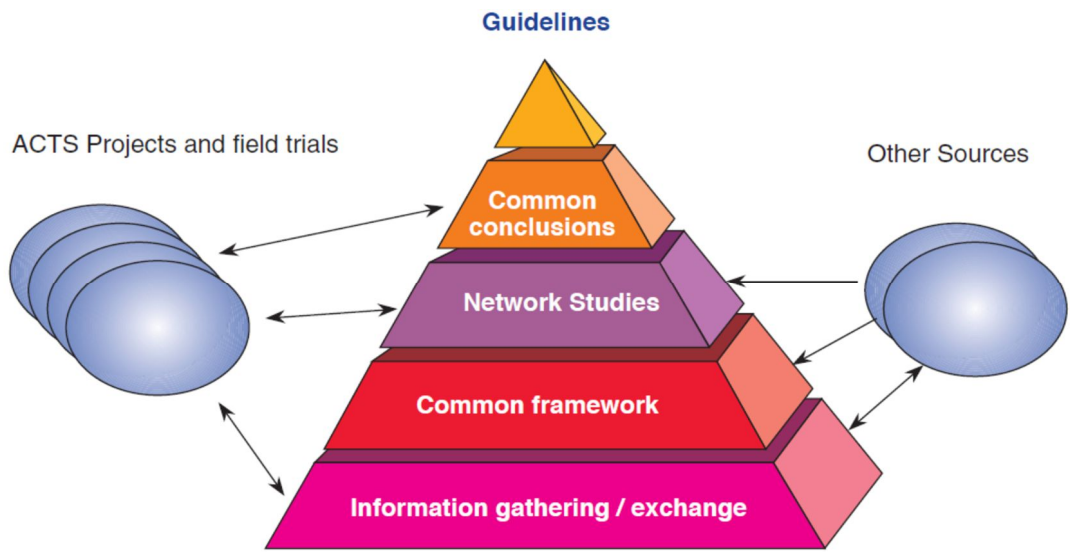

*Figure 2.5: Approach of the TERA project for the generation of techno-economic development lines (guidelines) [33].*

- (1998-2002) IST-25172 TONIC [33] [36] [23] (TechnO-econNomICs of IP optimized networks and services): This project evolves the tool and methodology of the TITAN and OPTIMUM projects for the evaluation of the entire network and communications services, in order to give maximum flexibility to the user andit focuses on the architecture of the entire telecommunications network, not just the access network. It is also capable of providing advanced telecommunications services such as multimedia. Its main objective is to evaluate the introduction of advanced communications services in fixed and mobile networks, assessing different business cases for the introduction of IP services in mobile networks as well as the introduction of broadband in competitive and non-competitive environments, facilitating the choice of the more appropriate infrastructure based on economic costs and benefits,providing relevant recommendations to regulators, telecommunications operators and service providers.

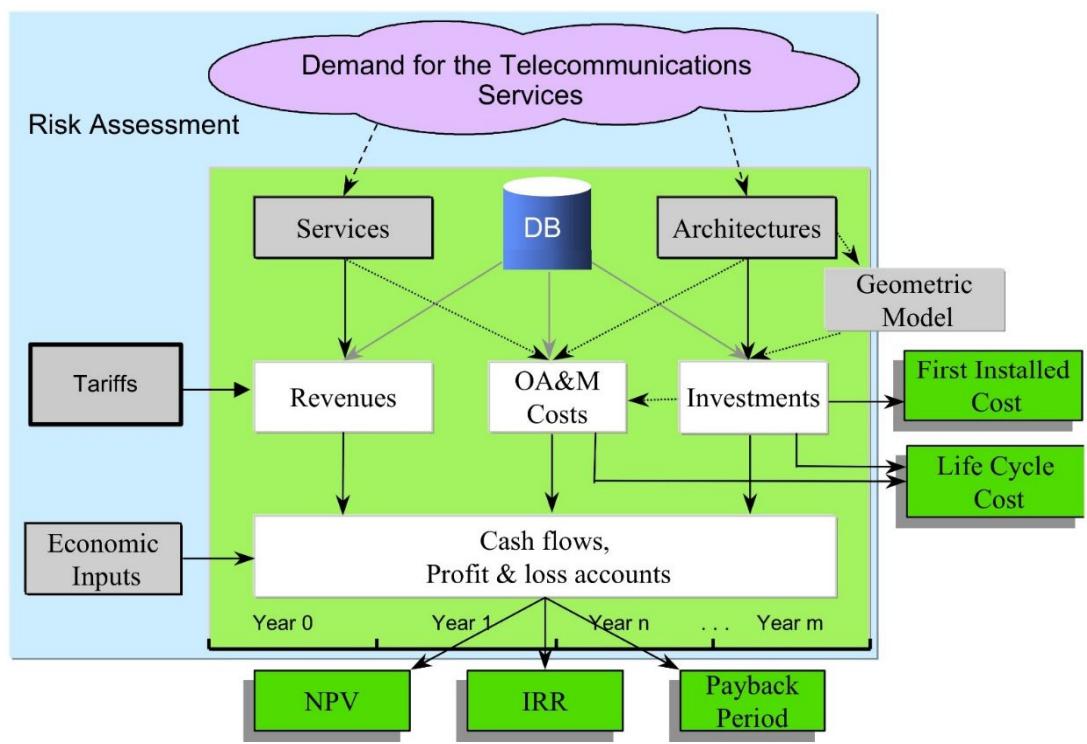

*Figure 2.6: TONIC [21] techno-economic evaluation model.*

- (1999-2001) EURESCOM P901 (Extended investment analysis of telecommunication operator strategies) [99]. This project uses the tool and methodology developed by TITAN and improved by the OPTIMUM, TERA and TONIC projects.
- (2003-2006) IST BROADWAN (Broadband services for everyone over fixed wireless access networks: Broadband services for everyone on fixed and wireless access networks). Within the 6th European Framework Program (EU FP-6), this project has three objectives: to develop an economically viable network architecture to provide true broadband services for all of Europe, to lead the European industry at the forefront of wireless solutions in new generation, and promote the advanced use of broadband services at all levels of society, carrying out demonstrations and tests in rural areas [22]. The BROADWAN project analyzes deployment costs based on the TONIC [35] model and methodology.

- (2004-2006) IST BREAD (Broadband in Europe for all: Broadband in Europe for all): The BREAD Project within the aforementioned IST program, uses a multidisciplinary approach with a coordinated action aimed at the materialization of the concept "Broadband for all " in Europe. The main objective of the BREAD project to achieve this objective of the European Union is to integrate and coordinate multiple disciplines (social, economic, regulatory and technological), in order to develop new strategies and recommendations of good practices in the field of "Broadband for everybody". The BREAD consortium also studied the techno-economic aspect of this concept for the access network and the backbone network, relying on the techno-economic results of the TONIC, TERA, ECOSYS and BROADWAN projects,

- (2004-2007) IST ePhoton / One: (Towards Bandwidth Manageability and Cost Efficiency: Towards Bandwidth Manageability and Cost Efficiency). The ePhoton / One Network of Excellence initiative seeks to integrate European knowledge on optical networks, favoring coordination among participants to reach a consensus regarding technical alternatives for the deployment of optical networks, in order to provide information to standardization and the pertinent recommendations to operators [26] [30]. This project includes both the access network, metro, trunk and the 'in-home' infrastructure, specifying the economic study in investment costs (CapEx). The ePhoton / One community subsequently spawned the BONE [47] project.

- (2004-2007) ECOSYS [33] [44] [49]: The ECOSYS project introduces as a novelty a methodology of technical-economic evaluation integrated in a specific software tool (Most of the previous models were implemented in spreadsheets). ECOSYS incorporates a new version of the cost prediction model developed by the TITAN project, based on a combination of learning curves and logistic models. The original TITAN methodology and its tool were improved in the OPTIMUM, TERA and TONIC projects to contemplate complex multimedia services. ECOSYS improves the methodology especially with regard to the definition of services and evaluation of OAM costs (Operation, Administration & Maintenance: Operation, Administration and Maintenance).Thus, it also incorporates not only the calculation of NPV (Net Present Value or NPV: Net Present Value) based on the classic analysis of DCF (Discounted Cash Flows: Depreciated Cash Flows), but also incorporates analysis of Real Options (ROA), based on the theory of financial options - a class of so-called financial derivatives - in order to consider flexibility and volatility in the course of projects [66]. In this context, flexibility is the right to take an action and volatility is the uncertainty of the value of the project's assets.[33]. Figure 2.3 schematically shows the ECOSYS methodology, since the structure of the technical-economic model is identical to that shown in Figure 2.3 [99].

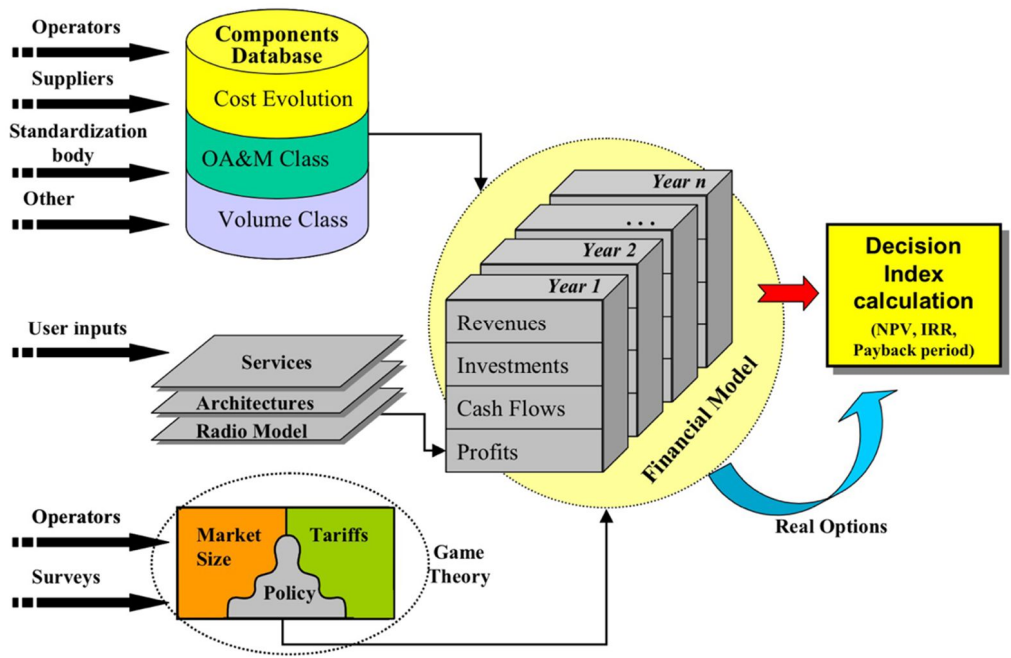

*Figure 2.7. Methodology for the technical-economic evaluation of the ECOSYS project [33].*

- (2004-2007) MUSE (Multi-Service Access Everywhere: Multiservice Access Anywhere): The overall objective of the European MUSE project is the research and development of a future low-cost multiservice network, which can provide secure connectivity between terminals of end user and the nodes that connect the aggregation network with the backbone network (edge nodes) [48]. MUSE contemplates in its objective not only the access network, but also the aggregation network, in such a way that it contemplates the entire path from the end user to the edge-node or PoP Point of Presence (Point of Presence) of the Service Provider , contemplating the Quality of Service (QoS: Quality of Service). MUSE uses and extends TONIC in this regard, and recovers the global economic analysis based on Cash Flow, considering, therefore, income, investments (CapEx), and operating costs (OpEx) [29]. The MUSE model is shown in Figure 2.4. As can be seen, its structure is practically the same as that of the TONIC and ECOSYS models, except that it does not expressly include risk analysis.

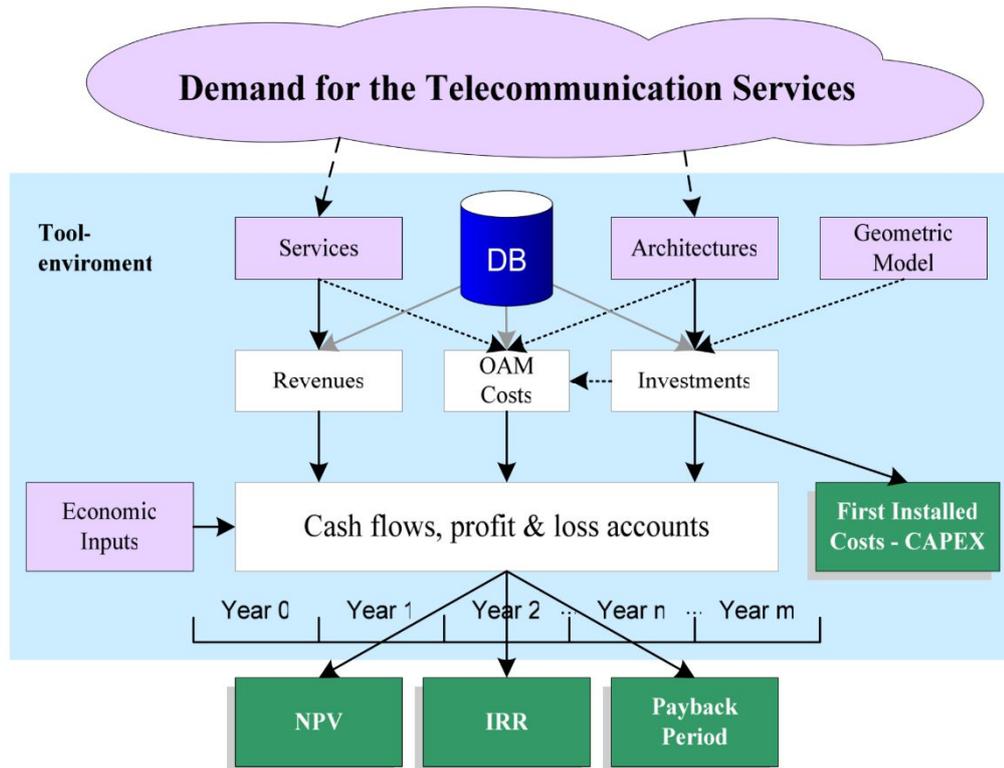

*Figure 2.8: MUSE [29] techno-economic evaluation model.*

- (2007-2013) OASE (Optical Access Seamless Evolution: Unlimited Evolution of Optical Access). The OASE project uses a multidisciplinary approach integrating European operators, FTTH technology manufacturers and European universities in order to provide a set of technological solutions in the field of FTTH access, within the FP-7 (Seventh Framework Program: 7th Framework Program)[111] [102] [108] [109]. At OASE, it was decided to use TONIC as a framework tool, to which other demand models, topology, architecture, dimensioning, evaluation and operation were added, either created by project partners, or already existing and created by third parties. The facility to add new business models and migration tools is also incorporated. All this in order to take advantage of external 'know-how', as shown in Figure 2.9 and Figure 2.10. As mentioned in section 2.2.1, the OASE project proposes a specific methodology for technical-economic evaluation oriented to the deployment of networks called Scope, Model, Evaluate, Refine (Visualize, Model, Evaluate, Refine), which is shown in Figure 2.11.

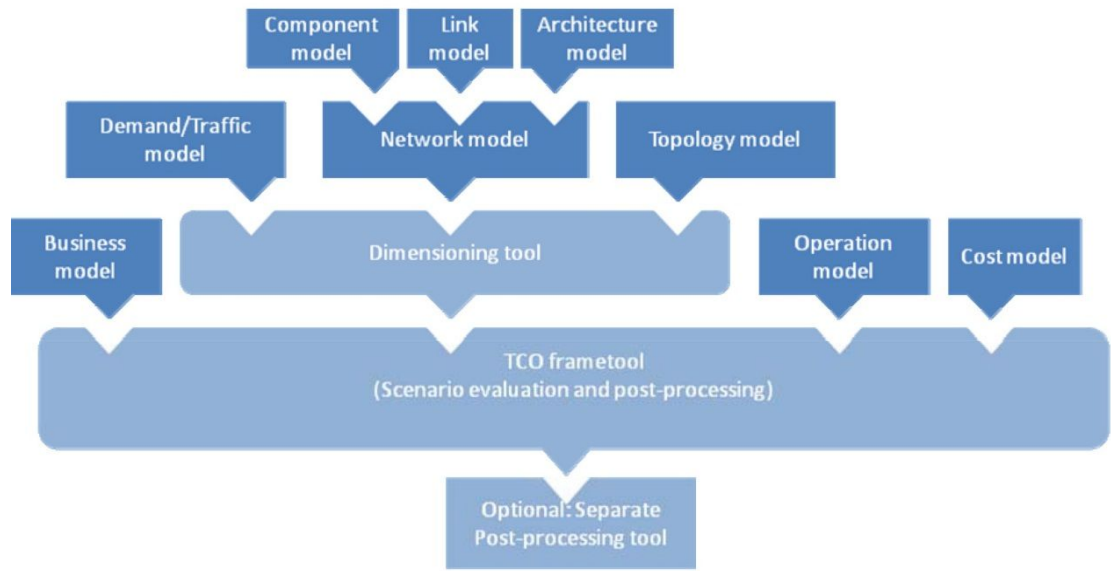

*Figure 2.9: OASE [63] framework scheme.*

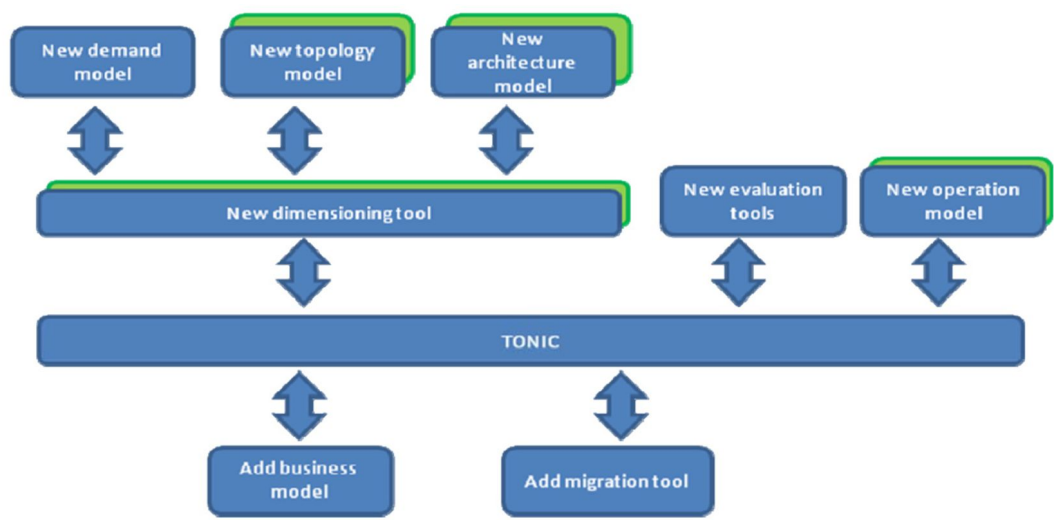

*Figure 2.10: Example of TCO analysis using TONIC as a frame-tool in the OASE project. The tools and methods developed by different partners are shown for some cases (in green) [63].*

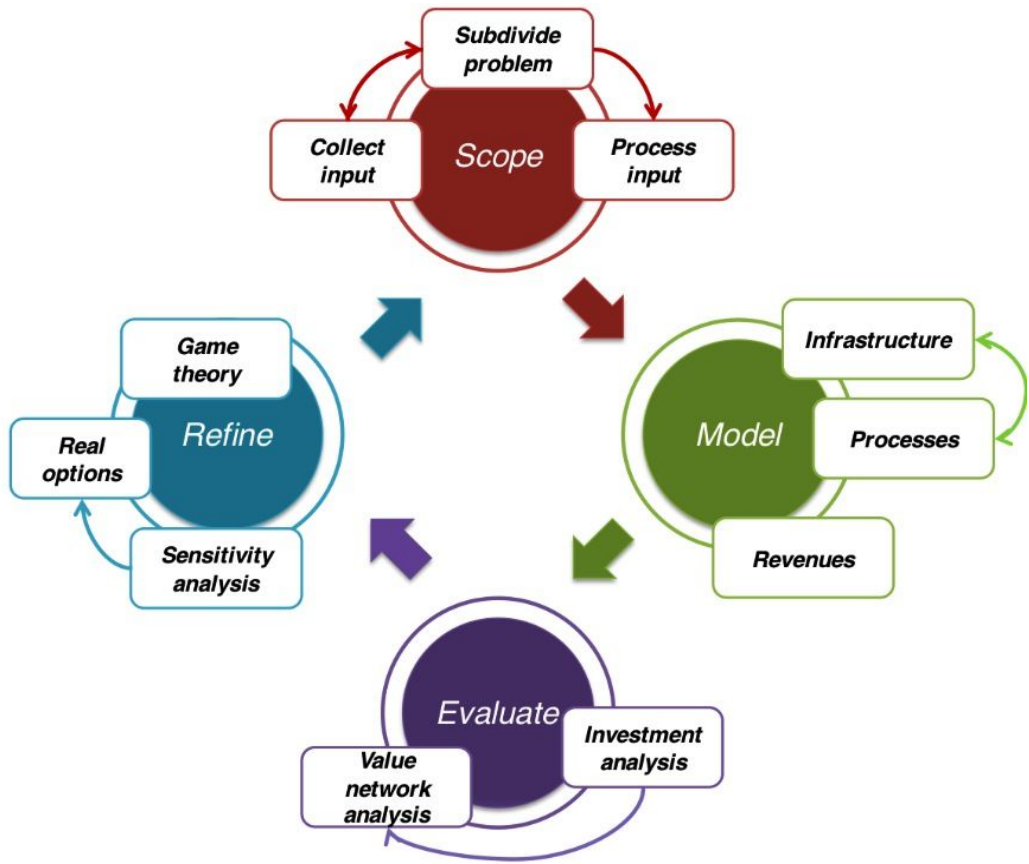

*Figure 2.11: OASE methodology of technical-economic evaluation for the deployment of networks [59].*

- (2008-2011) BONE (Building the future Optical Network in Europe: Building the future Optical Network in Europe) [72] [47] [54] [70] [71] [86]. This project, which represents an evolution of the e-Photon / One [47] project, aims to exchange and consolidate the latest research and developments in access systems that use optical technology to provide broadband connections to fixed and mobile users, establishing a platform that allows the comparison of the different optical access technologies in order to provide recommendations for the deployment of the most optimal in the different European scenarios. The novelty, as can be seen, in Figure 2.8 is the introduction of a performance analysis of the optical access / metro network, which is not finally developed,

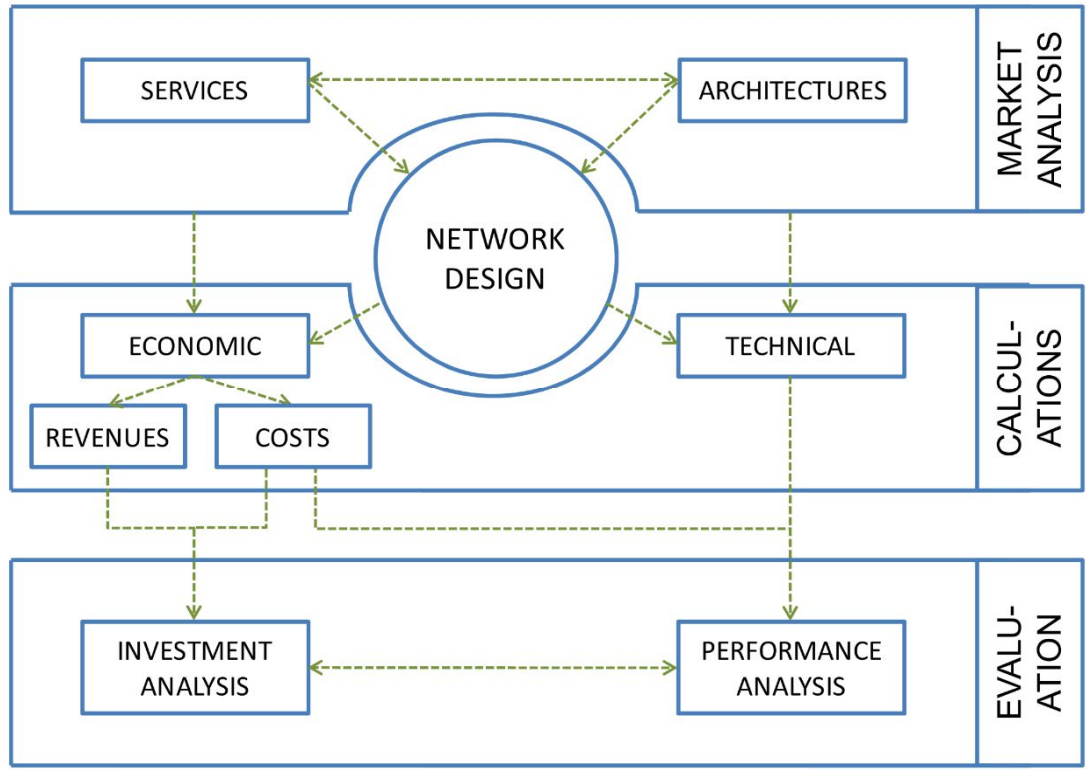

*Figure 2.12: Flowchart of technical-economic evaluation in BONE [68].*

- (2010-2011) CELTIC-AWARE (Aggregation of WLAN Access Resources: Aggregation of Access Resources of Wireless Local Area Network). The objective of this project is to add private WLANs and allow agents in the telecommunications sector to exploit this aggregation. AWARE intends that mobile broadband networks can take advantage of said aggregation in urban environments, encouraging different agents in the telecommunications sector to deploy their services on this new network infrastructure [62] [88].
- The ICON TERRAIN project uses the OASE [77] tool and methodology. In [77], it is stated verbatim: "'Techno-economic' research complements technological evolution and products with an economic analysis in order to indicate different equilibrium points and provide recommendations", in line with the traditional concept of technical modeling -economic of Smura [99].
- (2010-2013) ACCORDANCE. Within the 7th Framework Program of the European Union (EU FP-7), the STREP ACCORDANCE project investigates a new paradigm for the access network: The introduction of OFDMA technology (Orthogonal Frequency Division Multiple Access: Multiple Access by Orthogonal Frequency Division ) in PON networks (Passive Optical Network: Passive Optical Network), simultaneously offering optical access for wireless networks and copper networks [90], [61].
- (2012-2015) CONTENT (Convergenge of Wireless Optical Network and IT REsources IN Support of Cloud Services: Convergence of Optical and Wireless Networks and Information Technology Resources to support Cloud Services). The ICT-CONTENT project within the 7th European Research Framework Program (EU FP-7) addresses the development of a new generation convergent and ubiquitous network infrastructure model, based on the Infrastructure as a Service (IaaS: Infrastructure as a Service), in order to provide a technological platform interconnecting geographically distributed computing resources that can support Cloud Services for fixed and mobile users [93].

- (2013-2016) COMBO (COnvergence of fixed and Mobile BrOadband access / aggregation networks: Convergence of fixed and mobile broadband access and aggregation networks). It is an integrated project with funds from the 7th European Framework Program (EU FP-7), which investigates the convergence of fixed and mobile broadband access and aggregation networks (FMC: Fixed / Mobile Converged) (FMC) in different settings (dense urban, urban, rural). The convergent architectures proposed by the COMBO project are based on the innovative concept of the Next Generation Point of Presence (NG-POP) [104].

- (2013-2016) ICT-DISCUS (The DIStributed Core for unlimited bandwidth supply for all Users and Services: The Distributed Core for unlimited supply of bandwidth for all Users and Services). This project, within the 7th European Framework Program (EU FP-7), aims to define an end-to-end architecture for a future-oriented optical network that is economically viable, energy efficient, and sustainable, developing the opportunity offered by the LR -PONs (Long Reach Passive Optical Networks: Long Range Passive Optical Networks) and flat optical trunk networks, eliminating the traditional limits in the optical network relative to metropolitan, regional, trunk and access areas [105] [127] [123 ] [126].

- (2013-2016) UNIFY (Unifying Cloud and Carrier Networks: Unifying the Cloud and Networks). The ICT-UNIFY project, financed in the 7th European Framework Program (EU FP-7), has as its objective the complete virtualization of the network and services in order to provide flexibility and agility in the creation of services on the network, gaining efficiency operational. The UNIFY consortium researches and develops means to orchestrate the provision of end-to-end services over a virtualized network infrastructure that integrates physical networks and Data Centers, proposing a universal node architecture. Among its objectives is to generate a specific techno-economic model on the possible future universal node architecture, which is expected to see the light after the completion of the project scheduled for April 2016 [117].

- (2009-2017) CELTIC 4GBB (4th Generation Broadband Systems: 4th Generation Broadband Systems). It is a trilogy of projects, the first 4GBB project (2009-2012) aimed to conceive a state-of-the-art DSL technology that would occupy the space between VDSL2 and the long-term FTTH scenario. Comprised of European manufacturers and universities, it developed advanced cable models, high-performance broadband transmission techniques, as well as multi-channel communication and resource management methods, and initiated the creation of the G.fast standard approved by the ITU: International Union Telecommunications (ITU: International Telecommunications Union) in 2014 from the second HFCC / G.fast project (2013-2014). The G.fast standard corresponds to the FTTdp concept (Fiber To The Distribution Point: Fiber to the Distribution Point), and allows speeds between 150 Mbits / s and 1 Gbit / s depending on the length of the copper pair, for user loops of length equal to or less than 250 meters [51] [121]. The third project called GOLD (2015-2017) (Gigabits over the legacy drop: Gigabits over the old loop) is currently underway with the aim of improving standards to open a potential mass market for G.fast, promoting its use in dense urban areas, even replacing the fiber backhaul of the first G.fast standard with copper and increasing the maximum speed above 1 Gigabit / s [130] [57]. The 4GBB project develops a specific model for FTTdp whose diagram is shown in Figure 2.9. The output of the model is a single economic parameter called DNPV (Delta Net Present Value: Delta Net Present Value), which expresses the difference in economic value added (EBIT: Earnings Before Interest and Taxes, ie: Earnings before payment of Interest and Taxes) between two chosen deployment alternatives, such as: deploying FTTH or deploying FTTdp . The 4GBB model uses 5 sub-models, 1) to calculate the investment costs (CapEx) of the network, 2) the connections combining speed of deployment and conversion rate from homes passed to connected homes, 3) the annualized customers by technology, 4 ) operating costs (OpEx), and 5) additional revenues per extra customer per year [113]. such as: deploy FTTH or deploy FTTdp. The 4GBB model uses 5 sub-models, 1) to calculate the investment costs (CapEx) of the network, 2) the connections combining speed of deployment and conversion rate from homes passed to connected homes, 3) the annualized customers by technology, 4 ) operating costs (OpEx), and 5) additional revenues per extra customer per year [113]. such as: deploy FTTH or deploy FTTdp. The 4GBB model uses 5 sub-models, 1) to calculate the investment costs (CapEx) of the network, 2) the connections combining speed of deployment and conversion rate from homes passed to connected homes, 3) the annualized customers by technology, 4 ) operating costs (OpEx), and 5) additional revenues per extra customer per year [113].

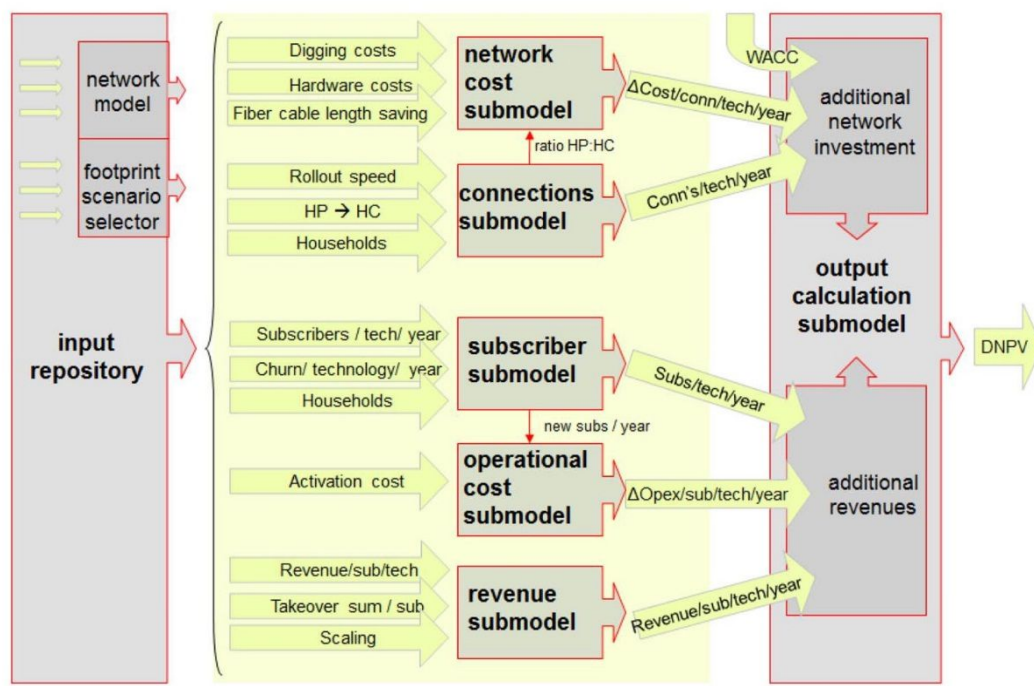

*Figure 2.13: Diagram of the technical-economic model 4GBB [113].*

- (2014-2020) Initiative 5G-PPP (The 5G Infrastructure Public Private Partnership: The Public Private Consortium of 5G Infrastructure) within the 8th Framework program of the European Union called Horizon 2020 for the development of 5G networks. This Consortium promoted by the European Commission, equipment manufacturers, telecommunications operators, service providers, SMEs and researchers, will provide solutions, architectures, technologies and standards for the next ubiquitous generation of communications infrastructures of the next decade, with the following objectives [118]:

  o *"To provide wireless networks with a 1000 times higher capacity and more varied service capacities, compared to those existing in 2010.*
  o *Save up to 90% of energy per service provided, with the main focus on reducing consumption in the mobile communications radio access network.*
  o *Reduce the average service creation cycle from 90 hours to 90 minutes.*
  o *Create a safe and reliable Internet with "zero" perceived downtime during service provision.*
  o *Facilitate very dense deployments of wireless communications links to connect more than 7 trillion devices wireless networks that will serve more than 7 billion people.*
  o *Ensuring for anyone and anywhere access to a greater range of services and applications at a lower cost. "*

  This initiative will integrate successive waves of projects. The use or development of techno-economic models has not yet been identified, since the first projects are still in their initial stages. The mention is included so that the reader is aware of this initiative, which also partially uses public funding.

All the projects identified with public funding are European, without any projects with public funding being identified on other continents. References to the literature from other continents, on the other hand minority, come from private companies and some universities, as can be seen in Table 2.1 [3], [7], [53] [112].

According to the above, and in light of the number of research projects in this area, with funding from European Union institutions, there is an economic interest and public funding, which justifies continuing to deepen the development of technical evaluation models. economical for access technologies.

It is observed that, despite having developed technical-economic models for access technologies in the hands of projects with public funding, to date no universal model has been identified that allows the comparison of any access technologies in any configuration or combination; that it is aimed at any agent in the telecommunications sector; that allows the evaluation and comparison of technical and not only economic feasibility, and that is flexible, extensible and integrable with other techno-economic models.

## 2.3 Characteristics of a Universal Technical-Economic Model

The characteristics that are considered to have a universal and generalizable theoretical model for the technical-economic evaluation of access technologies are set out below. The definition of these characteristics is based on an in-depth study of the State of the Art, complemented by the author's professional experience as a telecommunications engineer in the design of innovative solutions in the field of the access network for different agents in the telecommunications sector:

- **Multiaccess universality**: it has to allow to compare different current and future access technologies.
- **Universality in Combination of technologies**: it must allow the technical and economic evaluation of accesses made up of a combination of different technologies.
- **Universality in the User Orientation of technology:** aimed at both telecommunications operators and clients and end users of telecommunications services, as well as any other agent in the telecommunications market, such as: the Regulator, Communications Service Providers (CSPs: Communication Service Providers ).
- **Universality in the incorporation of "micro" and "macro" approaches:** It must incorporate the "micro" ('bottom-up') (from the client or end user perspective) and "macro" ('top-down') approaches (from the deployment perspective), when evaluating techno -economically access technologies.
- **Guidance on Model User Requirements**: It must consider the requirements of the user of the model, be it a client, operator or any other agent in order to assess the access technologies. This characteristic refers to the application methodology of the model.
- **Geographic universality:** It must allow its application in any geographical area or geotype, whatever its population density, population segments (households, companies) and its distribution.
- **Universality** Technical and Economic: it has to provide technical and economic input and output parameters, in order to allow both a technical and economic assessment of the different access technologies.
- **Extensibility and flexibility:** the model must be extensible and flexible. It has to provide facility to add new input and output parameters contributing to its universality.
- **Technical and Economic Comparability**: it has to provide facility to compare both its technical results and its economic results with other models.
- **Predictive Capacity**: it must allow incorporating and making predictions in a given period of time.
- **Integration capacity with other models**: it must allow integration with other technical-economic models to favor an assessment as complete as possible and facilitate the evolution of Art.

## 2.4 Classification and analysis of technical-economic models for access networks

Next, we proceed to classify a selection of technical-economic models for access networks, identified in the literature, based on the characteristics presented for a universal and generalizable technical-economic model.

A sample of 14 articles has been chosen over the 40 shown in Table 2.1. The models object of the classification have been selected, choosing 7 articles corresponding to project models with public funding from the EU, and 7 corresponding to proprietary models (for which no public funding from the EU has been identified).

### 2.4.1 Multiaccess universality

The Multiaccess Universality feature refers to the model's ability to compare different types of access technologies, whether they are current: fixed, wireless, mixed or hybrid access technologies, or future, such as NFV (Network Function Virtualization) / SDN virtualized accesses. (Software Defined Networking), or any other type of future access technology.

In the literature we can distinguish (See Table 2.2):
- Models comparing only fixed access technologies.
- Models comparing only wireless access technologies.
- Models comparing fixed and wireless access technologies.
- Models comparing fixed, wireless and mixed access technologies: hybrid networks composed of the series combination of fixed access technology and wireless access technology (FiWi: Fixed & Wireless)

It should be noted that, in the literature review, no mention has been identified of hybrid networks composed of a parallel combination of fixed and wireless accesses, or techno-economic models that evaluate virtualized accesses, although with respect to the latter it is understood that the Results of the UNIFY project, which will end in April 2016, will provide literature on the matter.

### 2.4.1.1 Models comparing only fixed access technologies

Models [14], [56], [85], [100] have the ability to compare only fixed access technologies but not wireless or hybrid access technologies composed of fixed access technology and wireless access technology. Models [14] and [29] mention a wider range of fixed access technologies, compared to [56] which is limited to FTTH. [85] is limited to the comparison between HFC / DOCSIS and FTTH-GPON. The model [100] focuses on the comparison of FTTH technology deployments distinguishing between PON, ASN (Active Star Network) and HRN (Home Run Network).

[29] includes fixed hybrid access networkxDSL over Optics,consisting of the extension of the xDSL access network through a fiber optic link between the DSL NT (Network Termination: Network Termination) of the exchange and the remotely

located DSLAM multiplexer. It is therefore the series combination of xDSL and FTTCab (Fiber To The Cabinet: Fiber To The Node or Remote Cabinet) technologies.

## 2.4.1.2 Models comparing only wireless access technologies

From the analysis of the literature, [31] presents a technical-economic analysis of WiMAX technology based on the TONIC and ECOSYS models [34].

### 2.4.1.3 Models comparing fixed and wireless access technologies

In the literature, [19] presents a technical-economic evaluation of implementation models of fixed access technologies: ISDN, xDSL, HFC, FTTx and wireless access technologies: WLL, Satellite. The ECOSYS [34] model also provides such capability.[76] analyzes and compares ADSL and WiMAX access technologies for the implementation of Rural Broadband and identifies as future work the expansion of the techno-economic model to analyze the mobile access network. [74] presents the COSTA model that allows the technical-economic analysis for the deployment of Next Generation Access Networks (NGAN) and provides the comparison between different deployment alternatives: FTTN / VDSL, FTTH / GPON, FTTH / P2P, HFC / DOCSIS 3.0, WiMAX and LTE. Finally, [79] provides a comparison of the deployment of Next Generation Access Networks (NGAN) in rural surroundings, contemplating FTTH / GPON, FTTC / FTTB / VDSL, HFC / DOCSIS 3.0 and LTE (4G).

### 2.4.1.4 Models comparing fixed, wireless and mixed / hybrid access technologies

[58] distinguishes between static and nomadic layers depending on whether the location of the users is fixed or they are in mobility, providing techno-economic analysis of the FTTH (PON), xDSL, HFC and PLC technologies for the static layer and WiMAX for the nomadic layer, allowing the evaluation of mixed or hybrid access technologies composed of a series combination of different fixed and wireless technology (FiWi: Fixed & Wireless). The author Pereira in [42] includes not only WiMAX, but also Satellite and FWA (Fixed Wireless Access) accesses for the nomadic layer, preserving the ability to evaluate hybrid networks made up of a combination of fixed and wireless access technologies. [112] focuses on the analysis of hybrid fixed-wireless networks contemplating xDSL technologies in the static layer,

| Multiaccess universality | | | |
|---|---|---|---|
| | Technologies Fixed access | Wireless access technologies | Access technology Mixed (Hybrid) |
| Olsen et al. [14]. TITAN (1996) | ADSL, PON, CATV, ISDN, FTTx, HFC | | |
| **Jankovich et al. [19]. EURESCOM (2000)** | ISDN, xDSL, HFC, FTTx | WLL, Satellite | |
| **Smura [31]. WiMAX only. TONIC & ECOSYS (2005)** | | WiMAX | |
| **Olsen et al. ECOSYS [34] (2006)** | ISDN, B-ISDN (FITL), xDSL, HFC, FTTx | WLL, Satellite, WiMAX | |
| **Monath et al. [29]. MUSE (2005)** | FTTx, ADSL, SHDSL, VDSL | | xDSL over Optics |
| **Pereira & Ferreira [58] (2009)** | FTTH (PON), xDSL, HFC, PLC | WiMAX | Static Layer and Nomadic Layer with WiMAX |
| **Van der Merwe et al. [56]. FTTH only (2009)** | GPON, AON / Active Ethernet (AE), P2P Optical Networks | | |
| **Zagar et al. [76] (Rural Broadband in Croatia) (2010)** | ADSL | WiMAX | |
| **Pereira [42] (2007)** | xDSL, HFC, FTTH, PLC | WiMAX, Satellite, Fixed Wireless Access (FWA) | Not specifically mentioned but possible |
| **Vergara et al. [74]. COSTA model (2010)** | FTTH / GPON, FTTN / VDSL, FTTH / P2P, HFC / Docsis, | WiMAX, LTE | |
| **Feijoo et al. [79]. RURAL (2011)** | FTTH / GPON, FTTN / VDSL, HFC / Docsis | LTE (4G) | |
| **Martin et al. [85]. HFC only (2011)** | HFC / DOCSIS | | |
| **Van der Wee et al. [100]. Only FTTH. OASE (2012)** | FTTH / PON, FTTH / ASN (Active Star Network), FTTH / HRN (Home Run Network) | | |
| **Pecur [112] FiWi (2013)** | xDSL, FTTx, FSO | WiFi, WiMAX, LTE (4G) | FiWi - Series Combination |

*Table 2.2: Analysis of the literature based on the Multiaccess Universality characteristic.*

## 2.4.2 Universality in Combination of Access Technologies

The characteristic Universality in Combination of Access Technologies refers to the capacity of the model to allow the technical and economic evaluation of accesses made up of a series or parallel combination of different technologies or even the same technology.

In the literature, it can be distinguished (See Table 2.3):
- Models that allow combination of series of fixed access technologies
- Models that allow a combination of series of fixed and wireless access technologies
- Models that allow a parallel combination of different access technologies

No model is identified in the literature that allows a parallel combination of accesses of the same technology, or identical redundant accesses, in order to increase benefits such as bandwidth and availability for the customer or end user.

### 2.4.2.1 Combination series of fixed access technologies

In the literature, [14] evaluates serial combinations of different fixed access technologies such as FTTB PON for POTs / N-ISDN. [29] includes series combination of fixed access technologies: xDSL over Optics. The COSTA [74] model allows the technical-economic evaluation of the FTTN / VDSL technology composed of a series combination of fiber access network and VDSL access. [19], [34], [79] also allow the technical-economic evaluation of serial combinations in case of FTTx access technologies other than FTTH (FTTN / VDSL, FTTB / VDSL, FTTC / VDSL).

### 2.4.2.2 Combination series of fixed and wireless access technologies.

[58] presents a model specifically prepared for a combination of series of fixed and wireless access technologies, distinguishing between the static layer of fixed technologies (xDSL, FTTH, HFC, PLC) and the nomadic layer with WiMAX wireless technology. [112] presents a techno-economic analysis of a hybrid access network made up of a combination of a series of fixed technologies (xDSL, FTTx, FSO) and wireless (WiFi, WiMAX, LTE).

### 2.4.2.3 Parallel combination of different access technologies

[14] contemplates the parallel combination of HFC / TPON fixed access technologies consisting of HFC access and Telephony over PON. This parallel combination is not intended as a backup access for the same end user, but rather as a mixed deployment for a CATV cable operator, given the orientation of the model to the deployment of access technologies by the operators. This is an exception in the literature, since the rest of the models do not contemplate any combination of parallel access technologies.

| Universality in Combination of access technologies | | | |
|---|---|---|---|
| | Fixed technology series combination | Series combination of fixed and wireless technologies | Parallel combinations of different technology |
| Olsen et al. [14]. TITAN (1996) | FTTB PON for POTS, N-ISDN | | HFC / TPON (Telephony over PON) |
| Jankovich et al. [19]. EURESCOM (2000) | FTTC / VDSL | | |
| Smura [31]. WiMAX only. TONIC & ECOSYS (2005) | | | |
| Olsen et al. ECOSYS [34] (2006) | FTTC / VDSL | | |
| Monath et al. [29]. MUSE (2005) | xDSL over optics | | |
| Pereira & Ferreira [58] (2009) | | Static Layer (xDSL, FTTH, HFC, PLC) + Nomadic Layer (WiMAX) | |
| Van der Merwe et al. [56]. FTTH only (2009) | | | |
| Zagar et al. [76] (Rural Broadband in Croatia) (2010) | | | |
| Pereira [42] (2007) | | | |
| Vergara et al. [74]. COSTA model (2010) | FTTN / VDSL | | |
| Feijoo et al. [79]. RURAL (2011) | FTTN / VDSL | | |
| Martin et al. [85]. HFC only (2011) | | | |
| Van der Wee et al. [100]. Only FTTH. OASE (2012) | | | |
| Pecur [112] FiWi (2013) | | Fixed-Wireless. Fixed (xDSL, FTTx, FSO). Wireless (WiFi, WiMAX, LTE) | |

*Table 2.3: Analysis of the literature based on the Universality characteristic in Combination of access technologies*

## 2.4.3 Universality in User Orientation

The characteristic Universality in User Orientation refers to the fact that a universal and generalizable model must be oriented both to telecommunications operators and to clients and end users of telecommunications services, as well as to any other agent in the telecommunications market such as , for example, the Regulator or Communication Service Providers (CSPs).

In the literature, the following are distinguished (See table 2.4):
- Models oriented only to operators
- Models oriented to operators and customers / end users.
- Models oriented to operators and other agents.

### 2.4.3.1 Models oriented only to operators

In the literature, [14], [19], [31], [34], [29], [58], [56], [76], [42], [74], [79], [85 ], [100], [112] are oriented to the deployment of access technologies by telecommunication operators. This orientation occurs naturally, given the impact on the operators' income statement due to the high volume of CapEx investment and OpEx operating expenses required by access networks (the aforementioned 'last mile' problem).

### 2.4.3.2 Models oriented to operators and customers / end users

Only [42] and [58] contemplate as general input parameters the average bandwidth required by the user in transmission and reception, in addition to the operator's perspective in the deployment of access networks. [42] also contemplates QoS and peak-hour concurrency factor for end users. [79] provides cost results for different Guaranteed Data Rate per User values.

### 2.4.3.3 Models oriented to operators and other agents

[74] considers the regulatory authorities in the European Union in addition to the perspective of the operators regarding the deployment of access networks. [112] adds an express orientation to Investors / Lenders in order to compare the different capital strategies, without forgetting the perspective of telecommunications operators in the deployment of access networks.

| Universality in user orientation | | | |
|---|---|---|---|
| | **Operator Oriented (deployment KPIs)** | **Customer Oriented (KPIs of use)** | **Oriented to other agents** |
| **Olsen et al. [14]. TITAN (1996)** | YES | DO NOT | DO NOT |
| **Jankovich et al. [19]. EURESCOM (2000)** | YES | DO NOT | DO NOT |
| **Smura [31]. WiMAX only. TONIC & ECOSYS (2005)** | YES | DO NOT | DO NOT |
| **Olsen et al. ECOSYS [34] (2006)** | YES | DO NOT | DO NOT |
| **Monath et al. [29]. MUSE (2005)** | YES | DO NOT | DO NOT |
| **Pereira & Ferreira [58] (2009)** | YES | Required Avg. Bandwidth (Upstream and Downstream) | DO NOT |
| **Van der Merwe et al. [56]. FTTH only (2009)** | YES | DO NOT | DO NOT |
| **Zagar et al. [76] (Rural Broadband in Croatia) (2010)** | YES | DO NOT | DO NOT |
| **Pereira [42] (2007)** | YES | Required Avg. Bandwidth (Upstream and Downstream), QoS, Concurrency factor | DO NOT |
| **Vergara et al. [74]. COSTA model (2010)** | YES | DO NOT | EU Regulation Authorities |
| **Feijoo et al. [79]. RURAL (2011)** | YES | Guaranteed Data Rate per User | DO NOT |
| **Martin et al. [85]. HFC only (2011)** | YES | DO NOT | DO NOT |
| **Van der Wee et al. [100]. Only FTTH. OASE (2012)** | YES | DO NOT | DO NOT |
| **Pecur [112] FiWi (2013)** | YES | DO NOT | Investors / Lenders |

*Table 2.4: Analysis of the literature based on the Universality characteristic in user orientation.*

## 2.4.4 Universality in the incorporation of "micro" and "macro" approaches

The author Economides states in "The Economics of Networks" [11]: "We distinguish between results that do not depend on the underlying industrial microstructure (" macro "approach) and results that do depend (" micro "approach)".

Within the framework of this research, the "micro" approach or approach is called that which starts from the particular to the general ('bottom-up') from the perspective of the end user, and the "macro" approach or approach to that which starts from the general to the particular ('top-down'), from the perspective of the deployment of an access network by the operator, based on the characteristics of the geographic area to be covered.

It can be said that, when contemplating the perspective of the client or end user of the technology, a "micro" ('bottom-up') approach or approach is required in the technical-economic analysis of access technologies, depending on the results of the microstructure of the access network, as established in the previous statement by Economides [11] [105].

On the other hand, the perspective of the operators in order to evaluate the economic viability of the deployment of access technology in a specific geographical area, is carried out starting from the general to the particular ('top-down'), that is, , with a "macro" approach or approach, considering external aspects to the microstructure of the access network, such as the geographic area to be covered, but which in the case of access networks, does condition the results, contradicting the aforementioned statement by Economides [11] [105]

It is verified in the analysis of literature models that all the techno-economic evaluation models of access technologies incorporate the "macro" approach, suffering from the "micro" approach from the perspective of the customer or end user. In the literature, the term 'bottom-up' approach is mentioned for detailed cost analyzes from the perspective of the components that make up the access, but without incorporating the perspective of the end user [1] [34].

## 2.4.5 Orientation to User Requirements of the model

A universal and generalizable technical-economic model must take into account the Technical and Economic Requirements of the User of the model, be it a client, operator, regulator or any other agent, in order to assess access technologies technically and economically.

The literature review makes it possible to distinguish (See Table 2.5):

- Models that incorporate Technical Requirements of the model user
- Models that incorporate Economic Requirements of the model user

### 2.4.5.1 Models that incorporate Technical Requirements of the model user

In the literature, only [42] and [58] contemplate as input requirements the average bandwidth required by the user in transmission and reception. In [42] it also contemplates the quality of service QoS. [79] contemplates the Guaranteed Data Rate per User (the Guaranteed Data Rate per User).

### 2.4.5.2 Models that incorporate Economic Requirements of the model user

Economic requirements of the model user are not specified in the literature. It can be understood that all models try to maximize income and minimize OpEx costs and CapEx investment, maximizing Net Present Value (NPV) and IRR (Internal Rate of Return)allowing to use this criterion as a comparison between models.

| Orientation to User Requirements of the model |
|---|

|  | Model User Requirements (Economic Ranges) | Model User Requirements (Technical Ranges) |
|---|---|---|
| **Olsen et al. [14]. TITAN (1996)** | Do not | Do not |
| **Jankovich et al. [19]. EURESCOM (2000)** | Do not | Do not |
| **Smura [31]. WiMAX only. TONIC & ECOSYS (2005)** | Do not | Do not |
| **Olsen et al. ECOSYS [34] (2006)** | Do not | Do not |
| **Monath et al. [29]. MUSE (2005)** | Do not | Do not |
| **Pereira & Ferreira [58] (2009)** | Do not | Only in BW |
| **Van der Merwe et al. [56]. FTTH only (2009)** | Do not | Do not |
| **Zagar et al. [76] (Rural Broadband in Croatia) (2010)** | Do not | Do not |
| **Pereira [42] (2007)** | Do not | BW, QoS |
| **Vergara et al. [74]. COSTA model (2010)** | Do not | Do not |
| **Feijoo et al. [79]. RURAL (2011)** | Do not | Guaranteed Data Rate per User |
| **Martin et al. [85]. HFC only (2011)** | Do not | Do not |
| **Van der Wee et al. [100]. Only FTTH. OASE (2012)** | Do not | Do not |
| **Pecur [112] FiWi (2013)** | Do not | Do not |

*Table 2.5: Analysis of the literature based on the User Requirements Orientation characteristic of the model*

## 2.4.6 Geographical universality

A universal and generalizable techno-economic model for the comparison of access technologies must allow its application in any geographical area or geotype regardless of its population density, population segments (households, companies) and its distribution.

As has already been verified in section 2.4.3, all the models are aimed at the deployment of access technologies by telecommunications operators. Hence, all the techno-economic models analyzed include the description of the geographic area to be covered and the description of the situation of the existing infrastructures (pipelines / copper). Most of the models analyzed in the literature also allow the description of the population mix (Residential, SMEs, Large Companies - GGEE -), with the exception of [14] and [19]. [58] and [42] also distinguish the number of mobile users in the coverage area.

## 2.4.7 Technical and Economic Universality

A universal and generalizable techno-economic model for the comparison of access technologies must provide technical and economic universality, that is, it must incorporate technical and economic input parameters and output parameters in order to allow both technical and economic valuation of the access technologies. different access technologies.

For the purposes of the techno-economic evaluation of access technologies, the economic input / output parameters are of interest, and the technical input / output parameters related to the technical performance of the access technology, which are those contemplated in the present study. There are other technical input / output parameters related to marketing or marketing, used by most of the literature models, as a result of their orientation towards the deployment of access networks from the operator's perspective, such as the rate of penetration of an access technology in the market, which are not detailed in this document as they are not the subject of this investigation.

Regarding the economic parameters, a distinction can be made between parameters related to income (eg: fees), expenses (CapEx investment, OpEx operation) and financial (eg: interest rate, inflation, NPV, Cash Flow).

Next, we proceed to show the study of input parameters and the study of output parameters of the literature models.

## 2.4.7.1 Study of models based on input parameters

When carrying out the study of techno-economic models based on the input parameters, it is verified that all the models use economic input parameters, and are distinguished according to whether or not they contemplate technical input parameters (See Table 2.6):

- Models with technical and economic input parameters: we will see that they are more than half of the literature sample, although most use very few technical input parameters, and a wider range of economic input parameters.
- Models with only economic input parameters: which make up the rest of the analyzed sample.

### A- Models with technical and economic input parameters

[31] includes as input parameterstechnicians the frequency band, attenuation and gain of the system, andIt uses as economic input parameters the price of the equipment, the rates of the services, the ARPU, the OAM costs and the interest rate (discount rate).

[34] includes as technical input parameters the services to be provided, the architecture of the access networks, a component database, a radio model for wireless technologies, and a geometric model. [34] incorporates as economic input parameters, the cost of the components, the rates, the OAM costs and the evolution of the costs (OAM Class & Volume Class for Costs Evolution).

[29] works with the same technical input parameters as [34], with the exception of the radio model, as it is applied only to fixed access technologies. As economic input parameters, you use component cost, OAM costs, and rates.

[58] includes as input technical parameters, the components of the access network architecture (internal plant, external plant, wiring - feeder -), a geometric model for wired networks, and service characteristics (bandwidth average ascending and average descending bandwidth), and as economic input parameters, it incorporates the installation fee and the monthly subscription fee for the service as well as the interest rate.

[56] contemplates technical input parameters related to the design of FTTH access technologies: GPON, AON / AE and P2P (length of pipes in the power and distribution section, fiber length in the power and distribution section, no. control panels needed, no. outdoor cabins, no. multi-dwelling units). As economic input parameters, it considers the costs of the components, distinguishing between CapEx and OpEx. He calls them the investment price book (CapEx Price book) and the expense price book (OpEx Price book).

[76] incorporates the distance to the presence point as a technical input parameter. As economic input parameters, it also uses the costs of the components, distinguishing between CapEx and OpEx.

[42], regarding the technical parameters, distinguishes between two layers: LAYER 1 with the definition of the services to be offered (upstream and downstream bandwidth per user, QoS, concurrency during peak hours) and LAYER 2 with the specific input parameters for each access technology (number of modems, number of cabinets, number of ONUs optical network units, number of base stations, cable length, etc.). It incorporates economic input parameters in the so-called CAPA 1, in the definition of services (commercial parameters, installation and monthly fees), as well as the CapEx and OpEx databases of the components.

[112] considers as technical input parameters, bandwidth, traffic mix and backhauling (link to the access node) of the wireless access points / base stations, etc. It includes financial input economic parameters (inflation, investment plans, interest rates, economic prospects, etc.), related to the client (monthly fee, billing cycle, etc.), CapEx (equipment costs, interest rates, etc.), OpEx (salaries, telehousing, truck rolls, support calls, etc.) and others (withdrawal of equipment, outlet of equipment, etc.)

## B-  Models with only economic input parameters

As can be seen in Table 2.6, [14] incorporates as economic input parameters the cost of the components (civil works, wiring, cabinets, electronics, passive components, installation) and the OAM Operation, Administration and Maintenance costs. It also uses the rates as an input parameter to calculate the income from the estimate of the penetration of the services, using the Delphi Method or panel of experts, and multiplying it by the rates.

[19] uses as economic input parameters the price of the network elements, the evolution of prices, the OAM costs, and also considers the income, considering the tariffs of the operators as an input parameter.

On the other hand, [74] incorporates CapEx and OpEx as economic input parameters, noting that it uses public price lists of components for CapEx and that OpEx operating costs are estimated as a percentage of CapEx. It is a cost analysis model that does not include income.

[79] also uses CapEx and OpEx as economic input parameters, as does [85], mentioning as sources the public prices of components or network elements, without considering income.

[100] mentions as input parameters the costs of the components (CapEx), OpEx and the extension of the model incorporating the revenues.

### 2.4.7.2 Study of models based on the output parameters

In the literature review, after studying the models based on the output parameters, it is found (Table 2.6) that all the models in the literature present only economic output parameters, without any of them providing technical output parameters, neither perform any assessment of the technical performance of access technologies.

[14] uses the economic parameters of output Net Present Value (NPV), the internal rate of return (IRR), the cost of first installation (IFC), the Cash Flow (Cashflow), the investment (CapEx) and the cost per connection.

[19], [31], [34], [29], [56], [76], [42], [74], [112] provide only economic output parameters being the Net Present Value (NPV: Net Present Value) the one used by all of them.

As can be seen in Table 2.6, [19] also provides the Cost (Cost), Income (Income) and Profit (Profit). [31] adds to the NPV the operating costs (OpEx), the investment (CapEx), the internal rate of return TIR (IRR: Internal Rate of Return) and the amortization period (Payback Period). [34] adds to the output parameters of [19] and [31], the Life Cycle Cost. [29] adds to the output parameters of [19] and [31], the Installed First Cost.

[58] uses the output economic parameter Net Present Value (NPV) andalso the Internal Rate of Return (IRR), Cost of First Installation (IFC: Installed First Cost), cash balance (Cash balance), investment (CapEx), cost (OpEx) per user and the cost per household passed.

[56] incorporates investment per client and year (CapEx per subscriber and year) and operating costs per client and year (OpEx per subscriber and year), as well as the investment breakdown (CapEx) distinguishing between active network hardware, passive hardware Network and Civil Works and Installation. It also includes the breakdown in operating costs, distinguishing between fixed and variable operating costs.

[76] utiliza como parámetros de salida Valor Neto Presente (NPV), Período de Amortización (Payback Period) y tasa interna de retorno TIR (IRR).

The author Pereira in [42] adds to the economic output parameters of [19], [31] and [56], the average income per user (ARPU: Average Revenue per User), the past cost per household (Cost per home passed), the cost per Mbps, the cost of operation, administration and maintenance (OAM cost), the cost of installation and the cash balance (Cash Balance).

The COSTA [74] model uses the monthly cost per user for different take-up rates, CapEx, CapEx per past household, CapEx per customer, OpEx, internal IRR rate of return (IRR) and NPV for a period time under static circumstances.

[79] provides the economic output parameters: Net Present Value of CapEx investment, OpEx operating costs, per user, per geotype, for a given penetration rate, for a given data flow guaranteed per user and per geotype.
[85] provides the Present Value of the total CapEx per zone, the Present Value of the total Opex per zone, the Present Value of the total CapEx per user, and the Present Value of the total Opex per user.

[100] adds to the Net Present Value, the IRR internal rate of return (IRR), the deployment cost, the cost per customer based onof the flexibility points of the FTTH access network ie: street cabinets, patch cabinets, and the density of customers.

[112] uses the 5-year net profit margin for investors, distinguishing between investing the capital or lending it (investing vs. lending).

| | Technical and Economic Universality | | | |
|---|---|---|---|---|
| | Input Parameters | | Output Parameters | |
| | Technicians | Economical | Technicians | Economical |
| Olsen et al. [14]. TITAN (1996) | | Components cost (civil work, cable, enclosures, electronics, passivecomp., Installation), OAM Costs, Tariffs | | NPV, IRR, IFC, Cashflow, CapEx, Cost per connection |
| Jankovich et al. [19]. EURESCOM (2000) | | Network element prices, price evolution, OA&M Costs, Services revenues and tariffs | | Cost, income, profit, NPV |
| **Smura [31]. WiMAX only. TONIC & ECOSYS (2005)** | Frequency band, Path Loss, System gain | Equipment prices, Service tariffs, ARPU, OAM Costs, Discount rate | | NPV, OpEx, CapEx, IRR, Payback period |
| **Olsen et al. ECOSYS [34] (2006)** | Services, Architecture, Radio model, Components database | Components cost, Tariffs, OAM Class & Volume Class for Costs Evolution | | NPV, Payback period, IRR, Revenues, Cash Flow, Profit, CapEx, OpEx, Life Cycle Cost |
| **Monath et al. [29]. MUSE (2005)** | Services, Architecture, Components database, Geometric model | Components cost, OAM Costs, Tariffs | | NPV, IRR, Payback period, IFC, CapEx, Revenues, Cash flow, Profit |
| **Pereira & Ferreira [58] (2009)** | Access network architecture components (inside plant, outside plant, feeder), Geometric model for feeder networks,, Service characteristics (Avg. Downstream and Upstream bandwidth) | Pricing: one time activation / connection fee (€), subscription fee (€ / month), Discount rate | | Cost per user, Cost per homes passed, Payback period, NPV, IRR, Cash balance, CAPEX, OPEX |
| **Van der Merwe et al. [56]. FTTH only (2009)** | GPON design, AON / AE design, P2P design | CapEx Price book, OpEx Price book | | CapEx, OpEx, CapEx per subscriber and year, OpEx per subscriber and year, CapEx breakup, OpEx breakup |
| **Zagar et al. [76] (Rural Broadband in Croatia) (2010)** | Distance from user to PoP | Component Costs (CapEx, OpEx) | | NPV, Payback Period (PP), IRR |
| **Pereira [42] (2007)** | LAYER 1: Definition of services to be offered (bandwidth in SLA, QoS, Concurrency during peak hour). LAYER2: Specific input parameters for Access Technology | LAYER 1: Definition of services (commercial parameters, Activation and monthly fees), CapEx DB, OpEx DB | | CAPEX; OPEX; Subscriber costs; ARPU: Average Revenue Per User; Cost per subscriber; Cost per home passed; Mbps cost; OAM costs; Installation cost; Net Present Value (NPV); Internal Rate of Return (IRR); Payback Period; Revenues; Investments; Life |

| | | | | Cycle Cost; Cash balance. |
|---|---|---|---|---|
| **Vergara et al. [74]. COSTA model (2010)** | | CapEx public price lists, Component costs (CapEx and OpEx as a% of CapEx) | | Monthly cost per user for different take-up rates, total CAPEX, CAPEX per home passed, CAPEX per home connected, CAPEX per customer, OPEX, Internal Rate of Return (IRR) and Net Present Value for a period of time under static circumstances. |
| **Feijoo et al. [79]. RURAL (2011)** | | Network Element Public Prices | | CapEx, OpEx, per user, by geotype, for a given penetration, Cost for a given guaranteed data rate per user by geotype |
| **Martin et al. [85]. HFC only (2011)** | | Network Element Public Prices | | Present Value of Total CapEx by zone, Present Value of Total OpEx by Zone, Present Value of Total CapEx per user, Present Value of Total Cost per User |
| **Van der Wee et al. [100]. Only FTTH. OASE (2012)** | | Component Costs, Revenues | | NPV, IRR, Cost rollout, Cost per customer as function of flexibility points or customer density |
| **Pecur [112] FiWi (2013)** | back-end Internet bandwidth, traffic mix, backhauling of access points / base stations, etc.), bandwidth per customer | financial (inflation, investment schedules, interest rates, economic outlook, etc.), customer related (subscription charge, billing cycle, etc.), CAPEX (equipment costs, discounts, etc.), OPEX (salaries, telehousing, truck rolls, support calls, etc.), salvage (removal of equipment, end-of-life sale, etc.) | | Net Profit Margin 5 years investing vs. lending |

*Table 2.6: Analysis of the literature based on the technical and economic input and output parameters in order to evaluate the Technical and economic universality characteristic*

## 2.4.8 Extensibility and flexibility

A universal and generalizable technical-economic model must be extensible and flexible in such a way that it provides ease of adding new input and output parameters, contributing to its universality.

The analysis of the literature has been carried out considering:

- Flexibility for new input parameters: In the literature, no model is identified that allows adding new input parameters in a flexible and easy way.
- Flexibility for new output parameters: no models are identified that allow flexible and easy addition of new output parameters.

Since no model provides flexibility in the addition of input and / or output parameters, it is concluded that none is extensible.

## 2.4.9 Technical and economic comparability

A universal and generalizable technical-economic model must be comparable, ie: it must allow the comparison of its technical results and its economic results with other models.

The literature review distinguishes:

- Comparability of technical results: as we already saw in the previous section, [14] provided a penetration rate in the residential market, and [58] and [79] non-comparable sensitivity analysis, with which it is concluded that it is not found in the literature any model that allows the technical results to be compared.
- Comparability of economic results: All the technical-economic models in the literature provide the possibility of comparing the economic results at least based on the output parameter Net Present Value common in most of them. [56] does not include Value as an output parameter Net Present but CapEx and OpEx, the results of which can be compared with other models that also have these output parameters. [112] incorporates the net profit margin (Net Profit Margin) that can also be compared with other models by deducting said margin from income, expenses and investments.

## 2.4.10 Predictive Characteristic

A universal and generalizable technical-economic model must allow incorporating and making predictions in a given period of time.

The analysis of the literature distinguishes (See Table 2.7):

- Models that incorporate a study period as an input parameter.
- Models that allow input parameters with temporal prediction
- Models that perform temporal prediction on output parameters.

### 2.4.10.1 Models that incorporate a study period as an input parameter

In the literature, all the references except for [56], which does not expressly mention it, include a study period as an input parameter for the technical-economic analysis. We will see in the subsequent sections that although [56] does not expressly mention it, it must be considered, since it considers the temporal evolution of prices as an input parameter and incorporates CapEx and OpEx per year as output parameters.

### 2.4.10.2 Models that allow input parameters with temporal prediction

As can be seen in the analysis of the literature (See Table 2.7), [14] incorporates penetration rate prediction based on the Delphi method of expert surveys (panel of experts). [19] also incorporates penetration rate prediction. [31] also adds the evolution of equipment prices. [34] adds OAM cost evolution and equipment prices. [29] includes the evolution of OAM costs and component costs. [58] uses the trend (annual%) of evolution of the characteristics of the geographical area (households, population density, number of residential customers, SMEs, users in mobility) as well as the characteristics of the service (average bandwidth in reception and emission) and the evolution of prices (installation fee and monthly service fee). [56] seems to include price evolution although it does not mention it. [76], [79], [85] and [100] use price evolution. [42] also mentions the evolution of productivity. [74] includes the evolution of prices that it uses as a reference for the calculation of CapEx.

### 2.3.10.3 Models that perform temporal prediction on parameters of exit.

The techno-economic models in the literature carry out temporal prediction in the economic output parameters, the majority coinciding in the prediction of the Net Present Value evaluated in a given period of time (See Table 2.7). [14] also includes prediction of the Network First Installation Cost (IFC: Installed First Cost), Cash Flow and cost per connection. [19] adds cost and revenue prediction. [31] and [34] only predict NPV. [29] adds prediction of revenue, profit, CapEx and OpEx. [58] also includes the Internal Rate of Return (IRR), CapEx, OpEx and the cost per customer. [56] provides CapEx and OpEx but does not mention NPV. [76] includes NPV, IRR and the payback period. [42] provides CapEx, OpEx, costs per customer, ARPU, Cost per household passed, Cost per unit of bandwidth (Mbps), OAM costs, Installation cost, NPV, IRR, payback period, income, investments, Life Cycle Cost and Cash Balance. [74] includes the monthly cost per user for different adoption rates, CapEx, CapEx per past household, CapEx per connected household, CapEx per customer, OpEx, IRR and NPV for a period of time under static circumstances.

[79] uses prediction in output for CapEx, OpEx, per user, per geotype, for a given penetration rate, NPV, cost for a guaranteed bandwidth per user and per geotype. [85] includes present value of total CapEx per zone, present value of total OpEx per zone, per user, and present value of total cost per user. [100] uses NPV, IRR, Deployment cost, Cost per customer based on flex points or customer density. [112] provides the Net Profit Margin.

| Predictive Capacity | | | |
|---|---|---|---|
| | Study period as an input parameter | Allows input parameters with temporal prediction | Perform temporal prediction on output parameters |
| Olsen et al. [14]. TITAN (1996) | YES | Service penetration forecasting based on Delphi survey | NPV, IFC, Cash Flow, Cost per connection |
| Jankovich et al. [19]. EURESCOM (2000) | YES | Service penetration forecasting | Cost, income, profit, NPV |
| Smura [31]. WiMAX only. TONIC & ECOSYS (2005) | YES | Service penetration, Equipment price evolution | NPV |
| Olsen et al. ECOSYS [34] (2006) | YES | Service penetration, OAM Class and Volume Class for Equipment price and OAM Costs Evolution | NPV |
| Monath et al. [29]. MUSE (2005) | YES | Components cost evolution, OAM costs evolution | NPV, Revenues, Profit, CapEx, OpEx |
| Pereira & Ferreira [58] (2009) | YES | Trend (% per year) for Geographical area characteristics (Households, Population, HH / km2, Nr of residential subscribers, SME, Nomadic users) Service characteristics (Avg. Downstream and Upstream bandwidth) and Pricing (one time connection fee and subscription fee) | NPV, IRR, CapEx, OpEx, Cost per subscriber |
| Van der Merwe et al. [56]. FTTH only (2009) | Not mentioned | Price evolution (although not mentioned) | CapEx, OpEx (although not mentioned) |
| Zagar et al. [76] (Rural Broadband in Croatia) (2010) | YES | Price evolution | NPV, IRR, Payback Period |
| Pereira [42] (2007) | YES | Price evolution, Productivity evolution | CAPEX; OPEX; Subscriber costs; ARPU: Average Revenue Per User; Cost per subscriber; Cost per home passed; Mbps cost; OAM costs; Installation cost; Net Present Value (NPV); Internal Rate of Return |
| | | | (IRR); Payback Period; Revenues; Investments; Life Cycle Cost; Cash balance. |
| **Vergara et al. [74]. COSTA model (2010)** | YES | Price Evolution (CapEx) | Monthly cost per user for different take-up rates, total CAPEX, CAPEX per home passed, CAPEX per home connected, CAPEX per customer, OPEX, Internal Rate of Return (IRR) and Net Present Value for a period of time under static circumstances. |
| **Feijoo et al. [79]. RURAL (2011)** | YES | Price Evolution | CapEx, OpEx, per user, by geotype, for a given penetration, NPV, Cost for a given guaranteed data rate per user by geotype |
| **Martin et al. [85]. HFC only (2011)** | YES | Price Evolution | Present Value of Total CapEx by zone, Present Value of Total OpEx by Zone, Present Value of Total CapEx per user, Present Value of Total Cost per User |
| **Van der Wee et al. [100]. Only FTTH. OASE (2012)** | YES | Price evolution | NPV, IRR, Cost rollout, Cost per customer as function of flexibility points or customer density |
| **Pecur [112] FiWi (2013)** | 5 years | DO NOT | Net Profit Margin |

*Table 2.7: Analysis of the literature based on the predictive characteristic.*

## 2.4.11 Integration capacity with other models

A universal and generalizable technical-economic model must allow integration with other technical-economic models to favor synergies between models, which allow to take advantage of the strengths of each one.

In the literature, some models have been identified that integrate the output of another model as input. [31] mentions the input incorporation of a multipath loss model for WiMAX (Path Loss Model). [34] mentions a radio model (Radio model), an OAM Class & Volume Class for Price evolution model, as well as statistics or surveys to estimate the market penetration rate. [29] talks about a geometric model, a component cost evolution model and the evolution of Operation, Administration and Maintenance (OAM) costs. In some cases they are sub-models inherent to the model itself and in other cases they are inheritances from a previous model.

However, no models have been found in the literature whose logic allows any parameters of other models to be incorporated by default.

| Integration capacity with other models | | |
|---|---|---|
| | **It allows to integrate the output of another model as input** | **The model logic allows the incorporation of parameters from other models by default** |
| **Olsen et al. [14]. TITAN (1996)** | DO NOT | DO NOT |
| **Jankovich et al. [19]. EURESCOM (2000)** | DO NOT | DO NOT |
| **Smura [31]. WiMAX only. TONIC & ECOSYS (2005)** | Mention Path Loss Model | DO NOT |
| **Olsen et al. ECOSYS [34] (2006)** | Radio model, OAM Class & Volume Class for price evolution model, Statistics or surveys for market penetration | DO NOT |
| **Monath et al. [29]. MUSE (2005)** | Geometric model, Components cost evolution, OAM costs evolution | DO NOT |
| **Pereira & Ferreira [58] (2009)** | DO NOT | DO NOT |
| **Van der Merwe et al. [56]. FTTH only (2009)** | DO NOT | DO NOT |
| **Zagar et al. [76] (Rural Broadband in Croatia) (2010)** | DO NOT | DO NOT |
| **Pereira [42] (2007)** | DO NOT | DO NOT |

| Vergara et al. [74]. COSTA model (2010) | DO NOT | DO NOT |
|---|---|---|
| Feijoo et al. [79]. RURAL (2011) | DO NOT | DO NOT |
| Martin et al. [85]. HFC only (2011) | DO NOT | DO NOT |
| Van der Wee et al. [100]. Only FTTH. OASE (2012) | DO NOT | DO NOT |
| Pecur [112] FiWi (2013) | DO NOT | DO NOT |

*Table 2.8: Analysis of the literature based on the integration capacity of each model with other models.*

## 2.5 Global assessment and ranking

In order to be able to assess for each model in the literature the degree of compliance with the set of characteristics that is considered to have a universal, scalable, flexible and generalizable technical-economic model, the following is followed method, considering, for simplicity, that all items have the same weight:

- A value of 1 is granted for each column (item) of the aforementioned matrices in which compliance with each model is detected and a value of 0 in case of non-compliance.
- The valuation of each characteristic is calculated as the total sum of the values of the columns (items) of the same.
- The total evaluation for each model results from the sum of evaluations of the total characteristics

The result is shown in Table 2.9. Normalizing for each characteristic in base 100, the degree of compliance of each model from the literature is obtained with respect to the maximum possible score in Table 2.10.

| | Multiaccess universality | Universality in Combination of access technologies | Universality in user orientation | Universality in the incorporation of " micro" and " macro" approaches | Orientation to User Requirements of the model | Geographic universality | Technical and Economic Universality | Extensibility and Flexibility | Technical and economic comparability | Predictive Capacity | Integration capacity with other models | ASSESSMENT |
|---|---|---|---|---|---|---|---|---|---|---|---|---|
| Highest possible score | 3 | 4 | 3 | two | two | 1 | 4 | two | two | 3 | two | 25 |
| Pereira & Ferreira [58] (2009) | 3 | 1 | two | 1 | 1 | 1 | 3 | 0 | 1 | 3 | 0 | 16 |
| Pereira [42] (2007) | 3 | 0 | two | 1 | 1 | 1 | 3 | 0 | 1 | 3 | 0 | fifteen |
| Olsen et al. ECOSYS [34] (2006) | two | 1 | 1 | 1 | 0 | 1 | 3 | 0 | 1 | 3 | 1 | 14 |
| Monath et al. [29]. MUSE (2005) | two | 1 | 1 | 1 | 0 | 1 | 3 | 0 | 1 | 3 | 1 | 14 |
| Feijoo et al. [79]. RURAL (2011) | two | 1 | 1 | 1 | 1 | 1 | two | 0 | 1 | 3 | 0 | 13 |
| Vergara et al. [74]. COSTA model (2010) | two | 1 | two | 1 | 0 | 1 | two | 0 | 1 | 3 | 0 | 13 |
| Olsen et al. [14]. TITAN (1996) | 1 | two | 1 | 1 | 0 | 1 | two | 0 | 1 | 3 | 0 | 12 |
| Jankovich et al. [19]. EURESCOM (2000) | two | 1 | 1 | 1 | 0 | 1 | two | 0 | 1 | 3 | 0 | 12 |
| Smura [31]. WiMAX only. TONIC & ECOSYS (2005) | 1 | 0 | 1 | 1 | 0 | 1 | 3 | 0 | 1 | 3 | 1 | 12 |
| Zagar et al. [76] (Rural Broadband in Croatia) (2010) | two | 0 | 1 | 1 | 0 | 1 | 3 | 0 | 1 | 3 | 0 | 12 |
| Pecur [112] FiWi (2013) | 3 | 1 | 0 | 1 | 0 | 1 | 3 | 0 | 1 | two | 0 | 12 |
| Martin et al. [85]. HFC only (2011) | 1 | 0 | 1 | 1 | 0 | 1 | two | 0 | 1 | 3 | 0 | 10 |
| Van der Wee et al. [100]. Only FTTH. OASE (2012) | 1 | 0 | 1 | 1 | 0 | 1 | two | 0 | 1 | 3 | 0 | 10 |
| Van der Merwe et al. [56]. FTTH only (2009) | 1 | 0 | 1 | 1 | 0 | 1 | 3 | 0 | 1 | two | 0 | 10 |

*Table 2.9: Assessment and ranking of the literature based on the degree of compliance with the characteristics of a universal, generalizable, scalable and flexible technical-economic model.*

| | Multiaccess universality | Universality in Combination of access technologies | Universality in user orientation | Universality in the incorporation of "micro" and "macro" approaches | Orientation to User Requirements of the model | Geographic Universality | Technical and Economic Universality | Extensibility and Flexibility | Technical and economic comparability | Predictive Capacity | Integration capacity with other models | ASSESSMENT | %COMPLIANCE |
|---|---|---|---|---|---|---|---|---|---|---|---|---|---|
| Highest possible score | 100 | 100 | 100 | 100 | 100 | 100 | 100 | 100 | 100 | 100 | 100 | 1100 | 100% |
| Pereira & Ferreira [58] (2009) | 100 | 25 | 67 | fifty | fifty | 100 | 75 | 0 | fifty | 100 | 0 | 617 | 56% |
| Pereira [42] (2007) | 100 | 0 | 67 | fifty | fifty | 100 | 75 | 0 | fifty | 100 | 0 | 592 | 54% |
| Olsen et al. ECOSYS [34] (2006) | 67 | 25 | 3. 4 | fifty | 0 | 100 | 75 | 0 | fifty | 100 | fifty | 551 | fifty% |
| Monath et al. [29]. MUSE (2005) | 67 | 25 | 3. 4 | fifty | 0 | 100 | 75 | 0 | fifty | 100 | fifty | 551 | fifty% |
| Feijoo et al. [79]. RURAL (2011) | 67 | 25 | 3. 4 | fifty | fifty | 100 | fifty | 0 | fifty | 100 | 0 | 526 | 48% |
| Vergara et al. [74]. COSTA model (2010) | 67 | 25 | 67 | fifty | 0 | 100 | fifty | 0 | fifty | 100 | 0 | 509 | 46% |
| Olsen et al. [14]. TITAN (1996) | 3. 4 | fifty | 3. 4 | fifty | 0 | 100 | fifty | 0 | fifty | 100 | 0 | 468 | 43% |
| Jankovich et al. [19]. EURESCOM (2000) | 67 | 25 | 3. 4 | fifty | 0 | 100 | fifty | 0 | fifty | 100 | 0 | 476 | 43% |
| Smura [31]. WiMAX only. TONIC & ECOSYS (2005) | 3. 4 | 0 | 3. 4 | fifty | 0 | 100 | 75 | 0 | fifty | 100 | fifty | 493 | Four. Five% |
| Zagar et al. [76] (Rural Broadband in Croatia) (2010) | 67 | 0 | 3. 4 | fifty | 0 | 100 | 75 | 0 | fifty | 100 | 0 | 476 | 43% |
| Pecur [112] FiWi (2013) | 100 | 25 | 0 | fifty | 0 | 100 | 75 | 0 | fifty | 67 | 0 | 467 | 42% |
| Martin et al. [85]. HFC only (2011) | 3. 4 | 0 | 3. 4 | fifty | 0 | 100 | fifty | 0 | fifty | 100 | 0 | 418 | 38% |
| Van der Wee et al. [100]. Only FTTH. OASE (2012) | 3. 4 | 0 | 3. 4 | fifty | 0 | 100 | fifty | 0 | fifty | 100 | 0 | 418 | 38% |
| Van der Merwe et al. [56]. FTTH only (2009) | 3. 4 | 0 | 3. 4 | fifty | 0 | 100 | 75 | 0 | fifty | 67 | 0 | 410 | 37% |

*Table 2.10: Assessment and ranking of the literature based on the degree of compliance with the characteristics of a universal, generalizable, scalable and flexible technical-economic model (standardized compliance by characteristic based on 100).*

Table 2.10 is ordered from highest to lowest rating, providing the ranking of the literature models from highest to lowest degree of compliance with characteristics.

The maximum score corresponds to [58] with a 56% compliance, thus identifying a gap of 44 points to 100%, which shows the opportunity to deepen and investigate in the development of proposals that reach a higher degree compliance.

The ranking of models presented in Table 2.10 shows, taking as reference the model [58] with the highest degree of compliance, that the improvement path is concentrated on the following characteristics:

- Universality in Combination of access technologies
- Universality in User Orientation
- Universality in the incorporation of "micro" and "macro" approaches
- Orientation to User requirements of the model
- Technical and Economic Universality
- Extensibility and Flexibility
- Technical and Economic Comparability
- Ability to integrate with other models

## 2.6 Conclusions

In this chapter of the State of the Art, based on the main objective of the thesis, it has been presented:

- in section 2.2 a historical evolution of technical-economic models for access technologies, including a review and analysis of the literature and a chronology of projects with public funding from the EU that develop and / or use technical-economic evaluation models.
- in section 2.3. The characteristics of a universal, scalable, flexible and generalizable theoretical technical-economic model for access technologies are established.
- In section 2.4 a classification and analysis of the technical-economic models in the literature is elaborated, based on the characteristics of the universal and generalizable technical-economic model set out in section 2.2.
- In section 2.5 an overall assessment is made and a ranking of the technical-economic models in the literature is presented based on this classification.

After reviewing and classifying the literature, it has been found that all the models are oriented to deployment from the operator's perspective, and none are oriented to the end user, except for some exceptional wink in [58] that includes width as an input parameter. minimum transmission and reception band, and [42] that talks about QoS and concurrency factor. All incorporate the "macro" approach from the deployment perspective, but none incorporate the "micro" approach (end-user perspective).

No model develops and provides technical output parameters, with the exception of BONE [68] which suggests a network performance analysis, but does not develop it; It limits it exclusively to reliability in the field of optical networks. For this reason, the literature models do not allow the evaluation of the technical performance of access technologies, lacking technical comparability, in line with the traditional concept of Smura's technical-economic model [99].

Less than half of the sample models used for classification based on the characteristics of the theoretical model, address a combination of fixed technologies. No model addresses the parallel combination of the same or different access technologies, in order to increase the equivalent technical performance of the access, with the exception of a slight glimpse in [14] with HFC (CATV) in parallel with TPON. None contemplate the combination of fixed technologies + wireless technologies, with the exception of [58] and [112].

No model develops the incorporation of technical and economic requirements by the user of the model. There is some slight hint of incorporation of input information in [58] (minimum transmission and reception bandwidth), in [42] (QoS and concurrency factor) and in [79] that contemplates the Guaranteed Data Rate per User ( the Data Flow Guaranteed by User). However, no model develops this characteristic, incorporating, for example, a catalog or matrix of technical and economic requirements, since they are all oriented to the deployment of access technologies by operators.

No model is identified that allows adding new input or output parameters in a flexible and simple way, which is why it is concluded that they are not flexible or extensible, motivated by the fact that they all focus on the evaluation of economic viability.

No model includes in its logic the default incorporation of parameters from other models, thus limiting its ability to integrate with others. Of course, they all aim to assess economic viability.

Therefore, the review, classification and analysis of the literature presented in this chapter of the State of the Art shows that there is currently a path to developing models that meet the characteristics of a universal, flexible, generalizable and scalable techno-economic model that allows the analysis and comparison of access technologies.

According to the above, it makes sense to deepen and investigate the development of models of technical-economic evaluation of access technologies that achieve a higher degree of global compliance, and in each of the characteristics, thus approaching the universal theoretical technical-economic model and generalizable whose characteristics have been defined.

Page intentionally left blank

# Chapter 5

# VALIDATION

This chapter presents results of the proposed UTEM model in various scenarios, and its qualitative and quantitative validation in order to verify the degree of fulfillment of the double objective of defining a technical-economic model of universal, scalable, flexible and generalizable application that allows the comparison of multiple access technologies in different scenarios, and develop an application methodology for it.

The validation procedure that is followed is as follows:

- Qualitative validation:
    - o Functional validation based on the characteristics of the theoretical technical-economic model exposed in the State of the Art Chapter.
- Quantitative validation:
    - o Quantitative validation in Isolated Application Scenarios of the model
    - o Quantitative validation in Scenarios of Combined Application of the model.
    - o Quantitative validation of the model's predictability.
    - o Quantitative validation of model results with results from other models

Both qualitative and quantitative viewpoints are used in order to perform as complete a validation as possible.

In the qualitative validation, the proposed model is functionally validated with respect to the degree of compliance with the characteristics established for the theoretical technical-economic model in the State of the Art Chapter, comparing it with the model of the literature with a higher degree of compliance with them. as developed in section 5.1 of this Chapter.

The quantitative validation section 5.2 shows the results of the model in 9 scenarios that require the isolated application methodology of the model: ADSL, FTTH, WiMAX, 4G-LTE, FTTH with virtualized router, Dedicated point-to-point line, Redundant access 2 x ADSL, *ADSL in parallel with WiMAX and IEEE 802.11g WiFi access point in aggregate mode (without backup), and VDSL.* The aforementioned scenarios have been selected in order to have a representative sample of current access technologies. The dedicated point-to-point line has been included because it is related to the motivation of this research work [25], and FTTH technology with a virtualized router, since there are already some pilot tests in this regard by some operators. And two scenarios have been added with homogeneous and heterogeneous redundancy, respectively.

A comparative analysis of the 9 access technologies mentioned in 3 cases of technology user requirements is also presented in section 5.2: A - Residential type user with a minimum reception bandwidth of 30 Mbits / s (Data for 2015) according to the objectives of the European Digital Agenda 2020 [128], B - SME type user with a minimum reception bandwidth of 300 Mbits / s ( Data for the year 2015), C - Residential type user with a minimum reception bandwidth of 2 Mbits / s (Data for 2006) according to the objectives of the European Digital Agenda i2010 [67].

In section 5.3 it is shown for every scenario that requires the combined application methodology of the model ("macro" approach), that since the literature models only provide economic output parameters, the technical output results are obtained using the model with the isolated application methodology ("micro" approach), following the steps outlined in the Methodology Chapter.

Likewise, in section 5.4 the quantitative validation of the model's prediction capacity is carried out, contrasting its results with the prediction of the analysis signature Analysis Mason, showing that the saturation period of the evolution of over time for FTTH technology corresponds to decision periods for massive deployment by telecommunications operators in 3 European countries.$F_2$

Regarding the quantitative validation of the results of the UTEM model with results from other models, which is developed in section 5.5 of this Chapter, it should be noted that, since the literature models do not provide technical results but do provide economic results, the economic results of the proposed UTEM model will always be consistent with the economic results of the literature models, since in the economic output parameters, the UTEM model uses a universal formulation - for example, the formulation of the Net Present Value NPV or NPV (Net Present Value) - identical to the formulation of these economic parameters in the literature models. Therefore, equal to the input parameters of Revenue, CapEx and OpEx, the result will be exactly the same,

## 5.1 Qualitative Validation

The qualitative validation is based on the validation of the functionality of the model in order to verify the degree of compliance with the characteristics that were established in the State of the Art chapter for a technical-economic model of universal, scalable, flexible and generalizable application. .

The following table reproduces the top part of the ranking of compliance with the literature models, identifying the five models in the literature with the highest degree of global compliance.

As established in the State of the Art Chapter, all characteristics are considered to have the same weight. Each characteristic is made up of a variable number of items or columns, which is why all the characteristics have been normalized, with 100 being the maximum value for each of them, as can be seen in Table 5.1.

| | Multiaccess universality | Universality in Combination of access technologies | Universality in user orientation | Universality in the incorporation of "micro" and "macro" approaches | Orientation to User Requirements of the model | Geographic Universality | Technical and Economic Universality | Extensibility and Flexibility | Technical and economic comparability | Predictive Capacity | Integration capacity with other models | ASSESSMENT | %COMPLIANCE |
|---|---|---|---|---|---|---|---|---|---|---|---|---|---|
| Highest possible score | 100 | 100 | 100 | 100 | 100 | 100 | 100 | 100 | 100 | 100 | 100 | 1100 | 100% |
| Pereira & Ferreira [58] (2009) | 100 | 25 | 67 | fifty | fifty | 100 | 75 | 0 | fifty | 100 | 0 | 617 | 56% |
| Pereira [42] (2007) | 100 | 0 | 67 | fifty | fifty | 100 | 75 | 0 | fifty | 100 | 0 | 592 | 54% |
| Olsen et al. ECOSYS [34] (2006) | 67 | 25 | 3. 4 | fifty | 0 | 100 | 75 | 0 | fifty | 100 | fifty | 551 | fifty % |
| Monath et al. [29]. MUSE (2005) | 67 | 25 | 3. 4 | fifty | 0 | 100 | 75 | 0 | fifty | 100 | fifty | 551 | fifty % |
| Feijoo et al. [79]. RURAL (2011) | 67 | 25 | 3. 4 | fifty | fifty | 100 | fift y | 0 | fifty | 100 | 0 | 526 | 48% |

*Table 5.1: TOP 5 of the ranking of models in the literature.*

The model [58] is identified as the model with the highest global compliance and by characteristic.

Therefore, the functional validation of the proposed model will be carried out, comparing it with the literature model [58] with the highest degree of global compliance according to the ranking. This model presents a degree of global compliance of 56% with respect to the theoretical model.

## 5.1.1 Multiaccess Universality characteristic validation

Next, the validation of the model with respect to the Multiaccess Universality characteristic is presented together with the model from the literature that shows the highest global compliance.

| Multiaccess universality | | | |
|---|---|---|---|
| | Fixed access technologies | Wireless access technologies | Mixed access technologies (Hybrids) |
| Proposed model (UTEM) | YES (Any) | YES (Any) | YES (Any) |
| Pereira & Ferreira [58] (2009) | FTTH (PON), xDSL, HFC, PLC | WiMAX | Static Layer and Nomadic Layer with WiMAX |

*Table 5.2 Validation of the proposed model regarding the multi-access universality characteristic.*

As can be seen in the table, the UTEM model allows the technical-economic evaluation of Fixed, Wireless, and Mixed or Hybrid Access Technologies.

The UTEM model also allows the techno-economic evaluation of virtualized access technologies. In fact, in section 5.2 of Quantitative Validation of this doctoral thesis, results of the model are presented with scenarios corresponding to fixed, wireless, mixed and virtualized access technologies. [58] does not show the possibility of evaluating virtualized networks explicitly since it is a model published in 2008, although it is likely that it could allow it with slight adaptations.

## 5.1.2 Characteristic validation Universality in Combination of technologies

Next, the validation of the UTEM model with respect to the Universality in Combination of Technologies characteristic is presented together with the model from the literature with the highest global compliance.

| Universality in Combination of access technologies | | | | |
|---|---|---|---|---|
| | Fixed technology series combination | Series combination of fixed and wireless technologies | Parallel combinations of different technology | Parallel combinations of the same technology |
| **Proposed model (UTEM)** | **YES** | **YES** | **YES** | **YES** |
| **Pereira & Ferreira [58] (2009)** | DO NOT | Static Layer (xDSL, FTTH, HFC, PLC) + Nomadic Layer (WiMAX) | DO NOT | DO NOT |

*Table 5.3 Validation of the proposed model regarding the Universality characteristic in Combination of Technologies*

Table 5.3 shows that the UTEM model allows evaluating homogeneous and heterogeneous serial and parallel combinations of fixed, wireless and mixed technologies, thanks to the Serial Submodel and the Parallel Submodel, which make up the Access Technologies Characterization Module, as It is described in Chapter 3 Proposed Model of this doctoral thesis.

## 5.1.3 Characteristic validation Universality in User Orientation

The following shows the validation of the proposed model with respect to the Universality characteristic in User Orientation.

| Universality in user orientation |
|---|

| | Operator Oriented (deployment KPIs) | Customer Oriented (KPIs of use) | Oriented to other agents |
|---|---|---|---|
| **Proposed model (UTEM)** | **YES** | **YES** | **YES[1]    (Regulatory Authorities, Investors    / Lenders,    Local Public Administration, ...)** |
| **Pereira & Ferreira [58] (2009)** | YES | It only mentions the Average Bandwidth Required (Emission and Reception) | DO NOT |

[1]A compliance value of 0.5 is assigned to this item for the UTEM model since no model in the literature fully meets this characteristic and new agents could emerge.

*Table 5.4 Validation of the proposed model regarding the Universality characteristic in User Orientation.*

The proposed UTEM model uses KPIs output parameters aimed not only at evaluating the economic viability of possible access network deployments with a given technology, but also uses technical KPIs output parameters aimed at satisfying the needs of use of an access Internet or data networks of a given technology, as well as economic KPIs oriented to the end user or customer. Likewise, the KPIs used by the UTEM model are aimed at any other agent in the telecommunications market (Market Regulatory Authorities, Investors, Public Administrations, etc.). As can be seen,

## 5.1.4 Characteristic validation Universality in the incorporation of "micro" and "macro" approaches

The following shows the validation of the proposed UTEM model with respect to the Universality characteristic in the incorporation of "micro" and "macro" approaches.

| Universality in the incorporation of "micro" and "macro" approaches | | |
|---|---|---|
| | "Macro" approach | Micro approach |
| **Proposed model (UTEM)** | **YES** | **YES** |
| **Pereira & Ferreira [58] (2009)** | YES | DO NOT |

*Table 5.5: Validation of the proposed model regarding the Universality characteristic in the incorporation of "micro" and "macro" approaches.*

The proposed UTEM model incorporates both the "micro" approach with the Isolated Application methodology, as well as the "macro" approach with the Combined Application methodology, as set out in sections 4.1 and 4.2 of the Methodology Chapter, of in such a way that it fulfills the Universality characteristic in the incorporation of "micro" and "macro" approaches. [58] is eminently oriented to the deployment of access technologies by telecommunications operators, using a "macro" approach based on the dimensions and characteristics of the geographic area to be covered, and suffers from the incorporation of the "micro" approach .

## 5.1.5 Characteristic Validation Oriented to User Requirements of the model

The following shows the validation of the proposed model with respect to the User Requirements Oriented feature of the Model.

| Orientation to User Requirements of the model | | |
|---|---|---|
| | Model User Requirements (Economic Nature)) | Model User Requirements (Technical Ranges) |
| **Proposed model (UTEM)** | YES (all parameters) | YES (all parameters) |
| **Pereira & Ferreira [58] (2009)** | Do not | Bandwidth only |

*Table 5.6: Validation of the proposed model with respect to the User Requirements Oriented feature of the model.*

The proposed model UTEM is oriented to the user requirements of the model, be they economic or technical in nature, as can be seen in the Customer Requirements and Decision Criteria section of the Proposed Model Chapter. [58] does not mention economic requirements by the user of the model and works only with bandwidth requirements by the user of the model. Although the output parameters of [58] are economic, no criteria are explicitly established.

## 5.1.6 Characteristic validation Geographical universality

It continues by presenting the validation of the proposed UTEM model regarding the geographical universality characteristic.

| Geographic universality | | | |
|---|---|---|---|
| | Allows the description of the geographic area to be covered (Surface, Volume and population density) | It allows the description of the situation of existing infrastructures (pipes, copper) | It allows the description of the population mix to be covered (Residential, SMEs, GGCC, Users in mobility) |
| Proposed model (UTEM) | YES | YES | YES |
| Pereira & Ferreira [58] (2009) | YES | YES | YES |

*Table 5.7: Validation of the proposed model with respect to the Geographical universality characteristic.*

The proposed UTEM model incorporates the Geographical Universality characteristic by using the Combined application methodology of the model, which allows it to incorporate as inputs in the characterization module, the output parameters resulting from applying external topological models. Therefore, the proposed UTEM model imports the Geographical Universality characteristic through the use of external topological models. [58] uses a geometric model to calculate the amount of cabling required in the outside plant, the number of network elements and the associated civil works costs. In [110] a discussion is shown regarding the use of geometric models (based on approximate mathematical models) vs. geographic models (based on geospatial map data) for estimating an access network deployment based on FTTH. The UTEM model allows the incorporation of information from external topological models, whether they are based on geometric models, geographic models or any other type of future modeling, with the aim of taking advantage of the models that are estimated to be more accurate at the time of their application.

## 5.1.7 Characteristic validation Technical and economic universality

The following shows the functional validation of the proposed model with respect to the technical and economic universality characteristic together with the model from the literature with the highest overall compliance [58].

| Technical and Economic Universality | | | | |
|---|---|---|---|---|
| | Input Parameters | | Output Parameters | |
| | Technicians | Economical | Technicians | Economical |
| **Proposed model (UTEM)** | **Emission and Reception Bandwidth, MTTR, MTBF, Availability, Distance, QoS, Redundancy, LOS, Frequency band used, Max. Users, Concurrency, Geotype, Attenuation, Health Risk Reluctance, Learning Curves for component costs, Study period, Access Network Architecture** | **Revenue (ARPU over time), Component Costs and OAM (CapEx, OpEx over time)** | **Emission and Reception Bandwidth (max, min, average), Availability, Distances, QoS Capacity, LOS required ?, License ?, Ubiquity, Reluctance Health Risk** | **ARPU, NPV, IRR, Revenue, Total CapEx, Total OpEx** |
| **Pereira & Ferreira [58] (2009)** | Access network architecture components (inside plant, outside plant, feeder), Geometric model for feeder networks, Study Period, Geographical area characteristics (Households, Population, HH / km2, Nr of residential subscribers, SME, Nomadic users) Service characteristics (Avg. Downstream and Upstream bandwidth) | Pricing: one time activation / connection fee (€), subscription fee (€ / month), Discount rate | DO NOT | Cost per user, Cost per homes passed, Payback period, NPV, IRR, Cash balance, CAPEX, OPEX, Sensitivity Analysis |

*Table 5.8 Validation of the proposed model regarding the Technical and Economic Universality characteristic*

The UTEM model incorporates, as described in the Proposed Model Chapter, technical and economic input and output parameters of the nature shown in said Chapter and summarized in Table 5.7. [58] uses the technical and economic input parameters shown in Table 5.7, while the output parameters are only economic. [58] shows sensitivity analysis with respect to certain parameters in the field of marketing, the UTEM model also allowing said sensitivity analysis, which is why it is concluded that the UTEM model presents greater compliance with the characteristic Technical and economic Universality with respect to [ 58].

## 5.1.8 Extensibility and Flexibility characteristic validation

The following shows the functional validation of the proposed UTEM model with respect to the Extensibility and flexibility characteristic.

| Extensibility and Flexibility | | |
|---|---|---|
| | Flexibility for new input parameters (technical and economic) | Flexibility for new output parameters (technical and economic) |
| Proposed model (UTEM) | YES[two] | YES[two] |
| Pereira & Ferreira [58] (2009) | DO NOT | DO NOT |

[two] Given the lack of extensible and flexible technical-economic models in the literature, and the need to incorporate the formulation of the new parameters in the UTEM model, a compliance value of 0.5 will be assigned to each item..

*Table 5.9 Validation of the UTEM model regarding the Extensibility and Flexibility characteristic*

As explained in section 4.1.2, the UTEM model allows you to easily add new input parameters, both technical and economic, in order to easily incorporate new output parameters of a technical or economic nature, simply by incorporating the corresponding formulation, facilitating its extensibility and flexibility. This characteristic is not appreciated or mentioned in any of the models in the literature.

## 5.1.9 Characteristic validation Technical and economic comparability

This section presents the functional validation of the proposed UTEM model with respect to the Technical and economic comparability characteristic.

| Technical and economic comparability | | |
|---|---|---|
| | Does it allow you to compare the financial results? | Does it allow you to compare the technical results? |
| Proposed model (UTEM) | YES | YES[3] |
| Pereira & Ferreira [58] (2009) | YES | DO NOT |

[3] Given the current lack of techno-economic models with which to compare technical results, a compliance of 0.5 will be assigned to this item.

*Table 5.10 Validation of the proposed model with respect to the characteristic Technical and economic comparability.*

As a consequence of the technical or economic nature of the output parameters of the UTEM model and the exclusively economic nature of the output parameters of the literature models, it follows that the UTEM model allows both economic results to be compared, as shown in the Quantitative Validation section of this Chapter, such as the technical results in the event that models arise that provide output parameters of this nature.

## 5.1.10 Predictive characteristic validation

The functional validation of the proposed UTEM model with respect to the Predictive characteristic is shown below.

| Predictive Capacity | | | |
|---|---|---|---|
| | Study period as an input parameter | Allows input parameters with temporal prediction | Perform temporal prediction on output parameters |
| **Proposed model (UTEM)** | YES | **YES (ARPU, CapEx, OpEx)** | **YES (ARPU, NPV, IRR, Revenue, Total CapEx, Total OpEx, Cost per user)** |
| **Pereira & Ferreira [58] (2009)** | YES | Trend (% per year) for Geographical area characteristics (Households, Population, HH / km2, Nr of residential subscribers, SME, Nomadic users) Service characteristics (Avg. Downstream and Upstream bandwidth) and Pricing (one time connection fee and subscription fee) | NPV, IRR, CapEx, OpEx, Cost per subscriber |

*Table 5.11 Functional validation of the proposed UTEM model with respect to the Predictive characteristic.*

Table 5.10 shows that the UTEM model, as explained in the Proposed Model Chapter, incorporates the study period as an input parameter, allowing input parameters with temporal prediction, specifically the temporal vectors of ARPU, CapEx and OpEx, as well as making the corresponding temporal prediction in the output parameters: ARPU, NPV, IRR, Revenue, CapEx, OpEx, cost per user, therefore the UTEM model fulfills the Predictive characteristic. Similarly, table 5.10 shows the fulfillment of this characteristic for model [58].

## 5.1.11 Integrable characteristic validation

Table 5.11 shows, as stated in the Methodology Chapter of this doctoral thesis, that the proposed UTEM model allows the output of another model to be integrated as input. The analysis of the literature shows economic outlets for the models analyzed. It would also be possible to incorporate in the event that there were technical outputs from other models as technical inputs, taking advantage, if necessary, of the extensibility and flexibility of the proposed model. Likewise, the logic of the proposed UTEM model and specifically the Technology Comparison Module, allows the incorporation of parameters from other models by default, in the calculation of the figures of merit of technical and / or economic performance and efficiency. This characteristic is neither appreciated nor mentioned in the literature models. $F_1 F_2$

| Integration capacity with other models | | |
|---|---|---|
| | It allows to integrate the output of another model as input | The model logic allows the incorporation of parameters from other models by default |

| Proposed model (UTEM) | YES | YES |
|---|---|---|
| Pereira & Ferreira [58] (2009) | DO NOT | DO NOT |

*Table 5.12 Functional validation of the proposed UTEM model with respect to the Integrable characteristic*

## 5.1.12 Summary

Below is the summary table of the functional validation carried out of the UTEM model, comparing it with the model with the best global compliance [58].

| | Multiaccess universality | Universality in Combination of access technologies | Universality in user orientation | Universality in incorporating " micro" and " macro" approaches | Orientation to User Requirements of the model | Geographic Universality | Technical and Economic Universality | Extensibility and Flexibility | Technical and economic comparability | Predictive Capacity | Integration capacity with other models | ASSESSMENT | %COMPLIANCE |
|---|---|---|---|---|---|---|---|---|---|---|---|---|---|
| Highest possible degree of compliance | 100 | 100 | 100 | 100 | 100 | 100 | 100 | 100 | 100 | 100 | 100 | 1100 | 100% |
| Proposed model (UTEM) | 100 | 100 | 84 | 100 | 100 | 100 | 100 | fifty | 75 | 100 | 100 | 1009 | 92% |
| Pereira & Ferreira [58] (2009) | 100 | 25 | 67 | fifty | fifty | 100 | 75 | 0 | fifty | 100 | 0 | 617 | 56% |

*Table 5.13: Summary of the qualitative validation of the proposed UTEM model.*

As can be seen in Table 5.13 and Table 2.11, the proposed UTEM model presents a higher degree of compliance than the literature models, in each and every one of the characteristics defined in the State of the Art Chapter for a technical-economic model evaluation of access technologies, universal, flexible, generalizable and scalable, reaching a global degree of compliance of 92%, leaving, therefore, the proposed UTEM model qualitatively validated.

## 5.2 Quantitative Validation in Scenarios of Isolated Application of the Model

### 5.2.1 Scenarios contemplated

The results of the model in a "micro" approach ('bottom-up'), therefore using the Isolated Application methodology of the model, are shown considering the following scenarios:

- Scenario 1: ADSL
- Scenario 2: FTTH (Fiber to the Home: Fiber to the Home)
- Scenario 3: WiMAX
- Scenario 4: 4G-LTE
- Scenario 5: FTTH access with virtualized router
- Scenario 6: Dedicated point-to-point line
- Scenario 7: 2 x ADSL redundant access
- Scenario 8: ADSL in parallel with WiMAX and IEEE 802.11g WiFi access point in aggregate mode (no backup).
- Scenario 9: VDSL

The choice of scenarios has been made considering the access technologies most widely used since 2006 to date, by telecommunications operators in the world, including ADSL and FTTH fixed access technologies, continuing with access technologies. WiMAX wireless and UMTS / 4G mobiles, which allow the mobility of the end users, and incorporating an FTTH access with the virtualized router function, since some operators and manufacturers are conducting pilot tests in this regard. The dedicated line scenario is added because it is part of the original motivation for this doctoral thesis, since it emanates from the need to seek cheaper solutions with equal or better bandwidth and availability benefits than point-to-point lines. [25], as explained in the Introduction Chapter. This also adds a scenario with homogeneous redundancy: 2 x ADSL redundant access and a scenario with mixed redundancy: ADSL in parallel with WiMAX + WiFi IEEE 802.11g, as well as a scenario with VDSL technology.

The technological evolution from 2006 to the present, not only does not question but endorses the proposed model, and its application for current and future technologies. Consider, for example, that in 2006 there was no 4G technology, nor new Fiber Optic technologies nor virtualized accesses, which in 2015 are still in the development and testing phase by manufacturers and telecommunications operators .

In the application of the proposed model to any scenario, the bandwidth values used are at the service level provided by an operator, not with respect to the theoretical or practical maximum value of the access technology. This is because the bandwidth defined in the access always conditions the dimensioning of the backbone network in order to support data traffic. This criterion is established whatever the perspective of the user of the model: operator, end customer, regulator, etc.

Scenarios 7 and 8 require highlighting that two accesses are used in parallel in an aggregate way, that is, their bandwidth is added for the purpose of applying the model proposed in this scenario. There are different mechanisms regarding the distribution of bandwidth load between various accesses. Their study is outside the scope of this doctoral thesis, constituting a future line of research, including the fact that said load balancing function may be virtualized.

The access technology scenarios contemplated to show the results of the proposed model in a "micro" approach have been obtained from [58].

In order to facilitate the application of the model and the obtaining of results, an Excel tool has been developed that incorporates the entire formulation of the proposed model. The results shown in this Validation Chapter are supported on said Excel tool. Likewise, a Web tool has been developed that implements the proposed model.

The results of the proposed model presented in this Validation Chapter, consider as an example certain end customer requirements and decision criteria, regardless of who or what type of market agent is the user of the proposed model (end user of the technology, operator telecommunications - infrastructure area, economic control area, pre-sales technical support area, etc. -, regulator, etc.).

## 5.2.2 General considerations

The general considerations of the quantitative validation carried out for the 9 scenarios are presented below, taking into account the customer requirements and the decision criteria or user preferences that are established as an example in this specific validation case.

### 5.2.2.1 Customer requirements

The following table shows the customer requirements (range of values Umin k, Umax k for the output parameters and k that wish to be considered for this purpose), established for this validation, based on which the minimum number R is calculated of redundant accesses for each technology from which these requirements would be met. In this calculation process, the Redundancy Module of the proposed model intervenes, as described in section 3.4.

For the presentation of results in this Chapter, it is considered, as an example, a residential user with an access that allows connection to the Internet and the reception of Ultra High Definition TV contents UHD TV 4K, so it must be treated of an Ultra Broadband access (30Mbits / s in reception) according to [135] and in line with the minimum bandwidth objectives set out in the European Digital Agenda for 2020 [128]. It is considered a requirement of maximum availability of Carrier Grade access (99.9999%). The following table shows the customer requirements for said residential type user, as an example from which the results are shown in the following sections of this chapter.

| | Parameters | U min k | Umax k |
|---|---|---|---|
| **RECEPTION SPEED** | AVERAGE Bandwidth (Mbits / s per user) | 30 | 100 |
| **EMISSION SPEED** | AVERAGE Bandwidth (Mbits / s per user) | 3 | 10 |
| **AVAILABILITY** | Availability | 0.9999 | 0.999999 |
| **DISTANCE** | Minimum distance to be covered by user to access point (meters) | twenty | 30,000 |
| | Total minimum distance to cover from user to access node (meters) | twenty | 30,000 |
| **COST** | It is considered in this example CAPEX + OPEX (year 1) (€) | N / A | 12000 |
| **QoS** | QoS Capability (TRUE / FALSE) | N / A | TRUE |
| **THE** | Are systems that require LOS from user to access point supported? (Line of Sight Needed?) | N / A | TRUE |
| | Are systems that require LOS from access point to access node required? | N / A | TRUE |
| **LICENSE** | Are systems that require a License allowed? (TRUE FALSE) | N / A | TRUE |
| **Environment** | Environment (URBAN / SUBURBAN / RURAL) | N / A | SUBURBAN |
| **Weather attenuation** | Do you want to consider the influence of rain attenuation? | N / A | YES |
| | Do you want to consider the influence of fog attenuation? | N / A | DO NOT |
| | Do you want to consider the influence of snow attenuation? | N / A | DO NOT |
| **Ubiquity** | Is ubiquity required at the customer's address? | N / A | YES |
| **Health** | Is the probability of arousing reluctance due to health risks admitted? (0 = NONE; 1 = LOW; 2 = MEDIUM; 3 = HIGH) | N / A | 3 |

*Table 5.14: Customer requirements for a residential user who requires access with an Internet connection and reception of 4K UHD TV content according to [135]*

### 5.2.2.2 User preferences

As defined in the Proposed Model Chapter, the model allows the user, in its Access Technologies Comparison Module, to establish specific user preferences, from which the UTEM model calculates the output parameters of the different access technologies in different scenarios, allowing the user to opt for one or another technology.

The user enters his preferences (decision criteria), assigning the desired weights ak and bp to each of the output parameters and k, to obtain the two figures of merit: technical-economic performance and technical-economic efficiency, as described in section 3.5. $F_1 F_2$

A typical user will be considered that lacks a degree of deep knowledge about access technologies and that follows the standard recommendation established in the Methodology Chapter of setting the ak and bp parameters as follows. A value of -1 is assigned for the ak corresponding to parameters and k of direct vision requirements LOS, License and probability of arousing reluctance due to health risk, since if the access technology under study requires them, they represent a decrease in its technical performance. Weight ak = 0.1 is established in the Availability parameter so that the product of the weight ak by the standardized Availability parameter (), remains in an order of magnitude similar to that of the rest of the parameters, given that the difference of the maximum availability with these requirements: 99, 9999% and the minimum availability required in this case: 99.99% is equal to 0.999999-0.9999 = 0.000099, which causes a multiplier effect on normalized availability. Depending on the range of real parameters, some iteration may be required to adjust the value assigned to said weight, depending on the priority that the user wishes to give Availability with respect to the rest of the parameters. A +1 value is assigned to the coefficients corresponding to the rest of the technical performance parameters. = 0 is established for economic cost and = 1 only for economic cost, being = 0 for the rest of the parameters, in order to calculate the efficiency or economic performance of each access technology. Depending on the range of real parameters, some iteration may be required to adjust the value assigned to said weight, depending on the priority that the user wishes to give Availability with respect to the rest of the parameters. A +1 value is assigned to the coefficients corresponding to the rest of the technical performance parameters. = 0 is established for economic cost and = 1 only for economic cost, being = 0 for the rest of the parameters, in order to calculate the efficiency or economic performance of each access technology. Depending on the range of real parameters, some iteration may be required to adjust the value assigned to said weight, depending on the priority that the user wishes to give Availability with respect to the rest of the parameters. A +1 value is assigned to the coefficients corresponding to the rest of the technical performance parameters. = 0 is established for economic cost and = 1 only for economic cost, being = 0 for the rest of the parameters, in order to calculate the efficiency or economic performance of each access technology. A +1 value is assigned to the coefficients corresponding to the rest of the technical performance parameters. = 0 is established for economic cost and = 1 only for economic cost, being = 0 for the rest of the parameters, in order to calculate the efficiency or economic performance of each access technology. A +1 value is assigned to the coefficients corresponding to the rest of the technical performance parameters. = 0 is established for economic cost and = 1 only for economic cost, being = 0 for the rest of the parameters, in order to calculate the efficiency or economic performance of each access technology. $\bar{y}_k a_k a_k b_p b_p F_2$

The results offered by the proposed model, which are presented below for the different scenarios mentioned, have been obtained from the preferences established by the typical user, which are shown in Table 5.15. The minimum and maximum dimensions Umin k, Umax k come from Table 5.14 of customer requirements. The final column "SUM (ak> 0)" is used to normalize the value of the figures of merit and, as explained in section 3.5.1. $F_1 F_2$

.

**USER PREFERENCES**

| | Output Parameters | ak ON | bp ON | U min k | Umax k | SUM (ak> 0) |
|---|---|---|---|---|---|---|
| RECEPTION SPEED | AVERAGE Bandwidth per user in access (Mbits / s per user) | 1 | 0 | 30 | 100 | 1 |
| EMISSION SPEED | AVERAGE Bandwidth per user in access (Mbits / s per user) | 1 | 0 | 3 | 10 | 1 |
| AVAILABILITY | Availability | 0.1 | 0 | 0.9999 | 0.999999 | 0.1 |
| DISTANCE | Distance user to access point (meters) | 1 | 0 | twenty | 30,000 | 1 |
| | Total distance user to access node (m) | 1 | 0 | twenty | 30,000 | 1 |
| COST | It is considered in this example CAPEX + OPEX (year 1) (€) | 0 | 1 | 0 | 12000 | 0 |
| QoS | QoS capability | 1 | 0 | 0 | 1 | 1 |
| THE | LOS from user to access point (Line of Sight Necessary?) | -1 | 0 | 0 | 1 | 0 |
| | LOS from access point to access node? | -1 | 0 | 0 | 1 | 0 |
| LICENSE | Do you need a license? | -1 | 0 | 0 | 1 | 0 |
| Ubiquity | Ubiquity at customer's address | 1 | 0 | 0 | 1 | 1 |
| Health | Probability of provoking reluctance due to health risk (0 = NONE; 1 = LOW; 2 = MEDIUM; 3 = HIGH) | -1 | 0 | 0 | 3 | 0 |
| | | | | | | 6.1 |

*Table 5.15: User preferences set as an example for quantitative validation.*

### 5.2.3 Results

In order to simplify the structure of this section and make it easier to read, the results of the UTEM model for the first two mentioned scenarios are presented below. The results of the seven remaining scenarios are included in the ANNEXES section of this doctoral thesis, for the reader's convenience.

#### 5.2.3.1 Scenario 1: ADSL

Next, we proceed to show the results of the model for a single ADSL access, considering the requirements of the residential type customer set out as an example in section 5.2.2.1, and the user preferences set out in section 5.2.2.2.

The following tables show the input parameters of the model that are used as an example in this scenario, as well as the output parameters obtained by applying the UTEM model for this scenario. Any variation in input parameters, customer requirements, or user preferences will lead to different results.

The technical and economic input parameters () that feed the UTEM model for this specific case are shown in Table 5.16, and are obtained from the different sources indicated in section 3.2. The first column shows a categorization of the input parameters to make them easier to read. The second column contains the names of the input parameters. The rest of the columns correspond to the components or elements of the access and constitute the PxN matrix of input parameters described in section 3.3. In this example, 4 components or elements of the access have been considered, so N = 4, including from left to right from the one closest to the end user (the WiFi interface of the end user PC has been considered in this example), to the furthest (the interface of the access node with the aggregation network has been considered). Given the flexibility of the UTEM model, the P x N dimension of the input parameter matrix may vary from one evaluation session to another, depending on the input parameters and access elements that the user of the model wishes to consider. $x_{ij}$

For this example, the input data of the ARPU vector for the 3 years considered as the study period have been obtained considering an average of the EU countries according to the analysis firm Analysys Mason [133]. For simplicity, the values are assigned to the component closest to the user. The CAPEX data for each access component has been obtained according to data for the EMEA region (Europe, Middle East and Africa) from the consulting firm OVUM [136] [137]. The OPEX data, in this example, has been estimated as the product of the Availability of each component of the access by its CAPEX. The user could choose any other method, or even incorporate those of another model.

In Table 5.17, the data output of the model is presented. Column "" shows the results of applying the formulation set forth in section 3.3.1.2 - Formulas (3.1) to (3.21) -, in this case for the Series Submodel, as it is the only one that intervenes in this scenario. In the next two columns, the products for this specific case are shown, as intermediate calculations for the calculation of the figures of merit and using the formulas (3.68) and (3.69) of section 3.5.1. $y_k a_k \cdot \bar{y}_k$ y $b_P \cdot y_k F_1 F_2$

In Table 5.18, the set of output parameters that have established minimum and maximum limits in the matrix of customer requirements in Table 5.14 is shown in column "". Said output parameters coincide with those in Table 5.17. The column "Minimum value of to meet customer requirements" includes the intermediate calculation of for each output parameter, according to the formulation set out in section 3.4.2 - formulas (3.46) to (3.65) -, as an intermediate calculation and step prior to calculating the minimum number of redundant accesses R for ADSL access technology to meet user requirements, according to the application of formulas (3.43) to (3.45) and the flow chart in Figure 3.7. $y_k y_k r_k r_k y_k$

For the rest of the scenarios, the same procedure is followed. It should be noted that in the case of Scenarios 7 and 8, after the Series Submodel, the Parallel Submodel intervenes, according to the formulation set out in section 3.3.2.2 - formulas (3.22) to (3.42) - and the methodology in section 4.1, also illustrated in Figure 4.1

**INPUT PARAMETERS** $x_{ij}$ **of the UTEM MODEL**

SCENARIO NAME:                                                    ADSL

| | Input parameter | PC interface | Element 1 | Element 2 | Element 3 |
|---|---|---|---|---|---|
| **Element identification** | Element name | Wireless 802.11b / g adapter US ROBOTICS USR805420 | 3COM OfficeConnect 812 Router | DSLAM (Alcatel 7300) | Aggregation network |
| | Element function | Wi-Fi PC adapter | Router at customer's home | Access interface | Aggregation interface |
| **Bandwidth** | Unitary Bandwidth (Mbits / s) (Reception) | 100 | 10 | 10 | 10 |
| | Unitary Bandwidth (Mbits / s) (Emission) | 100 | 0.82 | 1 | 1 |
| **Availability** | Availability | 99.9962% | 99.9644% | 99.9990% | 100.0000% |
| **Distance** | Distance (meters) | N / A | N / A | 4500 | N / A |
| **QoS** | QoS capability | N / A | TRUE | TRUE | TRUE |
| **Redundancy** | Redundancy (No. elements in parallel) | 1 | 1 | 1 | 1 |
| **THE** | LOS (Line of Sight Needed?) | FAKE | FAKE | FAKE | FAKE |
| **Frequency band** | Band (GHz) | 2.4 | 2.4 | N / A | N / A |
| **License** | Do you need a license? | FAKE | FAKE | FAKE | FAKE |
| **Users** | No. users: | 1 | 1 | N / A | N / A |
| **Concurrence** | Estimated average concurrency of users | N / A | 100.00% | N / A | N / A |
| **Technology** | Do you use wireless technology in any section? | DO NOT | DO NOT | DO NOT | DO NOT |
| **Environment** | Vector Environment (DENSE URBAN / URBAN / SUBURBAN / RURAL) | (1,1,1,1) | (1,1,1,1) | (1,1,1,1) | (1,1,1,1) |
| **Attenuation by meteorology** | Total decrease in Reception Bandwidth due to meteorological effects (Mbits / s) | 0 | 0 | 0 | 0 |
| | Total decrease in Broadcast Bandwidth due to meteorological effects (Mbits / s) | 0 | 0 | 0 | 0 |
| **Ubiquity** | Ubiquity at customer's address | YES | N / A | N / A | N / A |
| **Health** | Probability of provoking reluctance due to health risk (0 = NONE; 1 = LOW; 2 = MEDIUM; 3 = HIGH) | 1 | 1 | 0 | 0 |
| **K (interest rate)** | Type of interest | 1.00% | N / A | N / A | N / A |
| **ARPU year 1** | Average revenue per user (Year 1) | € 416.54 | € 0.00 | € 0.00 | € 0.00 |
| **ARPU year 2** | Average revenue per user (Year 2) | € 363.60 | € 0.00 | € 0.00 | € 0.00 |
| **ARPU year 3** | Average revenue per user (Year 3) | € 363.60 | € 0.00 | € 0.00 | € 0.00 |
| **CAPEX year 1** | Investments (Year 1) | € 15.00 | € 100.00 | € 100.00 | € 100.00 |
| **CAPEX year 2** | Investments (Year 2) | € 0.00 | € 0.00 | € 0.00 | € 0.00 |
| **CAPEX year 3** | Investments (Year 3) | € 0.00 | € 0.00 | € 0.00 | € 0.00 |
| **OPEX year 1** | Operating Expenses (Year 1) | € 0.00 | € 0.04 | € 0.001 | € 0,000 |
| **OPEX year 2** | Operating Expenses (Year 2) | € 0.00 | € 0.04 | € 0.001 | € 0,000 |
| **OPEX year 3** | Operating Expenses (Year 3) | € 0.00 | € 0.04 | € 0.001 | € 0,000 |

*Table 5.16: Input parameters in ADSL scenario*

**OUTPUT PARAMETERS $y_k$ and Figures of merit F1 and F2**

| | Output Parameters | $y_k$ | $a_k \cdot \overline{y}_k$ | $b_P \cdot y_k$ | Weighted Valuation | |
|---|---|---|---|---|---|---|
| | | | | | **F1** | **F2** |
| **RECEPTION SPEED** | AVERAGE Bandwidth per user in access (Mbits / s per user) | 10 | -0.2857 | 0.0000 | 22.51% | 71.46% / K € |
| **EMISSION SPEED** | AVERAGE Bandwidth per user in access (Mbits / s per user) | 3 | 0.0000 | 0.0000 | | |
| **AVAILABILITY** | Availability | 99.9597% | -0.3065 | 0.0000 | | |
| **DISTANCE** | Distance user to access point (meters) | 4,500 | 0.1494 | 0.0000 | | |
| | Total distance user to access node (m) | 4,500 | 0.1494 | 0.0000 | | |
| **ARPU year 1** | Average revenue per user (Year 1) | € 416.54 | 0.0000 | 0.0000 | | |
| **ARPU year 2** | Average revenue per user (Year 1) | € 363.60 | 0.0000 | 0.0000 | | |
| **ARPU year 3** | Average revenue per user (Year 1) | € 363.60 | 0.0000 | 0.0000 | | |
| **CAPEX year 1** | Investments (Year 1) | € 315.00 | 0.0000 | 0.0000 | | |
| **CAPEX year 2** | Investments (Year 2) | € 0.00 | 0.0000 | 0.0000 | | |
| **CAPEX year 3** | Investments (Year 3) | € 0.00 | 0.0000 | 0.0000 | | |
| **OPEX year 1** | Operating Expenses (Year 1) | € 0.04 | 0.0000 | 0.0000 | | |
| **OPEX year 2** | Operating Expenses (Year 2) | € 0.04 | 0.0000 | 0.0000 | | |
| **OPEX year 3** | Operating Expenses (Year 3) | € 0.04 | 0.0000 | 0.0000 | | |
| **NPV** | Net Present Value at interest rate K | € 809.77 | 0.0000 | 0.0000 | | |
| **Net Cash Flow** | Net Cash Flow (interest rate K is not taken into account) | € 828.63 | 0.0000 | 0.0000 | | |
| **Payback Period (years)** | Amortization period | 1.00 | 0.0000 | 0.0000 | | |
| **COST** | CapEx + OpEx (year 1) | € 315.04 | 0.0000 | 315.0371 | | |
| **QoS** | QoS capability | TRUE | 1,0000 | 0.0000 | | |
| **THE** | LOS from user to access point (Line of Sight Necessary?) | N / A | 0.0000 | 0.0000 | | |
| | LOS from access point to access node required? | N / A | 0.0000 | 0.0000 | | |
| **LICENSE** | Do you need a license? | FAKE | 0.0000 | 0.0000 | | |
| **Ubiquity** | Ubiquity at customer's address | YES | 1,0000 | 0.0000 | | |
| **Health** | Probability of provoking reluctance due to health risk (0 = NONE; 1 = LOW; 2 = MEDIUM; 3 = HIGH) | 1 | -0.3333 | 0.0000 | | |

*Table 5.17: Output parameters and F1 and F2 figures of merit in ADSL scenario.$y_k$*

**OBTAINING THE MINIMUM NUMBER OF REDUNDANT ACCESSES R TO MEET CUSTOMER REQUIREMENTS**

| | Output Parameters | $y_k$ | Minimum rk value to meet customer requirements |
|---|---|---|---|
| **RECEPTION SPEED** | AVERAGE Bandwidth per user in access (Mbits / s per user) | 10 | 3 |
| **EMISSION SPEED** | AVERAGE Bandwidth per user in access (Mbits / s per user) | 3 | 1 |
| **AVAILABILITY** | Availability | 99.9597% | two |
| **DISTANCE** | Distance user to access point (meters) | 4,500 | COMPLIES |
| | Total distance user to access node (m) | 4,500 | COMPLIES |
| **COST** | It is considered in this example CAPEX + OPEX (year 1) (€) | € 315.04 | COMPLIES |
| **QoS** | QoS capability | TRUE | COMPLIES |
| **THE** | LOS from user to access point (Line of Sight Necessary?) | N / A | COMPLIES |
| | LOS from access point to access node required? | N / A | COMPLIES |
| **LICENSE** | Do you need a license? | FAKE | COMPLIES |
| **Ubiquity** | Ubiquity at customer's address | YES | COMPLIES |
| **Health** | Probability of provoking reluctance due to health risk (0 = NONE; 1 = LOW; 2 = MEDIUM; 3 = HIGH) | 1 | COMPLIES |

| CONCLUSION |
|---|
| **YES IT COMPLIES WITH:** |
| **R = 3** |

*Table 5.18: Minimum number of redundant accesses R for ADSL technology to meet the established user requirements.*

### 5.2.3.2 Scenario 2: FTTH (Fiber to the Home: Fiber to the Home)

We proceed to show the results of the UTEM model for the FTTH scenario considering the requirements and user preferences established for this example in section 5.2.2 in a manner analogous to Scenario 1.

The following shows the input parameters of the model that are used as an example in this scenario, as well as the output parameters obtained by applying the UTEM model for this scenario. As already mentioned for Scenario 1, any variation in input parameters, customer requirements, or user preferences will lead to different results.

The technical and economic input parameters () that feed the UTEM model for this specific case are shown in Table 5.19, and are obtained from the different sources indicated in section 3.2. The first column shows a categorization of the input parameters to make them easier to read. The second column contains the names of the input parameters. The rest of the columns correspond to the components or elements of the access and constitute the PxN matrix of input parameters described in section 3.3. In this example, 4 components or elements of the access have been considered, so N = 4, including from left to right from the one closest to the end user (the WiFi interface of the end user PC has been considered in this example), to the furthest (the interface of the access node with the aggregation network has been considered). Given the flexibility of the UTEM model, the P x N dimension of the input parameter matrix may vary from one evaluation session to another, depending on the input parameters and access elements that the user of the model wishes to consider. $x_{ij}$

For this example, the input data of the ARPU vector for the 3 years considered as the study period have been obtained considering an average of the EU countries according to the analysis firm Analysys Mason [133]. For simplicity, the values are assigned to the access component closest to the user. The CAPEX data for each access component has been obtained according to data for the EMEA region (Europe, Middle East and Africa) from the consulting firm OVUM [136] [137]. The OPEX data, in this example, has been estimated as the product of the Availability of each component of the access by its CAPEX. The user could choose any other method, or even incorporate those of another model.

In Table 5.20, the data output of the model is presented. Column "" shows the results of applying the formulation set forth in section 3.3.1.2 - Formulas (3.1) to (3.21) -, in this case for the Series Submodel, as it is the only one that intervenes in this scenario. In the next two columns, the products for this specific case are shown, as intermediate calculations for the calculation of the figures of merit and using the formulas (3.68) and (3.69) of section 3.5.1. $y_k a_k \cdot \bar{y}_k$ $y$ $b_P \cdot y_k F_1 F_2$

In Table 5.21, the set of output parameters that have established minimum and maximum limits in the matrix of customer requirements in Table 5.14 is shown in column "". Said output parameters coincide with those in Table 5.20. The column "Minimum value of to meet customer requirements" includes the intermediate calculation of for each output parameter, according to the formulation set out in section 3.4.2 - formulas (3.46) to (3.65) -, as an intermediate calculation and step prior to calculating the minimum number of redundant accesses R for ADSL access technology to meet user requirements, according to the application of formulas (3.43) to (3.45) and the flow chart in Figure 3.7. $y_k y_k r_k r_k y_k$

**INPUT PARAMETERS** $x_{ij}$ **of the UTEM MODEL**

**SCENARIO NAME:**                                          **FTTH**

| | Input parameter | PC interface | Element 1 | Element 2 | Element 3 |
|---|---|---|---|---|---|
| **Element identification** | Element name | Wireless 802.11b / g adapter US ROBOTICS USR805420 | Router + FTTH ONU | Access node (OLT) | Aggregation Network |
| | Element function | Fast Ethernet card | Router at customer's home + Optical Network Unit | Access node (Optical access interface) | Aggregation interface |
| **Bandwidth** | Unitary Bandwidth (Mbits / s) (Reception) | 100 | 100 | 100 | 100 |
| | Unitary Bandwidth (Mbits / s) (Emission) | 100 | 10 | 10 | 10 |
| **Availability** | Availability | 99.9962% | 99.9760% | 99.9990% | 100.0000% |
| **Distance** | Distance (meters) | N / A | N / A | 15000 | N / A |
| **QoS** | QoS capability | N / A | TRUE | TRUE | TRUE |
| **Redundancy** | Redundancy (No. elements in parallel) | 1 | 1 | 1 | 1 |
| **THE** | LOS (Line of Sight Needed?) | N / A | N / A | N / A | N / A |
| **Frequency band** | Band (GHz) | N / A | N / A | N / A | N / A |
| **License** | Do you need a license? | FAKE | FAKE | FAKE | FAKE |
| **Users** | No. users: | 1 | 1 | N / A | N / A |
| **Concurrence** | Estimated average concurrency of users | N / A | 100.00% | N / A | N / A |
| **Technology** | Do you use wireless technology in any section? | DO NOT | DO NOT | DO NOT | DO NOT |
| **Environment** | Vector Environment (DENSE URBAN / URBAN / SUBURBAN / RURAL) | (1,1,1,1) | (1,1,1,1) | (1,1,1,1) | (1,1,1,1) |
| **Attenuation by meteorology** | Total decrease in Reception Bandwidth due to meteorological effects (Mbits / s) | 0 | 0 | 0 | 0 |
| | Total decrease in Broadcast Bandwidth due to meteorological effects (Mbits / s) | 0 | 0 | 0 | 0 |

| Ubiquity | Ubiquity at customer's address | YES | N / A | N / A | N / A |
|---|---|---|---|---|---|
| Health | Probability of provoking reluctance due to health risk (0 = NONE; 1 = LOW; 2 = MEDIUM; 3 = HIGH) | 1 | 1 | 0 | 0 |
| K (interest rate) | Type of interest | 1.00% | N / A | N / A | N / A |
| ARPU year 1 | Average revenue per user (Year 1) | € 704.88 | € 0.00 | € 0.00 | € 0.00 |
| ARPU year 2 | Average revenue per user (Year 1) | € 704.88 | € 0.00 | € 0.00 | € 0.00 |
| ARPU year 3 | Average revenue per user (Year 1) | € 704.88 | € 0.00 | € 0.00 | € 0.00 |
| CAPEX year 1 | Investments (Year 1) | € 15.00 | € 150.00 | € 150.00 | € 200.00 |
| CAPEX year 2 | Investments (Year 2) | € 0.00 | € 0.00 | € 0.00 | € 0.00 |
| CAPEX year 3 | Investments (Year 3) | € 0.00 | € 0.00 | € 0.00 | € 0.00 |
| OPEX year 1 | Operating Expenses (Year 1) | € 0.00 | € 0.04 | € 0.002 | € 0,000 |
| OPEX year 2 | Operating Expenses (Year 2) | € 0.00 | € 0.04 | € 0.002 | € 0,000 |
| OPEX year 3 | Operating Expenses (Year 3) | € 0.00 | € 0.04 | € 0.002 | € 0,000 |

*Table 5.19: FTTH scenario input parameters.*

**OUTPUT PARAMETERS** $y_k$ **and Figures of merit F1 and F2**

| | Output Parameters | $y_k$ | $a_k \cdot \overline{y}_k$ | $b_P \cdot y_k$ | F1 | F2 |
|---|---|---|---|---|---|---|
| | | | | | **Weighted Valuation** | |
| **RECEPTION SPEED** | AVERAGE Bandwidth (Mbits / s per user) | 100 | 1,0000 | 0.0000 | 73.38% | 142.48 % / K € |
| **EMISSION SPEED** | AVERAGE Bandwidth (Mbits / s per user) | 10 | 1,0000 | 0.0000 | | |
| **AVAILABILITY** | Availability | 99.9712% | -0.1898 | 0.0000 | | |
| **DISTANCE** | Distance user to access point (meters) | 15000 | 0.4997 | 0.0000 | | |
| | Total distance user to access node (m) | 15,000 | 0.4997 | 0.0000 | | |
| **ARPU year 1** | Average revenue per user (Year 1) | € 704.88 | 0.0000 | 0.0000 | | |
| **ARPU year 2** | Average revenue per user (Year 1) | € 704.88 | 0.0000 | 0.0000 | | |
| **ARPU year 3** | Average revenue per user (Year 1) | € 704.88 | 0.0000 | 0.0000 | | |
| **CAPEX year 1** | Investments (Year 1) | € 515.04 | 0.0000 | 0.0000 | | |
| **CAPEX year 2** | Investments (Year 2) | € 0.00 | 0.0000 | 0.0000 | | |
| **CAPEX year 3** | Investments (Year 3) | € 0.00 | 0.0000 | 0.0000 | | |
| **OPEX year 1** | Operating Expenses (Year 1) | € 0.04 | 0.0000 | 0.0000 | | |
| **OPEX year 2** | Operating Expenses (Year 2) | € 0.04 | 0.0000 | 0.0000 | | |
| **OPEX year 3** | Operating Expenses (Year 3) | € 0.04 | 0.0000 | 0.0000 | | |
| **NPV** | Net Present Value at interest rate K | € 1,563.03 | 0.0000 | 0.0000 | | |
| **Net Cash Flow** | Net Cash Flow (interest rate K is not taken into account) | € 1,599.53 | 0.0000 | 0.0000 | | |
| **Payback Period (years)** | Amortization period | 1.00 | 0.0000 | 0.0000 | | |
| **COST** | CapEx + OpEx (year 1) | € 515.04 | 0.0000 | € 515.04 | | |
| **QoS** | QoS capability | TRUE | 1,0000 | 0.0000 | | |
| **THE** | LOS from user to access point (Line of Sight Necessary?) | N / A | 0.0000 | 0.0000 | | |
| | LOS from access point to access node required? | N / A | 0.0000 | 0.0000 | | |
| **LICENSE** | Do you need a license? | FAKE | 0.0000 | 0.0000 | | |
| **Ubiquity** | Ubiquity at customer's address | YES | 1,0000 | 0.0000 | | |
| **Health** | Probability of provoking reluctance due to health risk (0 = NONE; 1 = LOW; 2 = MEDIUM; 3 = HIGH) | 1 | -0.3333 | 0.0000 | | |

*Table 5.20: Output parameters and figures of merit F1 and F2 in FTTH scenario.$y_k$*

**OBTAINING THE MINIMUM NUMBER OF REDUNDANT ACCESSES R TO MEET CUSTOMER REQUIREMENTS**

| | Output Parameters | $y_k$ | Minimum rk value to meet customer requirements | |
|---|---|---|---|---|
| RECEPTION SPEED | AVERAGE Bandwidth per user in access (Mbits / s per user) | 100 | 1 | |
| EMISSION SPEED | AVERAGE Bandwidth per user in access (Mbits / s per user) | 10 | 1 | |
| AVAILABILITY | Availability | 99.9712% | two | |
| DISTANCE | Distance user to access point (meters) | 15,000 | COMPLIES | CONCLUSION |
| | Total distance user to access node (m) | 15,000 | COMPLIES | YES IT COMPLIES WITH: |
| COST | CapEx + OpEx (Year 1) | € 515.04 | COMPLIES | |
| QoS | QoS capability | TRUE | COMPLIES | R = 2 |
| THE | LOS from user to access point (Line of Sight Necessary?) | N / A | COMPLIES | |
| | LOS from access point to access node required? | N / A | COMPLIES | |
| LICENSE | Do you need a license? | FAKE | COMPLIES | |
| Ubiquity | Ubiquity at customer's address | YES | COMPLIES | |
| Health | Probability of provoking reluctance due to health risk (0 = NONE; 1 = LOW; 2 = MEDIUM; 3 = HIGH) | 1 | COMPLIES | |

*Table 5.21: Minimum number of redundant accesses R for FTTH technology to meet the established user requirements.*

### 5.2.3.3 Comparative analysis of scenarios

The model offers the possibility of comparing the different access technologies based on the output data.Below is a comparative analysis of the 9 access technologies mentioned in 3 cases of technology user requirements:

- *Case A - Residential user with a minimum bandwidth in reception of 30 Mbits / s (Data for 2015) according to the objectives of the European Digital Agenda 2020 [128].*
- *Case B - SME-type user with a minimum reception bandwidth of 300 Mbits / s (Data for 2015).*
- *Case C - Residential type user with a minimum bandwidth in reception of 2 Mbits / s (Data for 2006) according to the objectives of the European Digital Agenda i2010 [67].*

### Case A. Requirements for a residential user client with a minimum reception bandwidth of 30 Mbits / s (Data for 2015)

Below is a table showing the results of the output parameters for each technology, the figures of merit calculated based on the preferences established by the user of the model, and the minimum R value of redundant accesses necessary to meet the requirements. customer requirements established with each of the access technologies.

| ACCESS TECHNOLOGIES | | ADSL | ADSL // 802.11g + WiMAX Backhaul with PC | Point to point 2 Mbps | FTTH with virtualized router | 4G-LTE | FTTH | VDSL | WiMAX Access Point + Backhaul WiMAX 802.16a | 2 x ADSL | Minimum customer requirements | Wei ghts ak | Wei ghts bp |
|---|---|---|---|---|---|---|---|---|---|---|---|---|---|
| | Output Parameters | Values | Values | Values | Values | Values | Values | Values | Values | Values | Values | | |
| RECEPTION SPEED | AVERAGE Bandwidth per user in access (Mbits / s per user) | 10 | 10.0912 | two | 100 | 24 | 100 | fifty | 2.7538 | twenty | 30 | 1 | 0 |
| EMISSION SPEED | AVERAGE Bandwidth per user in access (Mbits / s per user) | 3 | 3.0928 | two | 10 | 8 | 10 | 5 | 2,7641 | 6 | 3 | 1 | 0 |
| AVAILABILITY | Availability | 99.96% | 100.00% | 99.97% | 100.00% | 100.00% | 99.97% | 99.96% | 100.00% | 99.96% | 99.99% | 0.1 | 0 |
| DISTANCE | Distance user to access point (meters) | 4500 | Four. Five | 5000 | 15000 | 15000 | 15000 | 600 | 3000 | 4500 | twenty | 1 | 0 |
| | Total distance user to node of access to transport network (m) | 4,500 | 4,500 | 5,000 | 15,000 | 15,000 | 15,000 | 600 | 48,000 | 4,500 | twenty | 1 | 0 |
| COST | CapEx + OpEx (Year 1) | € 315.04 | € 327.04 | € 3,230.00 | € 390.00 | € 543.00 | € 515.00 | € 365.00 | € 370.00 | € 615.00 | N / A | 0 | 1 |
| QoS | QoS capability | TRUE | TRUE | TRUE | TRUE | TRUE | TRUE | TRUE | TRUE | TRUE | N / A | 1 | 0 |
| THE | LOS from user to access point (Line of Sight Necessary?) | N / A | TRUE | N / A | N / A | N / A | N / A | N / A | FAKE | N / A | N / A | -1 | 0 |
| | LOS from access point to transport network node required? | N / A | TRUE | N / A | N / A | N / A | N / A | N / A | TRUE | N / A | N / A | -1 | 0 |
| LICENSE | Do you need a license? | FAKE | TRUE | FAKE | FAKE | FAKE | FAKE | FAKE | TRUE | FAKE | N / A | -1 | 0 |
| Ubiquity | Ubiquity at customer's address | YES | YES | YES | YES | YES | YES | YES | YES | YES | N / A | 1 | 0 |
| Health | Probability of provoking reluctance due to health risk (0 = NONE; 1 = LOW; 2 = MEDIUM; 3 = HIGH) | 1 | two | 1 | 1 | 1 | 1 | 1 | 1 | 1 | N / A | -1 | 0 |
| F1 (Performance) | | 7.94% | -22.07% | 1.48% | 86.32% | 45.59% | 83.92% | 30.82% | 31.40% | 28.57% | | | |
| F2 (Economic efficiency) | | 25.20% | -67.49% | 0.46% | 221.34% | 83.96% | 162.96% | 84.45% | 84.86% | 46.45% | | | |
| R | | 3 | 3 | fifteen | two | two | two | two | 26 | two | | | |

*Table 5.22: Summary of the data output of the model for residential user requirements with a minimum bandwidth of 30Mbits / s according to [135] [78]*

Similarly, the model allows the graphical representation of the results in order to facilitate decision-making.

Below is the comparison between the different access technologies based on the value of the figure of merit (%) that accounts for the benefits of each technology. $F_1$

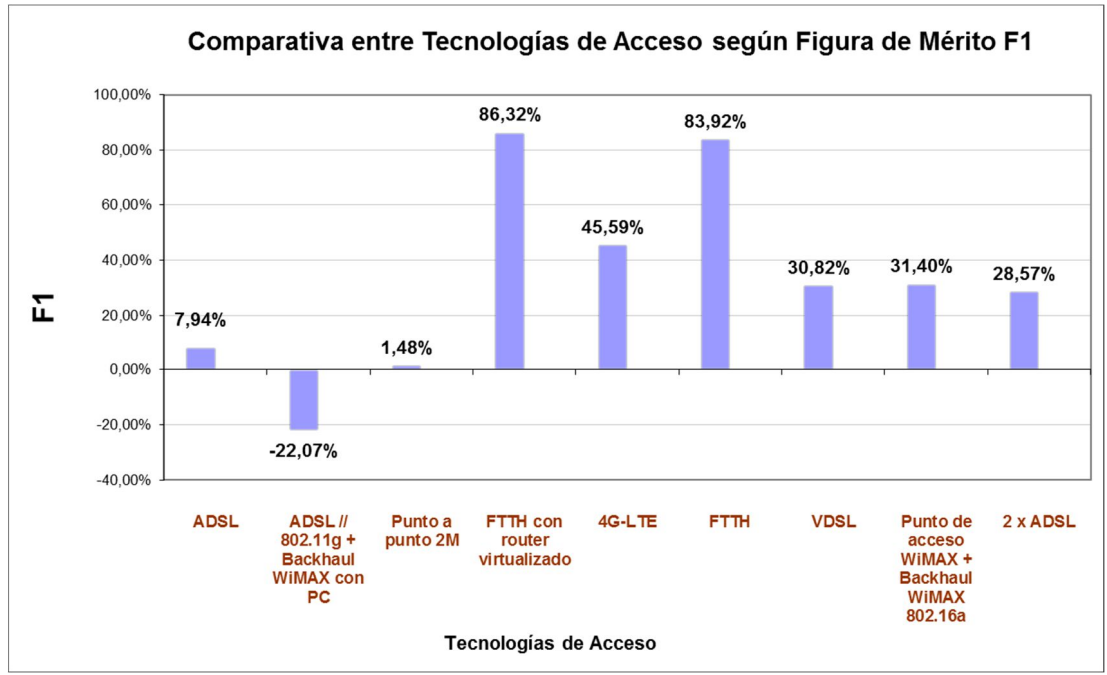

*Figure 5.1: Comparison between access technologies based on technical performance (Figure of Merit).* $F_1$

As can be seen, (%) can take negative values since it is a linear combination of the normalized coordinates of the difference vector between the technology vector and the vector of customer requirements uk. Depending on the weights established by the user and the value of each coordinate of the difference vector, negative values can be obtained. (%) is normalized with respect to the reference of benefits given by the sum of the positive weights established by the user. Values of greater than 100% can be given since there will be normalized coordinate values of the difference vector greater than unity, when - is greater than - . $F_1 y_k F_1 a_k F_1 y_k U_{mín\,k} U_{máx\,k} U_{mín\,k}$

The figure of merit (%) allows us to obtain a ranking of benefits. In this example, we are talking about a ranking of technical benefits given that = 0 for cost. In the previous figure, and for said example, the maximum value of technical performance is obtained for FTTH technology with virtualized router, obtaining a minimum value of technical performance for parallel ADSL access with WiMAX + Wifi 802.11g with a negative value due to which is penalized by the need for Direct Vision (LOS) and Licenses, as well as a MEDIUM probability of arousing reluctance due to health risk, as can be seen in the previous table (Table 5.22). Note that this is an example and that the input parameters may vary depending on the data sources and components used by the user of the model. $F_1 a_k F_1 F_1$

The following graph shows the comparison between the different access technologies based on the value of the figure of merit, which in this case, as defined in the decision criteria, accounts for the ratio of technical benefits / economic cost unit for each access technology (economic efficiency). $F_2$

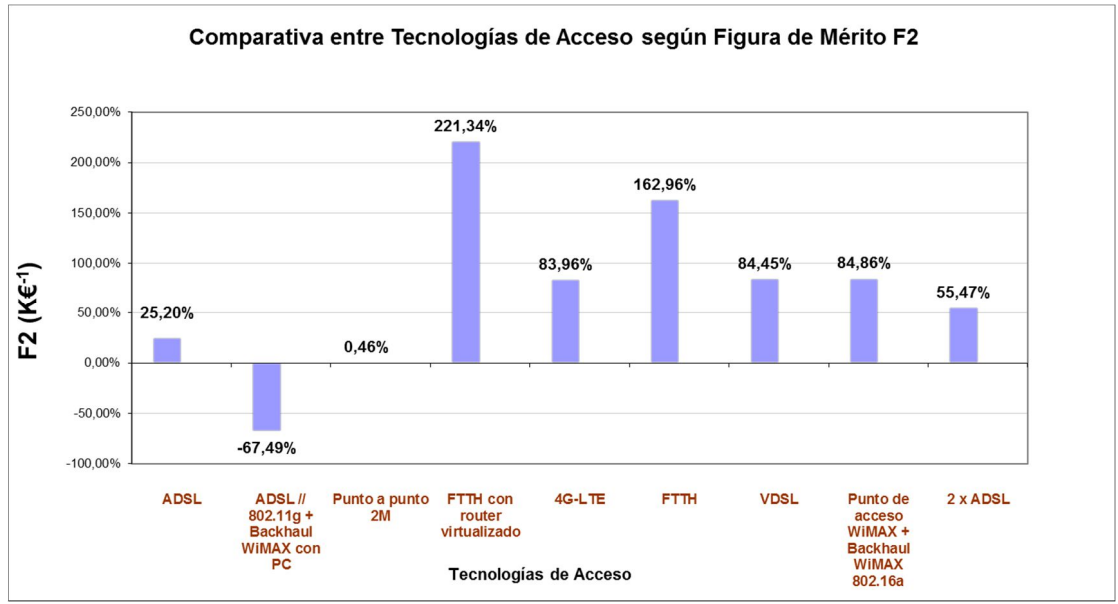

*Figure 5.2: Comparison between access technologies based on technical-economic efficiency. The units of in this case are% / K €, accounting for the technical performance per economic unit.$F_2F_2$*

The maximum levels of efficiency will be reached with performance values of 100% or higher (as specified when presenting the results of) / minimum cost requirement, in such a way that very high performance is achieved with more economical technology. In other words, with the economic efficiency parameter in this example, the user of the model seeks maximum technical performance at minimum cost. In this example, the ranking obtained provides maximum values for FTTH with virtualized router and FTTH (FTTH with virtualized router provides higher availability and lower cost (CAPEX and OPEX) than FTTH, which is why its is higher). The minimum value is given for ADSL in parallel with WiMAX + WiFi 802.11g because the value of technical performance obtained is negative, as can be seen in the graph of.$F_2F_1F_2F_2F_2F_1F_1$

The following figure shows the comparison between access technologies according to the average reception bandwidth.

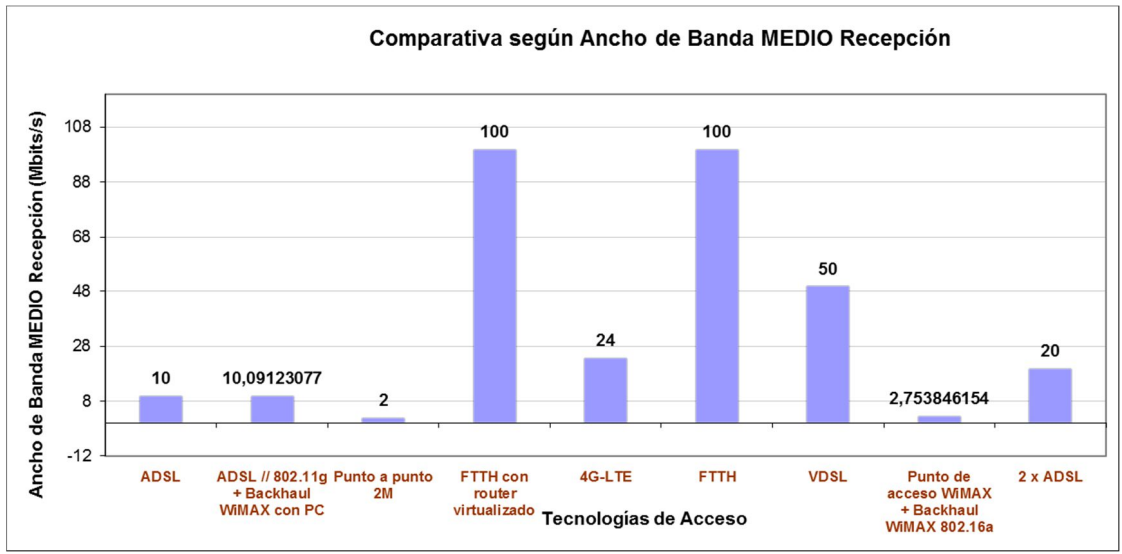

*Figure 5.3: Comparison between access technologies based on the Average Reception Bandwidth.*

Next, we proceed to compare the access technologies based on the average broadcast bandwidth.

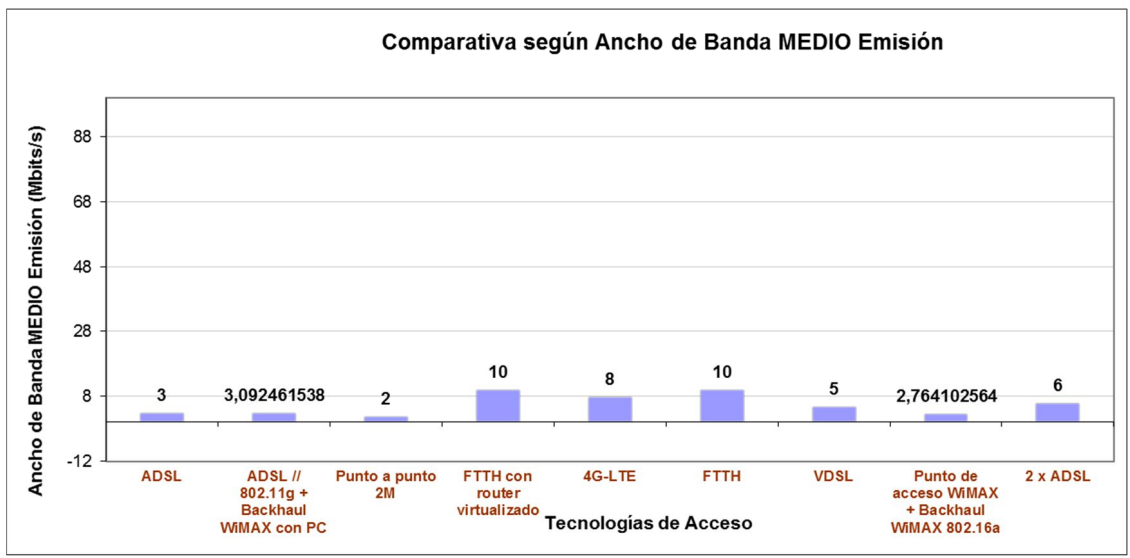

*Figure 5.4: Comparison between access technologies based on the Average Broadcast Bandwidth.*

Likewise, the comparison between access technologies is graphically represented using the availability criterion.

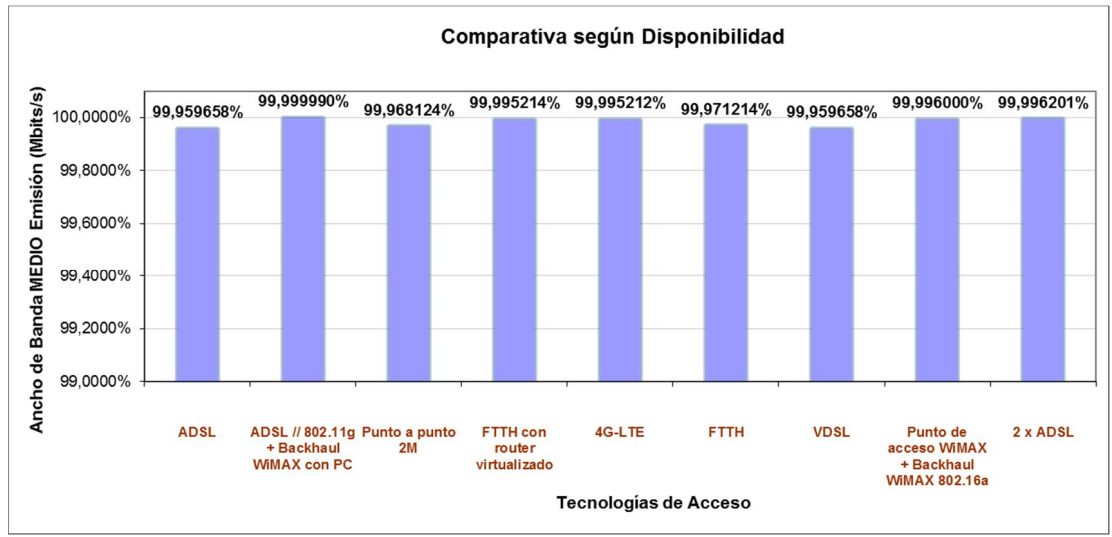

*Figure 5.5: Comparison between access technologies based on Availability.*

Finally, the graphical representation of the comparison between access technologies is shown based on the minimum number N of redundant accesses necessary to meet customer requirements with each access technology.

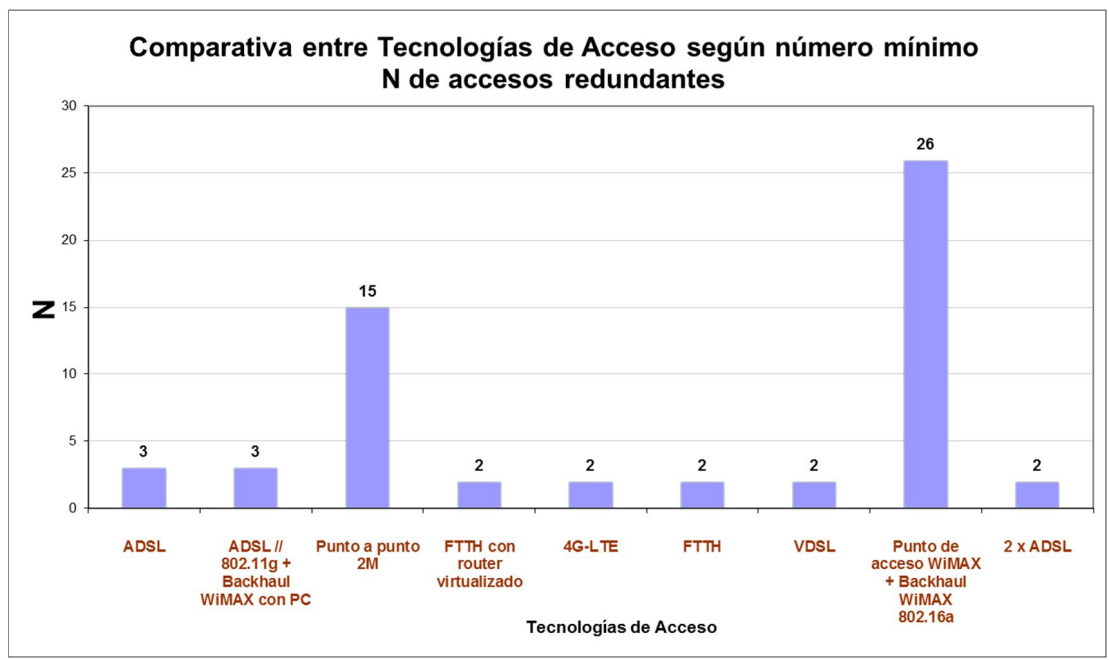

*Figure 5.6: Comparison between access technologies based on the minimum number of redundant accesses necessary to meet the customer requirements established in this use case as an example.*

Graphical comparisons have been shown using a single variable. It would also be possible from the output results table to present graphs that contemplate two and even more variables, in such a way that decisions can be made based on the quadrants or regions in which each technology is located.

Below are two graphs that contemplate the figure of merit (benefits) vs. Annual economic cost per user and the figure of merit (economic efficiency) vs. Annual economic cost per user, respectively.$F_1 F_2$

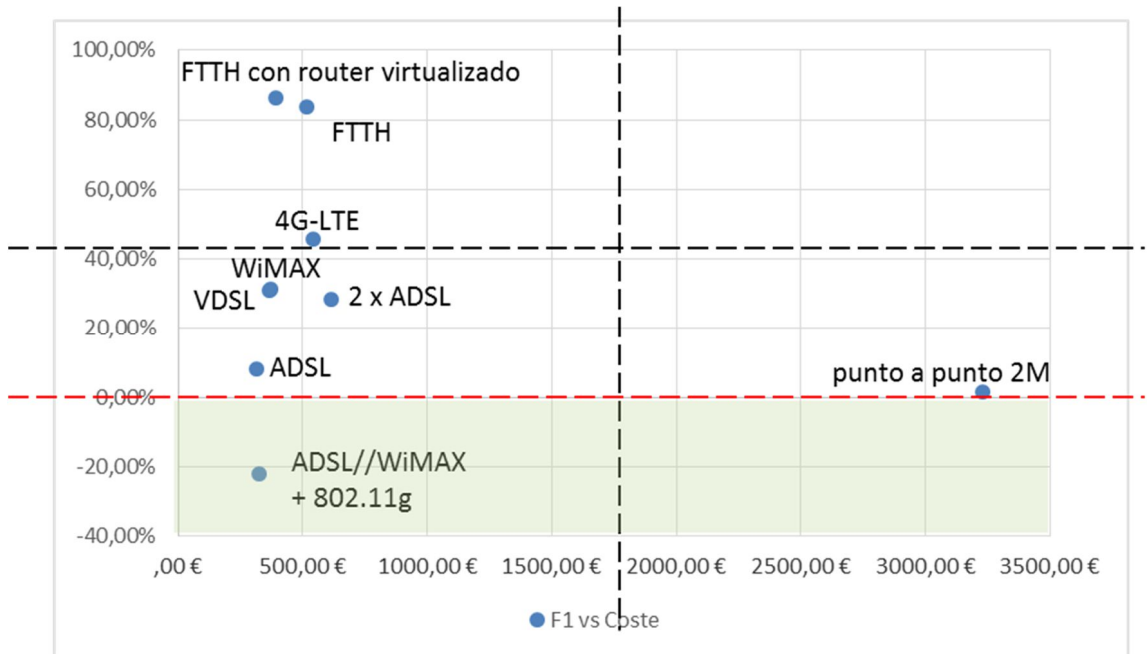

*Figure 5.7: Example of a graphical comparison of access technologies based on benefits (%) vs. Annual economic cost per user.$F_1$*

In the graph, the lines that determine the quadrants will be located in the middle of the range of positive real values for each axis in order to have a graphic reference. The region of negative values of is discarded (shaded in the figure). Therefore, the vertical is located at the economic value corresponding to the ADSL cost plus half the difference between the 2M point-to-point cost and the ADSL cost. The horizontal line is located in the percentage value corresponding to the value of ADSL plus half the difference between the value of for FTTH with virtualized router and that of for ADSL. Therefore, technologies can be classified according to the quadrant they occupy, taking into account that they are always being compared based on unique user criteria, in:$F_1 F_1 F_1 F_1$

- + performance and - cost: In this example, FTTH with virtualized router, FTTH, 4G-LTE
- - performance and - cost: In this example, VDSL, WiMAX, 2 x ADSL, ADSL. ADSL // WiMAX + 802.11g technology is discarded as it provides negative value of.$F_1$
- + benefits and + cost: In this example none.
- - performance and + cost: In this example, point-to-point 2M.

The optimal quadrant is the + performance and - cost quadrant. The least beneficial quadrant is - performance and + cost. Technologies with negative values of will be discarded by default.$F_1$

Below is the graphical comparison of access technologies based on economic efficiency vs. Annual economic cost per user.$F_2$

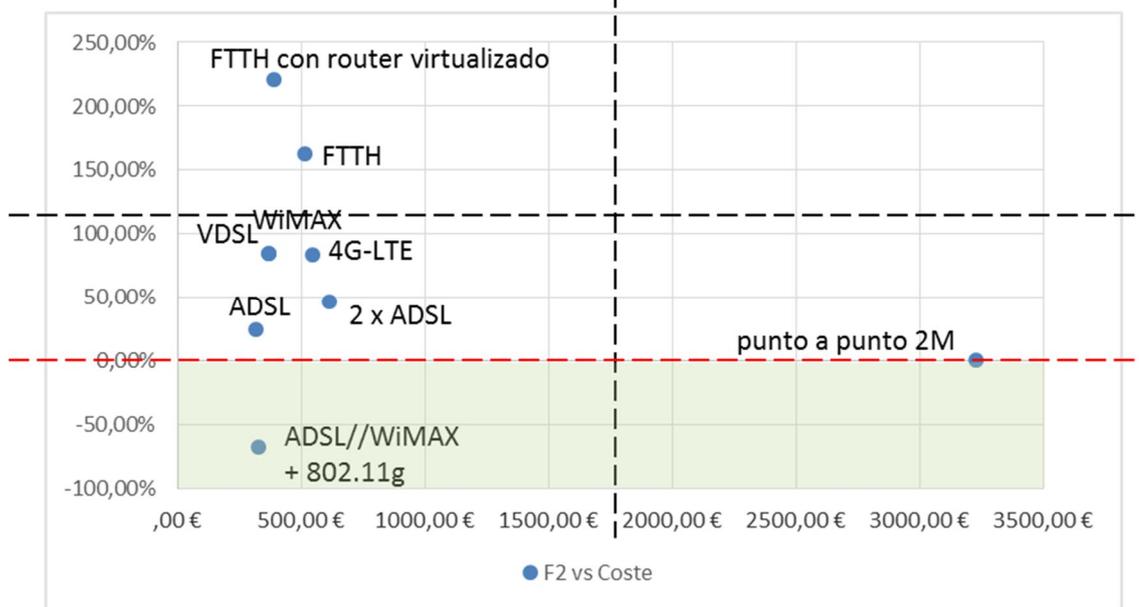

*Figure 5.8: Example of a graphical comparison of access technologies as a function of efficiency (% / €) vs. Annual economic cost per user.$F_2$*

The criterion for establishing the lines that delimit the quadrants is identical to that of the graph vs. Cost, allowing the following classification of access technologies based on the quadrant they occupy:$F_1$

- + efficiency and - cost: In this example, FTTH with virtualized router and FTTH.
- - efficiency and - cost: In this example, VDSL, WiMAX, 4G-LTE, ADSL and 2xADSL. ADSL // WiMAX + 802.11g is discarded for providing negative value of.$F_2$
- + efficiency and + cost: In this example none.
- - efficiency and + cost: In this example, point-to-point 2M.

The optimal quadrant is the + efficiency and - cost quadrant. The least beneficial quadrant is - efficiency and + cost. Technologies with negative values of will be discarded by default.$F_2$

This classification based on vs. Cost allows in a first iteration to identify the study quadrant in which to focus subsequent iterations. In this example, the optimal quadrant identifies FTTH technologies with virtualized router and FTTH. Therefore, the user of the model can propose a second iteration in order to discriminate between said technologies, either by proposing a variation in the access components or by resorting to other data sources in order to fine-tune the introduction of the input parameters and continue refining. .$F_2$

In this example, the refining to be carried out in the optimal quadrant considering as much as the cost could lead to decisions to deploy FTTH with a virtualized router where said technology is available due to its higher technical performance, FTTH in the event that there is no virtualization possible or even move towards the second most beneficial quadrant (- efficiency, - cost) by proposing WiMAX for mobile users and even due to its lower cost, VDSL for fixed users instead of FTTH, as is the case of the decision made by the operator TELENOR in Norway [135].$F_2$

In the bivariate graphical comparisons presented, it must be taken into account that the results are conditioned by the established decision criteria that determine the coefficients of the polynomials that make up the figures of merit and. It should be taken into account that, according to the criteria established in this case, the denominator polynomial in the figure of merit is the economic cost, which is why it is considered that, in this case, it is accounting for the economic efficiency of each technology access.$F_1 F_2 F_2 F_2$

The interpretation of the graphs will naturally vary depending on the agent who is using the model.

The model also allows changing the decision criteria and customer requirements, in such a way that the values of the figures of merit y, as well as the minimum value N of redundant accesses will vary. Likewise, the meaning of the figure of merit will be modified since it is a quotient of polynomials in whose denominator one or other variables can be activated as long as they are of the same nature in order to guarantee a consistent interpretation, such as and as established in the Model Chapter Proposals$F_1 F_2 F_2$

### Case B. Requirements for a SME-type user client with a minimum reception bandwidth of 300 Mbits / s (Data for 2015)

Below we will consider as an example some customer requirements of a medium SME type user in Europe, considering that said SME works in the service sector and requires high bandwidth requirements 3 times higher than those of a residential user with intensive use of 4K UHD TV due to the fact that it provides consulting and engineering services with mobile users who remotely access the desks of their PCs in the office and the need to send graphics and plans, thus establishing a minimum bandwidth requirement in reception. 300 Mbits / s.

Below is the comparative table of results offered by the proposed UTEM model for this example of a SME-type user.

| | ACCESS TECHNOLOGIES | ADSL | ADSL // 802.11g + WiMAX Backhaul with PC | Point to point 2M | FTTH with virtualized router | 4G-LTE | FTTH | VDSL | WiMAX Access Point + Backhaul WiMAX 802.16a | 2 x ADSL | Minimum customer requirements | Wei ghts ak | Wei ghts bp |
|---|---|---|---|---|---|---|---|---|---|---|---|---|---|
| | Output Parameters | Values | Values | Values | Values | Values | Values | Values | Values | Values | Values | | |
| RECEPTION SPEED | AVERAGE Bandwidth per user in access (Mbits / s per user) | 10 | 10.0912 | two | 100 | 24 | 100 | fifty | 2.7538 | twenty | 300 | 1 | 0 |
| EMISSION SPEED | AVERAGE Bandwidth per user in access (Mbits / s per user) | 3 | 3.0924 | two | 10 | 8 | 10 | 5 | 2,7641 | 6 | 30 | 1 | 0 |
| AVAILABILITY | Availability | 99.9597% | 100,000% | 99.9681% | 99.9952% | 99.9952% | 99.9712% | 99.9597% | 99.9960% | 99.9962% | 99.995% | 0.1 | 0 |
| DISTANCE | Distance user to access point (meters) | 4500 | Four. Five | 5000 | 15000 | 15000 | 15000 | 600 | 3000 | 4500 | twenty | 1 | 0 |
| | Total distance user to access node (m) | 4,500 | 4,500 | 5,000 | 15,000 | 15,000 | 15,000 | 600 | 48,000 | 4,500 | twenty | 1 | 0 |
| COST | CAPEX + OPEX (Year1) | € 315.04 | € 327.04 | € 3,230.00 | € 390.00 | € 543.00 | € 515.00 | € 365.00 | € 370.00 | € 615.00 | N / A | 0 | 1 |
| QoS | QoS capability | TRUE | TRUE | TRUE | TRUE | TRUE | TRUE | TRUE | TRUE | TRUE | N / A | 0 | 1 |
| THE | LOS from user to access point (Line of Sight Necessary?) | N / A | TRUE | N / A | N / A | N / A | N / A | N / A | FAKE | N / A | N / A | -1 | 0 |
| | LOS from access point to access network node required? | N / A | TRUE | N / A | N / A | N / A | N / A | N / A | TRUE | N / A | N / A | -1 | 0 |
| LICENSE | Do you need a license? | FAKE | TRUE | FAKE | FAKE | FAKE | FAKE | FAKE | TRUE | FAKE | N / A | -1 | 0 |
| Ubiquity | Ubiquity at customer's address | YES | YES | YES | YES | YES | YES | YES | YES | YES | N / A | 1 | 0 |
| Health | Probability of provoking reluctance due to health risk (0 = NONE; 1 = LOW; 2 = MEDIUM; 3 = HIGH) | 1 | two | 1 | 1 | 1 | 1 | 1 | 1 | 1 | N / A | -1 | 0 |
| | F1 (Performance) | -11.44% | -37.72% | -10.17% | 9.47% | 5.39% | 4.62% | -11.47% | -7.46% | -2.36% | | | |
| | F2 (Economic efficiency) | -36.32% | -115.35% | -3.15% | 24.27% | 9.93% | 8.96% | -31.43% | -20.15% | -3.84% | | | |
| | N | 30 | 30 | 150 | 3 | 13 | 3 | 6 | 253 | fifteen | | | |

*Table 5.23: Comparison of results offered by the proposed UTEM model for this example of a SME-type user.*

Below are the comparative graphs of technical performance and economic efficiency for the SME-type user example.$F_1$$F_2$

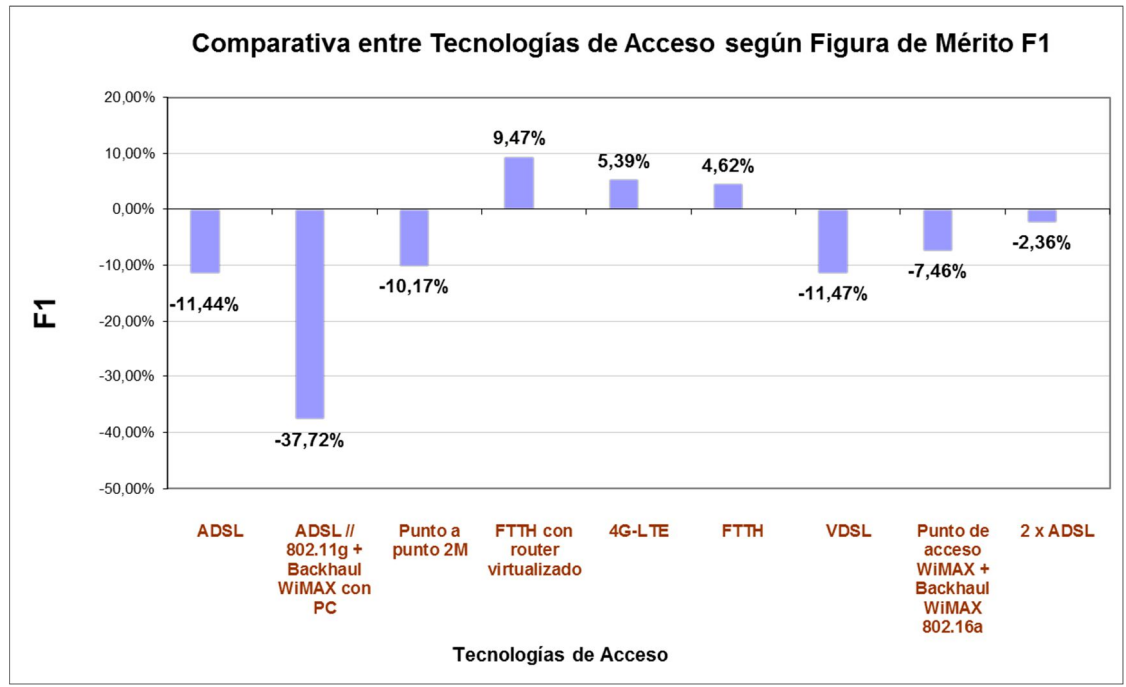

*Figure 5.9: Comparison of access technologies according to technical features for example of SME-type user client requirements*$F_1$

In this case, Figure 5.9 identifies the FTTH technologies with virtualized router, 4G-LTE and FTTH as the technologies that meet the customer requirements established in this example. The rest of the technologies present negative values for which they have to be discarded.$F_1$

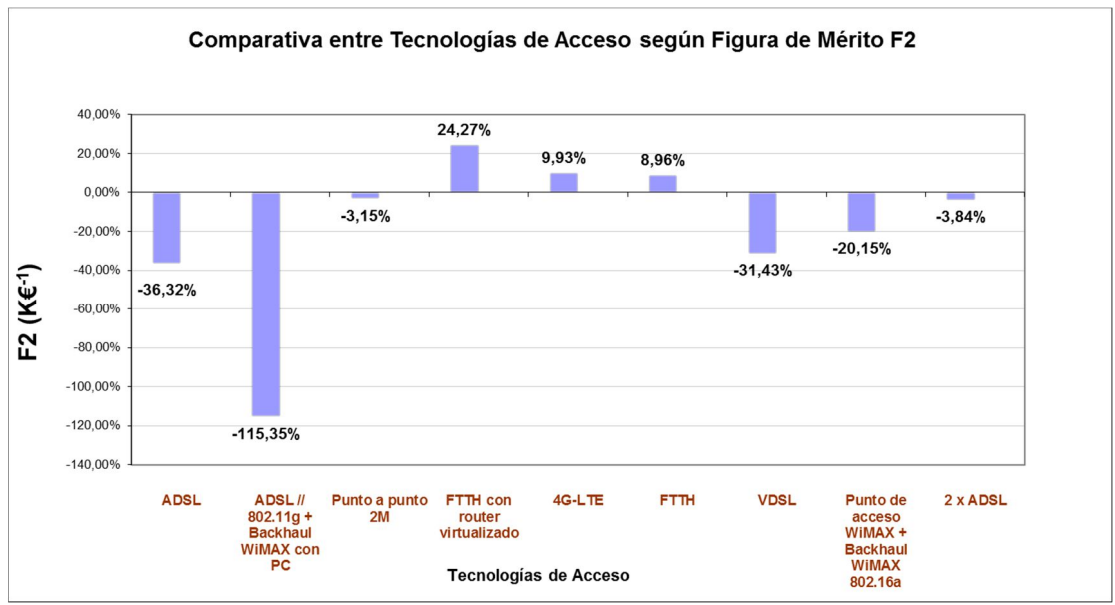

*Figure 5.10: Comparison of access technologies according to economic efficiency for example of SME-type customer user requirements$F_2$*

The comparison based on economic efficiency identifies the three technologies mentioned above with positive economic efficiency, highlighting FTTH with virtualized router as the most efficient in this example.$F_2$

## CASE C. Residential user client requirements with a minimum reception bandwidth of 2 Mbits / s (2006 data).

The following will be considered as an example the customer requirements of a Residential type user in 2006 that meets the requirements established according to the European Lisbon Strategy and the i2010 Digital Agenda with a minimum bandwidth in reception of 2 Mbits / s [ 40] [78].

Below is the comparison table of scenarios for this example of customer requirements. Note that 4G-LTE technology was not available in 2006. Even so, it remains in the study in order to allow its comparison with other technologies under these requirements established as an example.

| | ACCESS TECHNOLOGIES | ADSL | ADSL // 802.11g + WiMAX Backhaul with PC | Point to point 2M | FTTH with virtualized router | 4G-LTE | FTTH | VDSL | WiMAX Access Point + Backhaul WiMAX 802.16a | 2 x ADSL | Minimum customer requirements | Weights ak | Weights bp |
|---|---|---|---|---|---|---|---|---|---|---|---|---|---|
| | Output Parameters | Values | Values | Values | Values | Values | Values | Values | Values | Values | Values | | |
| RECEPTION SPEED | AVERAGE Bandwidth per user in access (Mbits / s per user) | 10 | 10.0912 | two | 100 | 24 | 100 | fifty | 2.7538 | twenty | two | 1 | 0 |
| EMISSION SPEED | AVERAGE Bandwidth per user in access (Mbits / s per user) | 3 | 3.0925 | two | 10 | 8 | 10 | 5 | 2,7641 | 6 | 0.2 | 1 | 0 |
| AVAILABILITY | Availability | 99.9597% | 100,000 % | 99.9681% | 99.9952% | 99.9952% | 99.9712% | 99.9597% | 99.9960% | 99.9962% | 99.99% | 0.1 | 0 |
| DISTANCE | Distance user to access point (meters) | 4500 | Four. Five | 5000 | 15000 | 15000 | 15000 | 600 | 3000 | 4500 | twenty | 1 | 0 |
| | Total distance user to access node (m) | 4,500 | 4,500 | 5,000 | 15,000 | 15,000 | 15,000 | 600 | 48,000 | 4,500 | twenty | 1 | 0 |
| COST | CAPEX + OPEX (Year 1) | € 315.04 | € 327.04 | € 3,230.00 | € 390.00 | € 543.00 | € 515.00 | € 365.00 | € 370.00 | € 615.00 | N / A | 0 | 1 |
| QoS | QoS capability | TRUE | TRUE | TRUE | TRUE | TRUE | TRUE | TRUE | TRUE | TRUE | N / A | 1 | 0 |
| THE | LOS from user to access point (Line of Sight Necessary?) | N / A | TRUE | N / A | N / A | N / A | N / A | N / A | FAKE | N / A | N / A | -1 | 0 |
| | LOS from access point to access node required? | N / A | TRUE | N / A | N / A | N / A | N / A | N / A | TRUE | N / A | N / A | -1 | 0 |
| LICENSE | Do you need a license? | FAKE | TRUE | FAKE | FAKE | FAKE | FAKE | FAKE | TRUE | FAKE | N / A | -1 | 0 |
| Ubiquity | Ubiquity at customer's address | YES | YES | YES | YES | YES | YES | YES | YES | YES | N / A | 1 | 0 |
| Health | Probability of provoking reluctance due to health risk (0 = NONE; 1 = LOW; 2 = MEDIUM; 3 = HIGH) | 1 | two | 1 | 1 | 1 | 1 | 1 | 1 | 1 | N / A | -1 | 0 |
| | F1 (Performance) | 75.83% | 47.07% | 47.31% | 350.35% | 191.93% | 347.95% | 172.27% | 173.69% | 145.49% | | | |
| | F2 (Economic efficiency) | 240.71% | 143.92% | 14.65% | 898.33% | 353.47% | 675.63% | 471.96% | 469.44% | 236.57% | | | |
| | N | two | two | two | two | two | two | two | two | two | | | |

Table 5.24. Comparison of scenarios for example with residential user client requirements and minimum bandwidth in reception of 2 Mbits / s according to criteria established by the European Lisbon Strategy and European Agenda i2010 [40] [78].

Below is the comparative graph of scenarios based on the figure of merit of technical performance (%).$F_1$

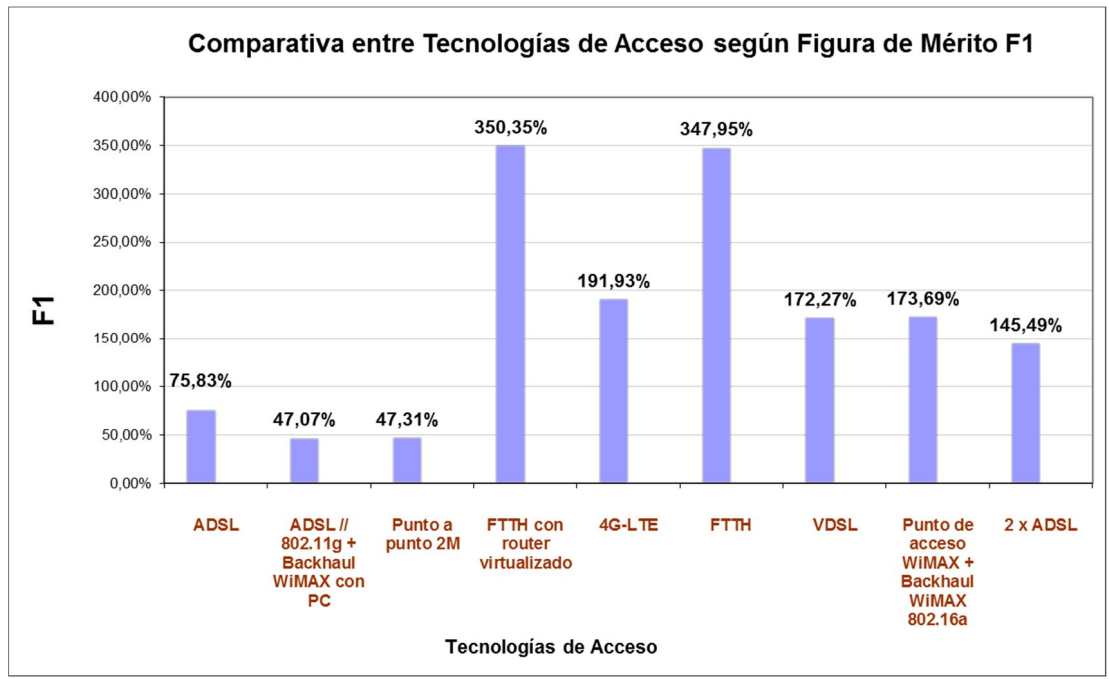

*Figure 5.11: Comparison of scenarios based on technical performance with example of Residential user customer requirements in 2006 according to the requirements of the European Agenda i2010 [40] [78].*$F_1$

We will discard in the graph the non-existent technologies in 2006 such as 4G-LTE and FTTH with virtualized router. It can be seen how the ranking of technical performance is topped by FTTH, followed by WiMAX, VDSL; 2xADSL and ADSL which were the existing deployment alternatives at that time.$F_1$

Below is the comparison of scenarios based on economic efficiency.$F_2$

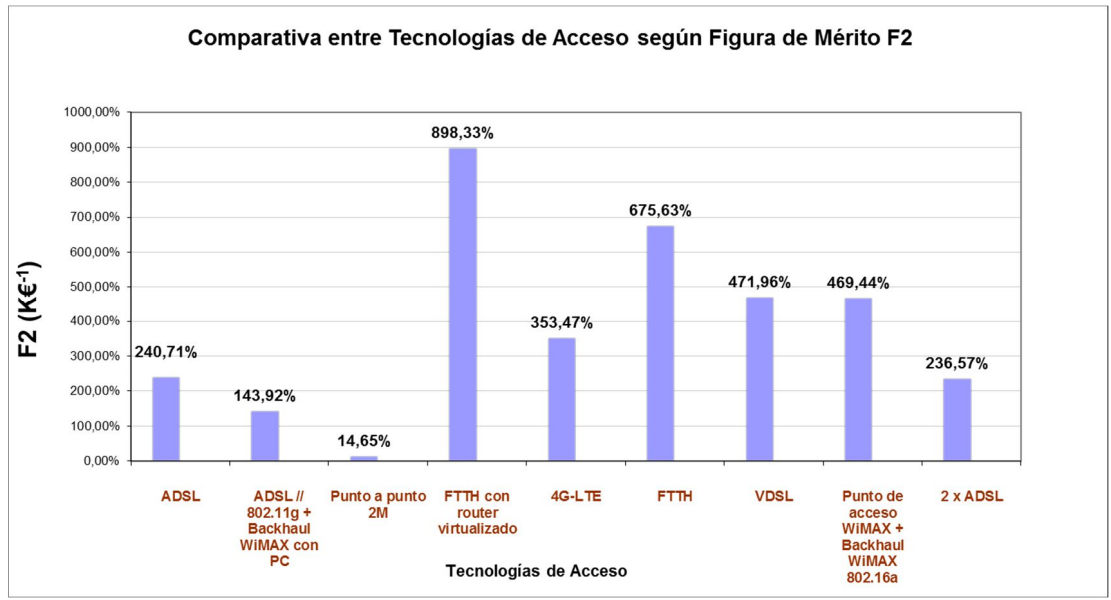

*Figure 5.12: Comparison of scenarios based on economic efficiency in the example of Case C (Year 2006).$F_2$*

As can be seen, discarding the FTTH scenarios with virtualized router and 4G-LTE as non-existent in 2006, the highest economic efficiency corresponds to FTTH followed by VDSL, WiMAX, ADSL and 2 x ADSL.

Note that in this case, for simplicity, the bandwidth requirements have been exclusively varied, keeping the rest of the requirements identical to Case A. Any incremental variation in FTTH costs to adjust to the real situation of FTTH deployment in 2015, would identify VDSL technology as the most likely to be developed in areas with deployed copper pairs and WiMAX technology in rural areas for reasons of economic efficiency, as was actually the trend during the 2006-2011 period corresponding to the European Agenda i2010 [ 40] [78].

### 5.2.3.4 Conclusions

According to the above, the proposed model allows establishing customer requirements in order to verify which access technologies and under what conditions they allow certain requirements to be met. For example, the proposed model allows to analyze the fulfillment of the access technologies requirements to support the prediction of the Analysis Mason analysis signature regarding UHD 4K / 8K TV as Killer Application [135] as it has been possible. check in the first example, having chosen as customer requirements those established by said prediction, which in turn are consistent with the objectives of the European Digital Agenda 2020 [78].

In this section, the results of the proposed model have been shown in a "micro" approach using the isolated application methodology described in the Methodology Chapter. It has been proven that it is applicable to technologies of different nature:

fixed and wireless, and to any combination of them by using the serial and parallel sub-models described in the Proposed Model Chapter.

As mentioned above, the model has been validated using fixed and wireless technologies that are weighing most telecom operators around the world. In any case, it is a universal model that can be perfectly applied to mobile technologies such as GSM, UMTS, ... satellite communications, free-space optics (FSO), etc., as well as virtualized access technologies.

As indicated at the beginning of this chapter, the results of the proposed model have been presented, considering as an example the final customer requirements and decision criteria set out, regardless of who or what type of market agent is the user of the proposed model ( end user of the technology, telecommunications operator - infrastructure area, economic control area, pre-sales technical support area, etc. -, regulator, etc.).

## 5.3 Quantitative Validation in Scenarios of Combined Application of the Model

In order to validate the results of the UTEM model in a "macro" ('top-down') approach, using the Combined Application methodology, the topological data of the literature model [58] are considered, for example, as it is the best positioned in the ranking presented in section 2.5.

Following the methodology developed for the integrated application of the model: Step 1: "Obtaining data from other models", when mapping the topological data from [58] in the UTEM model, identifying the different components of individual access and incorporating in Step 2: "Incorporate as inputs the outputs of other models", the information derived from the access network as a whole in the economic input parameters: income / ARPU and cost (CAPEX, OPEX), and apply Step 3: "Application of the Methodology Isolated from the Model ", it is concluded that the technical output results obtained in a" macro "approach are the same results of applying the model by itself in a" micro "approach following the steps outlined in the Methodology Chapter:

1. Systematize the access description
2. Application of the Access Technologies Characterization Module
3. Application of the Redundancy Calculation Module
4. Access Technology Comparison Module Application
5. Choice of technology and access configuration

Since the formulation of the economic parameters is universal and identical in all the models in the literature and in the proposed UTEM model, the economic results at the "macro" level will always coincide with those of the imported model, allowing the UTEM model to always complement the economic exit from any other model with the technical characterization of the access technology and configuration under study.

Now, if in the future models arose that developed the evaluation of technical feasibility, it would be possible to import them as described in the methodology, as well as to contrast the technical results resulting from the application of the UTEM model in both a "macro" and an approximation. "Micro", given that according to the above, the technical results in both cases are coincident.

## 5.4 Quantitative validation of the predictive capacity of the model

Next, Figure 5.13 shows the evolution of economic efficiency over time, obtained by incorporating in the proposed model, the evolution of annual cost per user for ADSL and FTTH access technologies. Source: OVUM 2015 [136] [137].$F_2$

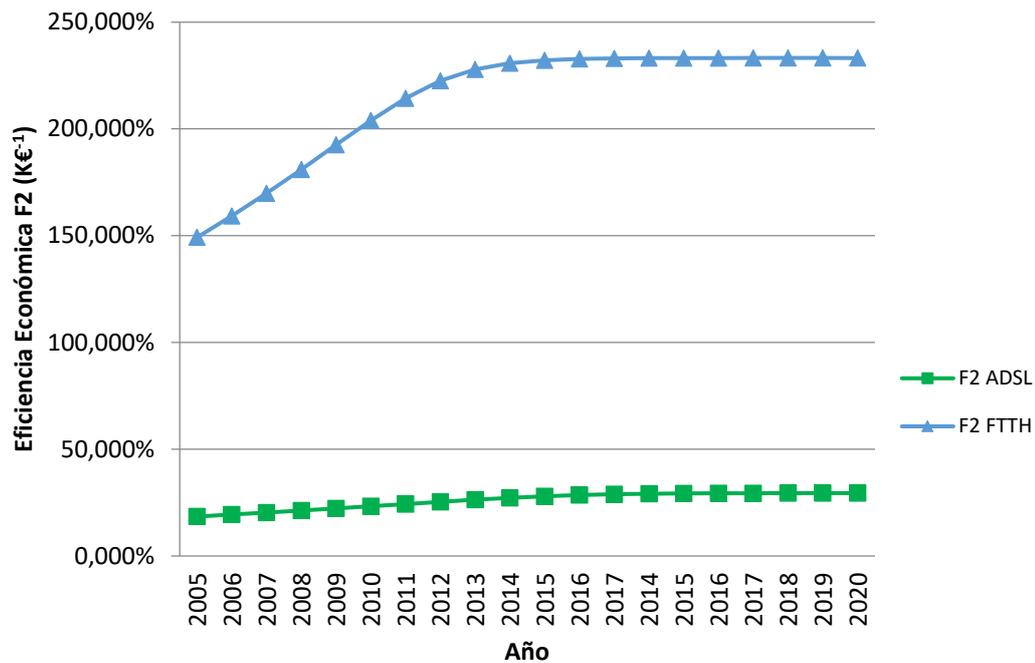

*Figure 5.13: Evolution of for ADSL and FTTH access technologies. Graph obtained by incorporating the evolution of annual cost per user for ADSL and FTTH according to data obtained from [136] [137].$F_2$*

Considering the evolution of economic efficiency for both ADSL and FTTH technologies and remembering that it accounts for technical performance per unit of cost, it appears from Figure 5.13 that the point from which it will be more profitable for a telecommunications operator to start Massive deployment of FTTH technology is the one from which the evolution curve reaches saturation (maximum benefits per unit of cost), which occurs practically in the 2014-2015 period.$F_2 F_2 F_2$

This hypothesis is compared with real data on the evolution of the mix of xDSL access technologies vs. FTTH in various countries.

Below is the real evolution and prediction until 2020 of the mix of xDSL accesses vs. FTTH in Sweden. Source: Analysis Mason 2015 [134].

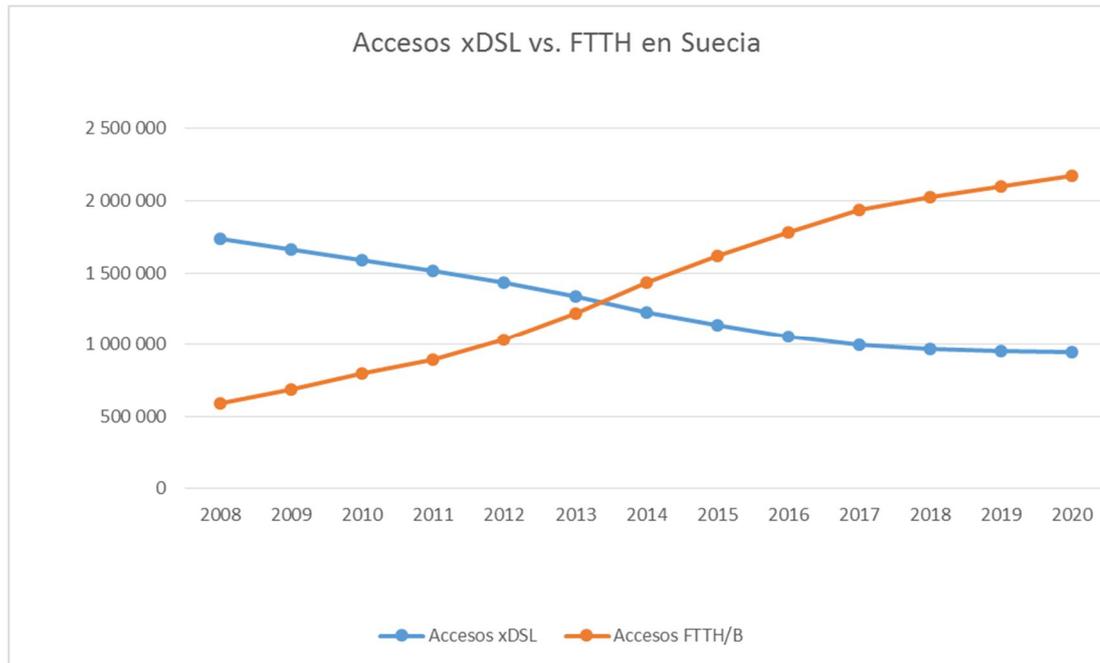

*Figure 5.14: Evolution of xDSL accesses vs. FTTH in Sweden. Source: Analysis Mason 2015 [134].*

It can be seen that the greatest growth slope of FTTH accesses in Sweden corresponds to the year 2014, at which time the number of FTTH accesses exceeds the decreasing number of xDSL accesses, given that the replacement of xDSL accesses by FTTH accesses is taking place. which is consistent with the saturation period identified for economic efficiency.$F_2$

The following figure shows said evolution of accesses in this case particularized for the United Kingdom. Source: Analysis Mason 2015 [134].

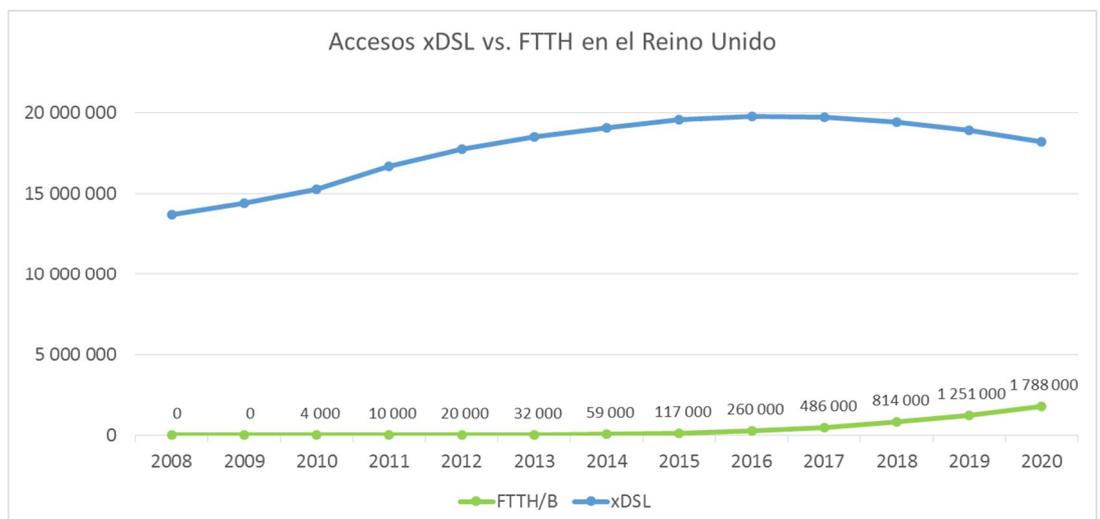

*Figure 5.15: Evolution of xDSL accesses vs. FTTH in the UK. Source: Analysis Mason 2015 [134]*

As can be seen in the figure, in the United Kingdom the deployment of FTTH is more modest than in Sweden, it begins in 2010 and begins to accelerate in 2014, 2015 and is expected in 2016 according to [134], which is consistent with the saturation period identified for economic efficiency.$F_2$

In the following figure, the evolution in Spain can be seen, accelerating the deployment of FTTH in 2014 with maximum slope in 2015 and anticipating a sustained rhythm, progressively replacing the xDSL plant with FTTH, which also coincides with the saturation period identified for efficiency economical.$F_2$

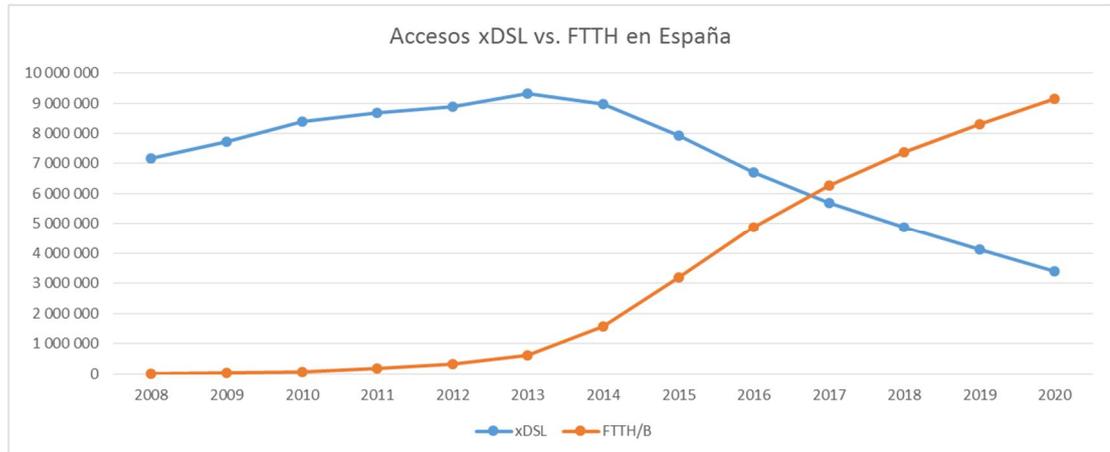

*Figure 5.16: Evolution of xDSL accesses vs. FTTH in Spain. Source: Analysis Mason 2015 [134]*

Therefore, it is concluded that it has been contrasted and validated with real data on the evolution of xDSL and FTTH access technologies, the hypothesis raised from the graph of economic efficiency evolution, obtained by incorporating the evolution in the proposed model annual costs per user according to [134$F_2$].

## 5.5 Validation of model results with results of other models

The proposed UTEM model provides technical results and economic results, in addition to the minimum number of redundant accesses necessary to meet customer requirements, as well as the figures of merit of techno-economic performance and techno-economic efficiency.$F_1F_2$

Given that the literature models do not provide technical results but economic results, we proceed to validate the hypothesis that the economic results of the proposed UTEM model are consistent with the economic results of the literature model with the best compliance in the ranking [58 ].

Given that the economic parameters of the UTEM model use a universal formulation, for example, the formulation of the NPV or NPV (Net Present Value), identical to the formulation of said parameter in the model [58], equal of input parameters of Income, CapEx and OpEx, the result will be exactly the same, thus being validated and demonstrated the hypothesis for any scenario.

It should be noted that the proposed UTEM model provides, in addition to the analysis and evaluation of the economic viability of access technologies, their technical characterization, technical analysis and comparison, and the evaluation of their viability from a technical point of view, considering the requirements established by the user of the model.

## 5.6 Conclusions

In this Chapter, the hypothesis raised in the introductory chapter has been contrasted, qualitatively and qualitatively validating the proposed UTEM techno-economic model and the application methodology developed.

A functional validation has been carried out based on the characteristics of the theoretical technical-economic model exposed in the State of the Art Chapter, concluding that the UTEM model presents a higher degree of compliance with these characteristics with respect to the literature models. Specifically, the UTEM model presents a compliance of 91% compared to 56% of the model with the highest degree of compliance in the literature [58].

A quantitative validation of the UTEM model has been carried out in 9 scenarios with a "micro" approach, using the Isolated Application methodology, *Y* A comparative analysis of the 9 access technologies mentioned in 3 different cases of technology user requirements has also been presented with data from 2015 and data from 2006, being at this point validated all the characteristics exposed for the UTEM model with the exception of the characteristic Geographical Universality, whose validation is demonstrated with the quantitative validation of the UTEM model in Combined Application scenarios.

Likewise, the prediction capacity of the model has been quantitatively validated by contrasting its results with the prediction of the analysis signature Analysis Mason, showing that the saturation period of the evolution of in time for FTTH technology corresponds to decision periods of massive deployment by telecommunications operators in 3 European countries.$F_2$

The quantitative validation of the results of the UTEM model with results of other models has also been demonstrated, since they use the same universal formulation in the economic output parameters.

The following table shows, as a summary, the validation of the UTEM model for each of the characteristics of the theoretical techno-economic model, in each of the stages of the validation procedure, throughout the sections of this Chapter.

| | Multi access univer sality | Unive rsality in Comb inatio n of acces s techn ologie s | Unive rsality in user orient ation | Unive rsality in incor porati ng "micr o" and "macr o" appro aches | Orient ation to User Requi remen ts of the model | Geogr aphic Unive rsality | Techn ical and Econ omic Unive rsality | Exten sibilit y and Flexib ility | Techn ical and econo mic comp arabili ty | Predi ctive Capac ity | Integr ation capac ity with other model s |
|---|---|---|---|---|---|---|---|---|---|---|---|
| **Functional validation of the UTEM model (section 5.1)** | ✔ | ✔ | ✔ | ✔ | ✔ | ✔ | ✔ | ✔ | ✔ | ✔ | ✔ |
| **Quantitative Validation Scenarios Isolated Application of the UTEM model (section 5.2)** | ✔ | ✔ | ✔ | ✔ | ✔ | • | ✔ | ✔ | ✔ | ✔ | ✔ |
| **Quantitative validation scenarios Combined application of the UTEM model (section 5.3)** | • | • | • | ✔ | • | ✔ | • | • | • | • | • |
| **Quantitative validation prediction capacity of the UTEM model (section 5.4)** | • | • | ✔ | • | • | • | • | • | • | ✔ | • |
| **Quantitative validation of results of the UTEM model with results of other models (section 5.5)** | • | • | • | • | ✔ | • | • | • | • | ✔ | ✔ |

*Table 5.25: Summary of validation of the UTEM model*

For all the above, it is concluded that the UTEM model is validated and contrasted the initial hypothesis raised in the introductory chapter, as an answer to the research question. Therefore, it is YES possible to define universally applicable, scalable, flexible and generalizable technical-economic models that make it possible to compare any access technologies in order to help the decision-making of the different market agents.

Consequently, the double objective of this research has been achieved:

- A universal, scalable, flexible and generalizable technical-economic model has been defined that allows the comparison of multiple access technologies in different scenarios.
- A methodology for applying the technical-economic model has been developed to facilitate its use by different market agents, providing guidelines for the application of the model and the proper interpretation of the results obtained.

Page intentionally left blank

# Chapter 6

# CONCLUSIONS

## 6.1 Synthesis of the research work

Throughout this doctoral thesis, the context in which this research work is inscribed has been presented: a very dynamic and changing telecommunications sector, made up of different agents whose technical and economic needs vary with great agility, and specifically the field of access networks, which, since it is the part of the telecommunications networks that requires the greatest capillarity in order to serve end users, is the part of the network with the highest volume of CapEx investment and OpEx operating costs consume.

As already indicated in the Introduction Chapter, access technologies have evolved in recent decades from dedicated point-to-point lines that allowed end users to connect with the points of presence of the operators' networks, towards access solutions and robust connectivity, cheaper than predecessor technologies, and continue to evolve in a dynamic and changing environment.

It has also been pointed out that there is a growing need for interconnection of devices of different nature and in very diverse environments coined in the Internet of Things, the Industrial Internet, connected vehicles or the Internet of Everything, which together with the improvement of video resolution in broadcast or streaming of Ultra High Definition 4K / 8K , causes the technical needs of bandwidth and availability, among others, to evolve, as well as the economic needs.

Given this evolution and dynamism of the telecommunications and access technologies sector, the need to define technical-economic models of universal application is detected to facilitate the choice of the most appropriate technical access network solution in each scenario, which satisfies the technical and economic needs of the different agents in the sector.

In this area, the technical-economic models of the literature have been studied, considering the following points relevant:

- Traditionally, technical-economic models are defined as a method used to evaluate the economic viability of complex technical systems [99], ignoring an authentic evaluation of technical viability that takes into account the requirements and preferences of users, which carries the risk to commit serious technical errors that can compromise the expected economic viability.

- The technical-economic models in the literature are eminently oriented towards the dynamics of deployment of access networks promoted by manufacturers and operators, ignoring the perspective of end users.
- Since the 90s, different projects with public funding have been developed within the framework of different European R&D programs with the aim of developing and evolving access networks, which have given rise to most of the existing literature regarding models of technical-economic evaluation of access technologies.

Given the limitations identified in the techno-economic models in the literature, the interest in finding optimal solutions for the access network, and his professional experience, the author has defined in this research work, eleven characteristics of a theoretical evaluation model technical-economic of access technologies, universal, generalizable, scalable and flexible: Universality multi-access, Universality in Combination of access technologies, Universality in user orientation, Universality in incorporation of "micro" and "macro" approaches, Orientation to Requirements User of the model, Geographical Universality, Technical and Economic Universality, Extensibility and Flexibility, Technical and Economic Comparability, Predictive Characteristic, Integrable with other models.

Based on the aforementioned characteristics of the defined theoretical model, a classification of the technical-economic models in the literature has been prepared and presented, showing as a summary a ranking that accounts for the degree of compliance of each model with respect to the characteristics of the theoretical model. , identifying a wide path or 'gap' between the model with the highest degree of compliance in the literature [58] and the defined theoretical model.

A new universal, flexible, scalable and generalizable access technology technical-economic evaluation model has been proposed and developed, called UTEM (Universal Techno-Economic Model: Universal Technical-Economic Model). The UTEM model has been defined in a modular way, composed of 3 blocks: 1) it obtains the vector of technical-economic characterization of any access technology in any configuration, 2) it calculates the minimum number of redundant accesses necessary for a technology to meet the requirements of client established by the user of the model, 3) allows the comparison of technologies, and the choice of the most appropriate, according to the preferences established by the user of the model.

Likewise, a Methodology for the application of the UTEM model has been defined that consists of two aspects: Isolated Application, and Combined or Integrated Application of the UTEM model with other models, for cases in which a "micro" approach is required (user perspective final), and a "macro" approach (deployment perspective), respectively, in order to allow the use of the UTEM model by any agents in the sector.

The definition and development of the proposed UTEM model and its application methodology have made it possible to maximize the degree of compliance of the UTEM model with respect to the characteristics of the defined theoretical model, as shown in the corresponding functional validation carried out.

Through the aforementioned functional validation and quantitative validation in scenarios of Isolated Application and Combined Application of the model, as well as its predictive capacity, it has been concluded that the model and methodology respond to expectations and meet the objectives of the present doctoral thesis.

## 6.2 Contributions

This doctoral thesis makes several original contributions that we group into three blocks.

On the one hand, the following contributions of conceptual redefinition and definition of characteristics are distinguished:

- The redefinition of the concept of the technical-economic model beyond the traditional scope, since it includes and underlines the inescapable need to carry out the evaluation of technical feasibility. From this perspective, this doctoral thesis redefines this concept as follows:

  *"The technical-economic models are a method that allows the evaluation of the technical and economic viability of complex technical systems."*

  This new definition emphasizes both the technical and economic aspects of modeling, considering the technical feasibility and the satisfaction of specific technical requirements and needs.
- The definition of the characteristics of a universal, flexible, generalizable and scalable theoretical technical-economic model for access technologies.

On the other hand, the State of the Art provides:

- The classification of the technical-economic models in the literature based on the characteristics defined for the theoretical technical-economic model.

Finally, as a result of the proposed model and methodology, the following contributions are made:

- The definition and development of a model and methodology for the technical-economic evaluation of access technologies, universal, generalizable, scalable and flexible, to a greater degree than the models in the literature, whose capacity for global technical and economic evaluation allows:
  - o the technical characterization of access technologies in any serial or parallel configuration
  - o obtain a figure of merit of technical and / or economic benefits for each technology
  - o Obtain a figure of merit for technical and / or economic efficiency for each technology.
  - o the possibility of introducing the technical and economic requirements of the client or end user.

- o the possibility of introducing the decision criteria or preferences of the model user by setting the weighting coefficients of the technical-economic parameters included in the formulation of the figures of merit.
- o identify the degree of compliance with the technical and / or economic customer requirements established by the user of the model for each technology
- o calculate the minimum number of redundant accesses that allows a given access technology to meet the established technical and / or economic customer requirements.
- o the technical-economic evaluation of redundant accesses and parallel combinations of the same or different access technology.
- o the extension of the concept of traditional technical-economic analysis beyond the evaluation of the economic viability of complex technical systems, adding and emphasizing the evaluation of technical viability, and therefore validating the proposed redefinition.
- o the prediction of the moment of decision of massive deployment of an access technology.
- o the comparison between any access technologies.

- The definition and methodology of application of the figures of merit of techno-economic benefits and techno-economic efficiency for the comparison of access technologies, conceived as a function of the linear combination of the coordinates of the difference vector between the vector of characterization of the technology of access under study and the vector of minimum user requirements. $F_1 F_2$
- The generalization of other models, through the Combined Application methodology of the UTEM model, to which the technical characterization of the access technology under study contributes, the minimum number of redundant accesses to meet customer requirements, as well as figures of merit of technical-economic performance and efficiency.
- The versatility in terms of the technical-economic evaluation of current and future access technologies.

## 6.3 Future lines of research

The present doctoral thesis leaves open, because they are out of its scope, several lines for future research, among which are those that seek to extend the application of the proposed model and methodology, such as:

- Extension of the model for technical-economic evaluation of alternatives in the trunk networks, which allow the aggregated traffic of the access networks to be carried.
- Extension of the model for technical-economic evaluation in other types of supply networks (gas, water, electricity, etc.).

And those collateral lines, which derive from contributions of this doctoral thesis, such as the concept of redundancy and parallel combinations of accesses:

- Load balancing systems between redundant accesses of identical access technology.
- Load balancing systems between parallel combinations of different access technologies.
- Design of virtualized network functions for the definition of load balancing systems.

Page intentionally left blank

# Glossary

TO

**AAPP:** See Public Administration.

**Public administration:** Set of State Administration bodies, whether at the central, regional or local level.

**ADSL (Asymetric Digital Subscriber Line, Asymmetric Digital Subscriber Line):** Transmission technology that allows conventional copper wires, initially used for telephony, to carry up to 2 Mbit / s over a medium-length subscriber pair. Like the rest of the xDSL solutions it does not have the need to replace existing cables, and it converts the copper pair that goes from the telephone exchange to the user into a medium for the transmission of multimedia applications.

**Bandwidth:** Technically it is the difference in hertz (Hz) between the highest and lowest frequencies of a transmission channel. Nevertheless,
This term is used very often to refer to the transmission speed.

**ASN**: Active Star Network: Active Star Network. One of the possible FTTH network architectures.

B

**Backbone:** Trunk network. Long-distance, high-capacity network to which smaller subsidiary networks are connected.

**Broadband:** This is the name given to communication channels whose transmission speed is much higher than that of a voice band channel. The definitions of Broadband have been changing as access technologies have provided higher transmission speeds. In the i2010 Digital Agenda in the EU, reception speeds greater than 2 Mbits / s in the access network were considered broadband.

**BB.DD .:** Databases.

**Bit (Binary unit):** Minimum unit of digital information, which is the discernment between two positions: affirmative or negative, 1 or 0, yes or no.

**Bit / s (Bits per second):** Unit of measurement of the transmission capacity of a telecommunication line.

**Client loop:**(Subscriber loop) Copper pair that connects the customer's home with the main dispatcher of the telecommunications operator's public exchange. See main dealer.

C

**Cable Modem:** Modulation and demodulation system for signals that broadcast by cable.

**CATV:**Cable TV. Formerly called Community Antenna Television (CATV). Communication system for the transmission of TV channels, original programming and services through coaxial cable.

**Loop quality.**Referred to xDSL. Result of the comparison between the theoretical or practical characterization of the customer loop with attenuation parameters in the frequency band used by the xDSL technology in question and the attenuation mask used to validate the suitability of the loop for said xDSL technology. In real measurements, the noise factor of the copper pair is considered to check if the S / N ratio is adequate. Theoretical validation results are usually expressed in the following terms: "POTENTIALLY VALID", "Doubtful", "NOT VALID". The results of practical validation are usually expressed in the following terms: "VALID", "NOT VALID" ..

**Carrier:** Physical infrastructure through which data, voice and image are transported. It also refers to the company that offers the signal transmission or conduction service.

**CATV:** Cable TV.

**CC.AA .:** Autonomous communities.

**Coaxial:** Signal conductor element, isolated and equipped with elements that minimize electromagnetic interference. Two copper conductors built around each other, separated by an insulating material and surrounded by an insulating cover. It is characterized by its significant bandwidth capacity and low susceptibility to interference.

**Coverage:** Geographical scope, space, surface in which signals whose physical medium is the radioelectric spectrum can be received .// Scope of a radioelectric emission .// Referred to xDSL: Area of influence of a public exchange of a telecommunications operator equipped with equipment xDSL technology. Despite the fact that there is coverage for a certain client, the situation may arise in which the quality of the client loop - copper pair - is not adequate for the xDSL technology used.

## D

**Data Center:** Infrastructure designed, implemented and specially conditioned in order to provide the operation with access to uninterrupted computer systems, one hundred percent secure, which allows hosting servers and / or content with the latest technologies as well as better connectivity, maintenance, etc.

**DSL (Digital Subscriber Line) (xDSL):** It encompasses the set of high-speed digital technologies to access the subscriber loop through the pair of copper wires.

**DTV (Digital Television)**: See digital television.

## AND

**EDGE (Enhanced Data for GSM Evolution):** It is a technology that provides the ability to handle services for the third generation of mobile telephony. EDGE was developed to allow the transmission of large amounts of data at higher speeds (384 kbps).

**Ethernet:** Communication standard that uses coaxial cable or unshielded twisted copper pair. It is the standard most used in Local Area Networks.

## F

**Optical fiber:** Communication line that allows the transmission of information by optoelectric techniques. It is characterized by high bandwidth (high capacity or transmission speed) and low signal loss.

**Frequency:** Number of cycles that a wave of the radioelectric spectrum carries out per second.

**FTTx (Fiber To The X):** General definition that refers to broadband technologies based on fiber optics.

## G

**GPRS (General Packet Radio Service):** (General Service Packages by Radio). Mobile phone communication service based on packet transmission. It can transmit at a speed of 114 kbit / s and allows connection to the Internet. It is a transition technology between GSM and UMTS systems, sometimes referred to as 2.5 G.

**GSM (Global System for Mobile communication):** (Global System for Mobile communications). Digital cellular telephone system for mobile communications developed in Europe with the collaboration of operators, Public Administrations and companies. European standard that operates in the 900 and 1,800 MHz bands. It constitutes the second generation of mobile telephony. In Europe it is identified with the 2G.

## H

**HDTV (High Definition Television):** (High definition television). Technology that defines a standard for the emission and reception of television signals with higher definition (around double) than the current one. Higher definition offers better quality and sharpness of images.

**Hertz:** Denomination of the frequency unit defined by the cycle / second ratio.

**HFC:** Hybrid Fiber Coax. Hybrid Fiber Coaxial access technology used in so-called cable networks.

**Hot Spot:** Also called wireless access points, they define coverage areas in which the Internet can be accessed using some type of wireless technology, such as Wi-Fi or Bluetooth, if you have the appropriate WLAN equipment (devices and cards).

**Housing:** Physical space rental service for hosting the servers of an organization, company or individual at the provider's facilities. The provider will guarantee physical and logical security, Internet connectivity and the inclusion of services in its monitoring system. The client, for his part, will remotely carry out the configuration and maintenance tasks of the hardware and software hosted.

**HRN:** Home Run Network. A possible FTTH network architecture.

**HTML (Hyper Text Mark-up Language):** Programming language in which the WWW service pages are written, which allows the use of hypertext.

## I

**ICT (Common Telecommunications Infrastructure.):** ICT regulations regulate the necessary facilities and infrastructures in homes that allow access to the services and applications that characterize the Information Society.

**R&D:** Investigation and development.

**R + D + i:** Research, development and innovation.

**Internet:** Digital packet-switched network, based on TCP / IP protocols. It interconnects smaller networks (hence its name), allowing data transmission between any pair of computers connected to these subsidiary networks.

**Interoperability:** Set of the characteristics of a system that allow operation on a variety of media and between equipment from different manufacturers.

**IPSEC:** Set of standard protocols developed by the IETF (Internet Engineering Task Force) to provide secure communications services over the Internet. It defines two IP packet security protocols as well as the Internet key exchange procedure.

**ISP (Internet Servide Provider):** (Internet Service Provider). Organization, usually for profit, that in addition to giving Internet access to individuals and / or legal entities, offers them a series of data between any pair of computers connected to these subsidiary networks.

**IST (Information Society Technologies):** European Union program within the R&D Framework programs.

**IT (Information Technology):** Information technology.

**ITU / UIT (International Telecommunications Union):** International Telecommunications Union.

.

## L

**LAN (Local Area Network):** According to the IEEE (Institute of Electrical and Electronics Enginners), a data communication system that allows a certain number of devices to communicate directly with each other within a reduced geographic area, using physical communication channels of moderate or high speed.

**LMDS (Local Multipoint Distribution System):** LMDS is a radio technology that has been developed for broadband wireless local access. Provides access to voice, data, Internet and video services. Use the 25 GHz (or higher) radio band.

## M

**Mbits (Megabits):** Measurement of the amount of information transmitted in a communication medium equivalent to 1,048,576 bits.

**MDF:** Acronym in English: Main Distribution Frame: literally Main Distribution Panel or Main Distributor. See main dealer.

**MHz (Megahertz):** Frequency measurement corresponding to 1,000 Hz

**MMDS (Multichannel Multipoint Distribution System):** Microwave Television Distribution. System that allows, in reduced geographical environments, to transmit several TV channels and support interactivity, which makes it possible to offer interactive audiovisual services. The return channel is made through an analog telephone line (Basic Telephone Service STB) or ISDN (Integrated Services Digital Network). It can be integrated with radio telephony in the same MMDS infrastructure.

**Modem:** Acronym for modulator / demodulator. Designates the device that converts digital signals to analog, and vice versa, and allows communication between two computers over a normal telephone line or a cable line (cable modem or cable modem).

**Multimedia:** Digitized information that combines various types of information, such as text, graphics, still or moving image, sound, etc.

## OR

**Telecommunications operator:** Company or entity that offers telecommunications services.

**OTT:** Over The Top. Content service providers that take advantage of the communications infrastructures that telecommunications operators provide.

## P

**Copper pair:** Communication line consisting of two copper conducting wires.

**PLC (Power Line Communications):** Internet access through the electrical network. See Chapter 2 of the State of the Art for more details.

**PUT:** Passive Optical Network: Passive Optical Network. FTTH network technology.

**SME:** Small and medium business.

## R

**RACE (Research and technical development in Advanced Communications technologies in Europe):** European program on research and technical development of Advanced Communications technologies.

**Radio link:** Radio equipment that allows the establishment of a set of communications between two fixed points.

**ISDN (Integrated Services Digital Network):** It combines voice and digital services through the network in a single medium, making it possible to offer customers digital data services as well as voice connections through a single "cable" (copper pair), through two channels. 64 Kbits / s.

**Access network:** Part of the telecommunications networks that connect each particular place (home, office, etc.) with the headquarters of the telecommunications operator to which it belongs, giving access to long-distance switching and transmission systems.

**Communications network:** It is the set of links and interconnections (made by copper pairs, coaxial cables, optical fibers, radio waves, infrared or any other means) between various electronic devices (including computers) that enables transmission, including , of both analog and digital signals.

**Long distance network:** See Transport network.

**Transport network:** Part of the telecommunications networks that connect some cities with others (or regions, or countries, even continents), also sometimes called a long-distance network. Users connect to it through the access network.

**Digital network:** Communications network through which information circulates in digital format (see Digital signal).

**Fixed network:** Communications network accessed from fixed locations whose situation does not vary over time.

**Main delivery person:** Main connection panel of a public exchange of a telecommunications operator in which the copper pairs coming from customer homes and terminal boxes located in the urban network terminate, as well as the interfaces of the exchange equipment. The interconnection bridges between the customer copper pair and the corresponding exchange equipment interface are laid on it to provide any service. It is called MDF in English. See MDF.

**VPN (Virtual Private Network):** It is a private network that extends through a process of encapsulation and encryption of data to different remote points through the use of public transport infrastructures, allowing you to enjoy the characteristics of a private network (confidentiality, security, access to corporate information) to through a public access.

**RTB (Basic Telephone Network):** National coverage network specially developed for the provision of telephone service, that is, for voice transmission.

**RTC (Switched Telephone Network):** Similar concept to RTB, but that places the emphasis on the circuit-switched technology on which it is based, as opposed to point-to-point data links. The PSTN concept encompasses both the Basic Telephone Network (RTB) and the Integrated Services Digital Network (ISDN).

## S

**Digital signal:** A signal is digital when it is discretized, that is, the signal's variation margins have both upper and lower limits and, furthermore, the signal cannot take any value between these limits, but only some specific ones. The most typical example is that of a signal converted to zeros and ones.

**Interactive Broadcasting services:** They involve the provision of

**Society of Information:** A stage of social development characterized by the ability of its members (citizens, companies and Administration) Share any information, instantly, from anywhere and in the way you prefer.

**Software:** (Logical components, programs). Programs or logical elements that operate, or run on, a computer or network, as opposed to the physical components of the computer or network.

See also Hardware.

## T

**3G telephony:** Third generation mobile telephony that is identified with the IMT2000 standards issued by the ITU, among which is UMTS.

**YOU:** Information technology.

**TIC:** Information and Communication Technologies.

.

## OR

**ITU / ITU:** International Telecommunications Union / International Telecommunications Union.

**UMTS (Universal Mobile Telecommunication System):** A high-speed broadband cellular mobile telephone standard developed by the ETSI (European Telecommunications Standard Institute) is a third-generation system intended to replace GSM.

**UNED:** National University of Distance Education.

**URL (Uniform Resource Locator):** (Uniform Resource Locator). Unified system for identifying resources on the network. The addresses are made up of protocol, FQDN and WWW address, Gopher, FTP, News, etc.

## V

**VDSL (Very high rate Digital Subcriber Line):** Transmission technology, evolution of ADSL, which uses fiber optics and, in the final section of the connection with the subscriber, conventional copper wires, allowing transport of up to 52 Mbit / s.

**Transmission speed:** Amount of data that can be sent in a given period of time through a given communication circuit. It is measured in bit / s or, more commonly, in its multiples. The term "bandwidth" is sometimes used as an equivalent, although it is more correct to use "transmission speed".

W

**WAN (Wide Area Network):** Wide area networks. These networks intercommunicate equipment in a very large geographical area. Its extensions can be national, supranational and international. Internet is a clear example of WAN.

**Website:** (Website). A collection of web pages that are accessed through a unique URL.

**WiFi (Wireless Fidelity):** Acronym under which the IEEE 802.11b standard, and later standards, is hidden for wireless local networks (WLAN) that operates in the 2.4 GHz band (free to use without a license) and allows a theoretical maximum transmission speed of 11 Mbit / s. Internet access services are generally provided with this technology through so-called *hot spots* or wireless access points.

**Wireless:** No wires, wireless. In general, it refers to access technologies that do not use cables as the transmission medium but rather Hertzian waves in microwave frequencies, infrared, etc., which propagate directly through the air.

**WLAN (Wireless Local Area Network):** They are Local Area Networks (see definition) that a user can access through a wireless connection such as Bluetooth or Wi-Fi.

**WWW (World Wide Web, literally "mesh that covers the world"):** Hypertext-based distributed information server created in the early 1990s by Tim Berners Lee, a researcher at CERN, Switzerland. The information can be of any format (text, graphic, audio, still or moving image) and is easily accessible to users through browser programs.

X

**xDSL (Digital Subscriber Line):** (Digital Subscription Line). Generic name of the family of technologies that offer wide bandwidth through the conventional copper pair initially deployed for telephone service.

Page intentionally left blank

Page intentionally left blank

# ANNEXES

Page intentionally left blank

# A Additional Scenarios of Isolated Application contemplated in the Quantitative Validation of the UTEM model

## A.1 Scenario 3: WiMAX 802.16 access point with WiMAX 802.16 backhaul

The following tables show the input parameters of the model that are used as an example in this scenario, as well as the output parameters obtained by applying the UTEM model for this scenario. Any variation in input parameters, customer requirements, or user preferences will lead to different results.

**INPUT PARAMETERS** $x_{ij}$ of the UTEM MODEL

**SCENARIO NAME:** WiMAX 802.16 with WiMAX 802.16 backhaul

| | Input parameter | PC interface | Element 1 | Element 2 | Element 3 |
|---|---|---|---|---|---|
| **Element identification** | Element name | **RedMAX Subscriber Unit SU-O** | **Redline Communications AN-100 Broadband Wireless System WiMAX 802.16a / 802.16-2004** | **Redline Communications AN-100 Backhaul WiMAX 802.16a / 802.16-2004** | **Redline Communications AN-100 Backhaul WiMAX 802.16a / 802.16-2004** |
| | Element function | **User Adapter** | **Access point** | **End 1 Link 1** | **End 2 Link 1** |
| **Bandwidth** | Unitary Bandwidth (Mbits / s) (Reception) | 2. 3 | 54 | 43 | 43 |
| | Unitary Bandwidth (Mbits / s) (Emission) | 2. 3 | 54 | 43 | 43 |
| **Availability** | Availability | 99.9990% | 99.9990% | 99.9990% | 99.9990% |
| **Distance** | Distance (meters) | N / A | 3000 | 45000 | N / A |
| **QoS** | QoS capability | N / A | TRUE | TRUE | TRUE |
| **Redundancy** | Redundancy (No. elements in parallel) | 1 | 1 | 1 | 1 |
| **THE** | LOS (Line of Sight Needed?) | N / A | FAKE | TRUE | TRUE |
| **Frequency band** | Band (GHz) | 3.5 | 3.5 | 3.5 | 3.5 |
| **License** | Do you need a license? | TRUE | TRUE | TRUE | TRUE |
| **Users** | No. users: | 1 | 30 | N / A | N / A |
| **Concurrence** | Estimated average concurrency of users | N / A | 65.00% | N / A | N / A |
| **Technology** | Do you use wireless technology in any section? | YES | YES | YES | YES |
| **Environment** | Vector Environment (DENSE URBAN / URBAN / SUBURBAN / RURAL) | (0,0,1,1) | (0,0,1,1) | (0,0,1,1) | (0,0,1,1) |
| **Attenuation by meteorology** | Total decrease in Reception Bandwidth due to meteorological effects (Mbits / s) | 0.1 | 0.3 | 0.4 | 0.5 |
| | Total decrease in Broadcast Bandwidth due to meteorological effects (Mbits / s) | 0.1 | 0.1 | 0 | 0 |
| **Ubiquity** | Ubiquity at customer's address | YES | N / A | N / A | N / A |
| **Health** | Probability of provoking reluctance due to health risk (0 = NONE; 1 = LOW; 2 = MEDIUM; 3 = HIGH) | 1 | 1 | 1 | 1 |
| **K (interest rate)** | Type of interest | 1.00% | N / A | N / A | N / A |
| **ARPU year 1** | Average revenue per user (Year 1) | € 748.88 | | | |
| **ARPU year 2** | Average revenue per user (Year 1) | € 540.00 | | | |
| **ARPU year 3** | Average revenue per user (Year 1) | € 540.00 | | | |
| **CAPEX year 1** | Investments (Year 1) | € 20.00 | € 116.67 | € 116.67 | € 116.67 |
| **CAPEX year 2** | Investments (Year 2) | | | | |
| **CAPEX year 3** | Investments (Year 3) | | | | |
| **OPEX year 1** | Operating Expenses (Year 1) | € 0.0002 | € 0.0012 | € 0.0012 | € 0.0012 |
| **OPEX year 2** | Operating Expenses (Year 2) | € 0.0002 | € 0.0012 | € 0.0012 | € 0.0012 |
| **OPEX year 3** | Operating Expenses (Year 3) | € 0.0002 | € 0.0012 | € 0.0012 | € 0.0012 |

*Table A.1: Input parameters in WiMAX 802.16 scenario with WiMAX 802.16 backhaul.*

**OUTPUT PARAMETERS $y_k$ and Figures of merit F1 and F2**

| | Output Parameters | $y_k$ | $a_k \cdot \bar{y}_k$ | $b_P \cdot y_k$ | Weighted Valuation F1 | Weighted Valuation F2 |
|---|---|---|---|---|---|---|
| **RECEPTION SPEED** | AVERAGE Bandwidth (Mbits / s per user) | 2.753846154 | -0.3892 | 0.0000 | 16.46% | 22.24% / K € |
| **EMISSION SPEED** | AVERAGE Bandwidth (Mbits / s per user) | 2.764102564 | -0.0337 | 0.0000 | | |
| **AVAILABILITY** | Availability | 99.9960% | 0.0606 | 0.0000 | | |
| **DISTANCE** | Distance user to access point (meters) | 3000 | 0.0994 | 0.0000 | | |
| | Total distance user to access node (m) | 48,000 | 1,6004 | 0.0000 | | |
| **ARPU year 1** | Average revenue per user (Year 1) | € 748.88 | 0.0000 | 0.0000 | | |
| **ARPU year 2** | Average revenue per user (Year 1) | € 540.00 | 0.0000 | 0.0000 | | |
| **ARPU year 3** | Average revenue per user (Year 1) | € 540.00 | 0.0000 | 0.0000 | | |
| **CAPEX year 1** | Investments (Year 1) | € 370.00 | 0.0000 | 0.0000 | | |
| **CAPEX year 2** | Investments (Year 2) | € 0.00 | 0.0000 | 0.0000 | | |
| **CAPEX year 3** | Investments (Year 3) | € 0.00 | 0.0000 | 0.0000 | | |
| **OPEX year 1** | Operating Expenses (Year 1) | € 0.00 | 0.0000 | 0.0000 | | |
| **OPEX year 2** | Operating Expenses (Year 2) | € 0.00 | 0.0000 | 0.0000 | | |
| **OPEX year 3** | Operating Expenses (Year 3) | € 0.00 | 0.0000 | 0.0000 | | |
| **NPV** | Net Present Value at interest rate K | € 1,428.60 | 0.0000 | 0.0000 | | |
| **Net Cash Flow** | Net Cash Flow (interest rate K is not taken into account) | € 1,458.87 | 0.0000 | 0.0000 | | |
| **Payback Period (years)** | Amortization period | 1.00 | 0.0000 | 0.0000 | | |
| **COST** | CapEx + OpEx (year 1) | € 370.00 | 0.0000 | 370,0037 | | |
| **QoS** | QoS capability | TRUE | 1,0000 | 0.0000 | | |
| **THE** | LOS from user to access point (Line of Sight Necessary?) | FAKE | 0.0000 | 0.0000 | | |
| | LOS from access point to access node required? | TRUE | -1.0000 | 0.0000 | | |
| **LICENSE** | Do you need a license? | TRUE | -1.0000 | 0.0000 | | |
| **Ubiquity** | Ubiquity at customer's address | YES | 1,0000 | 0.0000 | | |
| **Health** | Probability of provoking reluctance due to health risk (0 = NONE; 1 = LOW; 2 = MEDIUM; 3 = HIGH) | 1 | -0.3333 | 0.0000 | | |

*Table A.2: Output parameters and F1 and F2 figures of merit in WiMAX 802.16 scenario with WiMAX 802.16 backhaul.$y_k$*
**OBTAINING THE MINIMUM NUMBER OF REDUNDANT ACCESSES R TO MEET CUSTOMER REQUIREMENTS**

| | Output Parameters | $y_k$ | Minimum rk value to meet customer requirements | |
|---|---|---|---|---|
| **RECEPTION SPEED** | AVERAGE Bandwidth per user in access (Mbits / s per user) | 2.753846154 | eleven | |
| **EMISSION SPEED** | AVERAGE Bandwidth per user in access (Mbits / s per user) | 2.764102564 | two | |
| **AVAILABILITY** | Availability | 99.9960% | two | |
| **DISTANCE** | Distance user to access point (meters) | 3000 | COMPLIES | CONCLUSION |
| | Total distance user to access node (m) | 48,000 | COMPLIES | YES IT COMPLIES WITH: |
| **COST** | CapEx + OpEx (Year 1) | € 370.00 | COMPLIES | |
| **QoS** | QoS capability | TRUE | COMPLIES | R = 11 |
| **THE** | LOS from user to access point (Line of Sight Necessary?) | FAKE | COMPLIES | |
| | LOS from access point to access node required? | TRUE | COMPLIES | |
| **LICENSE** | Do you need a license? | TRUE | COMPLIES | |
| **Ubiquity** | Ubiquity at customer's address | YES | COMPLIES | |
| **Health** | Probability of provoking reluctance due to health risk (0 = NONE; 1 = LOW; 2 = MEDIUM; 3 = HIGH) | 1 | COMPLIES | |

*Table A.3: Minimum number of redundant accesses R for WiMAX 802.16 technology with WiMAX 802.16 backhaul to meet established user requirements.*

## A.2 Scenario 4: 4G-LTE

The following tables show the input parameters of the model that are used as an example in this scenario, as well as the output parameters obtained by applying the UTEM model for this scenario. Any variation in input parameters, customer requirements, or user preferences will lead to different results.

**INPUT PARAMETERS** $x_{ij}$ of the UTEM MODEL

**SCENARIO NAME:** 4G-LTE

|  | Input parameter | PC interface | Element 1 | Element 2 | Element 3 |
|---|---|---|---|---|---|
| **Element identification** | Element name | 4G adapter | 4G Base Station | MUX 4G | Aggregation network |
| | Element function | 4G adapter | 4G Base Station | Access Node (access interface) | Access Node (aggregation interface) |
| **Bandwidth** | Unitary Bandwidth (Mbits / s) (Reception) | 24 | 100 | 100 | 100 |
| | Unitary Bandwidth (Mbits / s) (Emission) | 8 | 10 | 10 | 10 |
| **Availability** | Availability | 99.9962% | 100.0000% | 99.9990% | 100.0000% |
| **Distance** | Distance (meters) | N / A | N / A | 15000 | N / A |
| **QoS** | QoS capability | N / A | TRUE | TRUE | TRUE |
| **Redundancy** | Redundancy (No. elements in parallel) | 1 | 1 | 1 | 1 |
| **THE** | LOS (Line of Sight Needed?) | N / A | N / A | N / A | N / A |
| **Frequency band** | Band (GHz) | 0.8 | 0.8 | N / A | N / A |
| **License** | Do you need a license? | FAKE | TRUE | FAKE | FAKE |
| **Users** | No. users: | 1 | 1 | N / A | N / A |
| **Concurrence** | Estimated average concurrency of users | N / A | 100.00% | N / A | N / A |
| **Technology** | Do you use wireless technology in any section? | YES | YES | DO NOT | DO NOT |

| Environment | Vector Environment (DENSE URBAN / URBAN / SUBURBAN / RURAL) | (1,1,1,1) | (1,1,1,1) | (1,1,1,1) | (1,1,1,1) |
|---|---|---|---|---|---|
| Attenuation by meteorology | Total decrease in Reception Bandwidth due to meteorological effects (Mbits / s) | 0.1 | 0.3 | 0.4 | 0.5 |
| | Total decrease in Broadcast Bandwidth due to meteorological effects (Mbits / s) | 0.1 | 0.1 | 0 | 0 |
| Ubiquity | Ubiquity at customer's address | YES | N / A | N / A | N / A |
| Health | Probability of provoking reluctance due to health risk (0 = NONE; 1 = LOW; 2 = MEDIUM; 3 = HIGH) | 1 | 1 | 0 | 0 |
| K (interest rate) | Type of interest | 1.00% | N / A | N / A | N / A |
| ARPU year 1 | Average revenue per user (Year 1) | € 330.00 | | | |
| ARPU year 2 | Average revenue per user (Year 1) | € 300.00 | | | |
| ARPU year 3 | Average revenue per user (Year 1) | € 300.00 | | | |
| CAPEX year 1 | Investments (Year 1) | € 15.00 | € 100.00 | € 100.00 | € 100.00 |
| CAPEX year 2 | Investments (Year 2) | | | | |
| CAPEX year 3 | Investments (Year 3) | | | | |
| OPEX year 1 | Operating Expenses (Year 1) | € 0.0006 | € 0.0000 | € 0.0010 | € 0.0000 |
| OPEX year 2 | Operating Expenses (Year 2) | € 0.0006 | € 0.0000 | € 0.0010 | € 0.0000 |
| OPEX year 3 | Operating Expenses (Year 3) | € 0.0006 | € 0.0000 | € 0.0010 | € 0.0000 |

*Table A.4: Entry parameters in 4G-LTE scenario.*

**OUTPUT PARAMETERS $y_k$ and Figures of merit F1 and F2**

| | Output Parameters | $y_k$ | $a_k \cdot \overline{y}_k$ | $b_P \cdot y_k$ | Weighted Valuation | |
|---|---|---|---|---|---|---|
| | | | | | F1 | F2 |
| **RECEPTION SPEED** | AVERAGE Bandwidth (Mbits / s per user) | 23.9 | -0.08714 | 0.00000 | 38.22 % | 121.34 % / K € |
| **EMISSION SPEED** | AVERAGE Bandwidth (Mbits / s per user) | 7.9 | 0.70000 | 0.00000 | | |
| **AVAILABILITY** | Availability | 99.9952% | 0.05265 | 0.00000 | | |
| **DISTANCE** | Distance user to access point (meters) | 15000 | 0.49967 | 0.00000 | | |
| | Total distance user to access node (m) | 15,000 | 0.49967 | 0.00000 | | |
| **ARPU year 1** | Average revenue per user (Year 1) | € 330.00 | 0.00000 | 0.00000 | | |
| **ARPU year 2** | Average revenue per user (Year 2) | € 300.00 | 0.00000 | 0.00000 | | |
| **ARPU year 3** | Average revenue per user (Year 1) | € 300.00 | 0.00000 | 0.00000 | | |
| **CAPEX year 1** | Investments (Year 1) | € 315.00 | 0.00000 | 0.00000 | | |
| **CAPEX year 2** | Investments (Year 2) | € 0.00 | 0.00000 | 0.00000 | | |
| **CAPEX year 3** | Investments (Year 3) | € 0.00 | 0.00000 | 0.00000 | | |
| **OPEX year 1** | Operating Expenses (Year 1) | € 0.00 | 0.00000 | 0.00000 | | |
| **OPEX year 2** | Operating Expenses (Year 2) | € 0.00 | 0.00000 | 0.00000 | | |
| **OPEX year 3** | Operating Expenses (Year 3) | € 0.00 | 0.00000 | 0.00000 | | |
| **NPV** | Net Present Value at interest rate K | € 600.11 | 0.00000 | 0.00000 | | |
| **Net Cash Flow** | Net Cash Flow (interest rate K is not taken into account) | € 615.00 | 0.00000 | 0.00000 | | |
| **Payback Period (years)** | Amortization period | 1.00 | 0.00000 | 0.00000 | | |
| **COST** | CapEx + OpEx (year 1) | € 315.00 | 0.00000 | € 315.0016 | | |
| **QoS** | QoS capability | TRUE | 1.00000 | 0.00000 | | |
| **THE** | LOS from user to access point (Line of Sight Necessary?) | N / A | 0.00000 | 0.00000 | | |
| | LOS from access point to access node required? | N / A | 0.00000 | 0.00000 | | |
| **LICENSE** | Do you need a license? | TRUE | -1.00000 | 0.00000 | | |
| **Ubiquity** | Ubiquity at customer's address | YES | 1.00000 | 0.00000 | | |
| **Health** | Probability of provoking reluctance due to health risk (0 = NONE; 1 = LOW; 2 = MEDIUM; 3 = HIGH) | 1 | -0.33333 | 0.00000 | | |

*Table A.5: Output parameters and F1 and F2 figures of merit in 4G-LTE scenario.$y_k$*

**OBTAINING THE MINIMUM NUMBER OF REDUNDANT ACCESSES R TO MEET CUSTOMER REQUIREMENTS**

| | Output Parameters | $y_k$ | Minimum rk value to meet customer requirements |
|---|---|---|---|
| **RECEPTION SPEED** | AVERAGE Bandwidth per user in access (Mbits / s per user) | 23.9 | two |
| **EMISSION SPEED** | AVERAGE Bandwidth per user in access (Mbits / s per user) | 7.9 | 1 |
| **AVAILABILITY** | Availability | 99.9952% | two |
| **DISTANCE** | Distance user to access point (meters) | 15000 | COMPLIES |
| | Total distance user to access node (m) | 15,000 | COMPLIES |
| **COST** | CapEx + OpEx (Year 1) | € 543.00 | COMPLIES |
| **QoS** | QoS capability | TRUE | COMPLIES |
| **THE** | LOS from user to access point (Line of Sight Necessary?) | N / A | COMPLIES |
| | LOS from access point to access node required? | N / A | COMPLIES |
| **LICENSE** | Do you need a license? | TRUE | COMPLIES |
| **Ubiquity** | Ubiquity at customer's address | YES | COMPLIES |
| **Health** | Probability of provoking reluctance due to health risk (0 = NONE; 1 = LOW; 2 = MEDIUM; 3 = HIGH) | 1 | COMPLIES |

CONCLUSION

YES IT COMPLIES WITH:

R = 2

*Table A.6: Minimum number of redundant accesses R for 4G-LTE technology to meet the established user requirements.*

## A.3 Scenario 5: FTTH access with virtualized router

The input parameters of the model are shown below, as well as the results obtained by applying the model for this scenario. The incidence of router virtualization with respect to a conventional FTTH access is reflected in greater availability, since an equipment is eliminated at the customer's premises (element connected in series) and the routing function is supported in the virtualization Datacenter , and lower cost (by reducing the cost of CapEx deployment and OpEx maintenance).

The following tables show the input parameters of the model that are used as an example in this scenario, as well as the output parameters obtained by applying the UTEM model for this scenario. Any variation in input parameters, customer requirements, or user preferences will lead to different results.

**INPUT PARAMETERS** $x_{ij}$ of the UTEM MODEL

**SCENARIO NAME:** FTTH with virtualized router

| | Input parameter | PC interface | Element 1 | Element 2 | Element 3 |
|---|---|---|---|---|---|
| **Element identification** | Element name | **Wireless 802.11b / g US ROBOTICS adapter** | **FTTH Subscriber Terminal** | **Access node (OLT)** | **Aggregation Network** |
| | Element function | **Fast Ethernet card** | **ONU + virtualized router** | **Access node (Optical access interface)** | **Aggregation interface** |
| **Bandwidth** | Unitary Bandwidth (Mbits / s) (Reception) | 100 | 100 | 100 | 100 |
| | Unitary Bandwidth (Mbits / s) (Emission) | 100 | 10 | 10 | 10 |
| **Availability** | Availability | 99.9962% | 100.0000% | 99.9990% | 100.0000% |
| **Distance** | Distance (meters) | N / A | N / A | 15000 | N / A |
| **QoS** | QoS capability | N / A | TRUE | TRUE | TRUE |
| **Redundancy** | Redundancy (No. elements in parallel) | 1 | 1 | 1 | 1 |

| THE | LOS (Line of Sight Needed?) | N / A | N / A | N / A | N / A |
|---|---|---|---|---|---|
| **Frequency band** | Band (GHz) | N / A | N / A | N / A | N / A |
| **License** | Do you need a license? | FAKE | FAKE | FAKE | FAKE |
| **Users** | No. users: | 1 | 1 | N / A | N / A |
| **Concurrence** | Estimated average concurrency of users | N / A | 100.00% | N / A | N / A |
| **Technology** | Do you use wireless technology in any section? | DO NOT | DO NOT | DO NOT | DO NOT |
| **Environment** | Vector Environment (DENSE URBAN / URBAN / SUBURBAN / RURAL) | (1,1,1,1) | (1,1,1,1) | (1,1,1,1) | (1,1,1,1) |
| **Attenuation by meteorology** | Total decrease in Reception Bandwidth due to meteorological effects (Mbits / s) | 0 | 0 | 0 | 0 |
| | Total decrease in Broadcast Bandwidth due to meteorological effects (Mbits / s) | 0 | 0 | 0 | 0 |
| **Ubiquity** | Ubiquity at customer's address | YES | N / A | N / A | N / A |
| **Health** | Probability of provoking reluctance due to health risk (0 = NONE; 1 = LOW; 2 = MEDIUM; 3 = HIGH) | 1 | 0 | 0 | 0 |
| **K (interest rate)** | Type of interest | 1.00% | N / A | N / A | N / A |
| **ARPU year 1** | Average revenue per user (Year 1) | € 704.88 | | | |
| **ARPU year 2** | Average revenue per user (Year 1) | € 704.88 | | | |
| **ARPU year 3** | Average revenue per user (Year 1) | € 704.88 | | | |
| **CAPEX year 1** | Investments (Year 1) | € 15.00 | € 25.00 | € 150.00 | € 200.00 |
| **CAPEX year 2** | Investments (Year 2) | | | | |
| **CAPEX year 3** | Investments (Year 3) | | | | |
| **OPEX year 1** | Operating Expenses (Year 1) | € 0.0006 | € 0.0000 | € 0.0015 | € 0.0000 |
| **OPEX year 2** | Operating Expenses (Year 2) | € 0.0006 | € 0.0000 | € 0.0015 | € 0.0000 |
| **OPEX year 3** | Operating Expenses (Year 3) | € 0.0006 | € 0.0000 | € 0.0015 | € 0.0000 |

*Table A.7: Input parameters in FTTH scenario with virtualized router.*

**OUTPUT PARAMETERS $y_k$ and Figures of merit F1 and F2**

| | Output Parameters | $y_k$ | $a_k \cdot \overline{y}_k$ | $b_p \cdot y_k$ | Weighted Valuation | |
|---|---|---|---|---|---|---|
| | | | | | F1 | F2 |
| **RECEPTION SPEED** | AVERAGE Bandwidth (Mbits / s per user) | 100 | 1,0000 | 0.0000 | 77.36 % | 198.35 % / K € |
| **EMISSION SPEED** | AVERAGE Bandwidth (Mbits / s per user) | 10 | 1,0000 | 0.0000 | | |
| **AVAILABILITY** | Availability | 99.9952% | 0.0527 | 0.0000 | | |
| **DISTANCE** | Distance user to access point (meters) | 15,000 | 0.4997 | 0.0000 | | |
| | Total distance user to access node (m) | 15,000 | 0.4997 | 0.0000 | | |
| **ARPU year 1** | Average revenue per user (Year 1) | € 704.88 | 0.0000 | 0.0000 | | |
| **ARPU year 2** | Average revenue per user (Year 1) | € 704.88 | 0.0000 | 0.0000 | | |
| **ARPU year 3** | Average revenue per user (Year 1) | € 704.88 | 0.0000 | 0.0000 | | |
| **CAPEX year 1** | Investments (Year 1) | € 390.00 | 0.0000 | 0.0000 | | |
| **CAPEX year 2** | Investments (Year 2) | € 0.00 | 0.0000 | 0.0000 | | |
| **CAPEX year 3** | Investments (Year 3) | € 0.00 | 0.0000 | 0.0000 | | |
| **OPEX year 1** | Operating Expenses (Year 1) | € 0.00 | 0.0000 | 0.0000 | | |
| **OPEX year 2** | Operating Expenses (Year 2) | € 0.00 | 0.0000 | 0.0000 | | |
| **OPEX year 3** | Operating Expenses (Year 3) | € 0.00 | 0.0000 | 0.0000 | | |
| **NPV** | Net Present Value at interest rate K | € 1,686.90 | 0.0000 | 0.0000 | | |
| **Net Cash Flow** | Net Cash Flow (interest rate K is not taken into account) | € 1,724.63 | 0.0000 | 0.0000 | | |
| **Payback Period (years)** | Amortization period | 1.00 | 0.0000 | 0.0000 | | |
| **COST** | CapEx + OpEx (year 1) | € 390.00 | 0.0000 | € 390.0021 | | |
| **QoS** | QoS capability | TRUE | 1,0000 | 0.0000 | | |
| **THE** | LOS from user to access point (Line of Sight Necessary?) | N / A | 0.0000 | 0.0000 | | |
| | LOS from access point to access node required? | N / A | 0.0000 | 0.0000 | | |
| **LICENSE** | Do you need a license? | FAKE | 0.0000 | 0.0000 | | |
| **Ubiquity** | Ubiquity at customer's address | YES | 1,0000 | 0.0000 | | |
| **Health** | Probability of provoking reluctance due to health risk (0 = NONE; 1 = LOW; 2 = MEDIUM; 3 = HIGH) | 1 | -0.3333 | 0.0000 | | |

*Table A.8: Output parameters and F1 and F2 figures of merit in FTTH scenario with virtualized router.$y_k$*

**OBTAINING THE MINIMUM NUMBER OF REDUNDANT ACCESSES R TO MEET CUSTOMER REQUIREMENTS**

| | Output Parameters | $y_k$ | Minimum rk value to meet customer requirements |
|---|---|---|---|
| RECEPTION SPEED | AVERAGE Bandwidth per user in access (Mbits / s per user) | 100 | 1 |
| EMISSION SPEED | AVERAGE Bandwidth per user in access (Mbits / s per user) | 10 | 1 |
| AVAILABILITY | Availability | 99.9952% | two |
| DISTANCE | Distance user to access point (meters) | 15000 | COMPLIES |
| | Total distance user to access node (m) | 15,000 | COMPLIES |
| COST | CapEx + OpEx (Year 1) | € 390.0021 | COMPLIES |
| QoS | QoS capability | TRUE | COMPLIES |
| THE | LOS from user to access point (Line of Sight Necessary?) | N / A | COMPLIES |
| | LOS from access point to access node required? | N / A | COMPLIES |
| LICENSE | Do you need a license? | FAKE | COMPLIES |
| Ubiquity | Ubiquity at customer's address | YES | COMPLIES |
| Health | Probability of provoking reluctance due to health risk (0 = NONE; 1 = LOW; 2 = MEDIUM; 3 = HIGH) | 1 | COMPLIES |

CONCLUSION
YES IT COMPLIES WITH:

R = 2

*Table A.9: Minimum number of redundant accesses R for FTTH technology with virtualized router to meet the established user requirements.*

## A.4 Scenario 6: Dedicated point-to-point line 34 Mbits / s

This scenario, as mentioned above, has to do with the motivation of this doctoral thesis, since it emanates from the need to seek cheaper solutions with the same bandwidth and availability benefits as point-to-point lines, as stated in the Introduction Chapter. The symmetric bandwidth of 34 Mbits / s corresponds to the E3 standard in Plesiochronous Digital Hierarchy (PDH).

The following tables show the input parameters of the model that are used as an example in this scenario, as well as the output parameters obtained by applying the UTEM model for this scenario. Any variation in input parameters, customer requirements, or user preferences will lead to different results.

**INPUT PARAMETERS** $x_{ij}$ of the UTEM MODEL

**SCENARIO NAME:** Dedicated point-to-point line 34 Mbits / s

| | Input parameter | PC interface | Element 1 | Element 2 | Element 3 |
|---|---|---|---|---|---|
| **Element identification** | Element name | Wireless 802.11b / g adapter US ROBOTICS USR805420 | CISCO 1841 Router G.703 interface | Valiant VLC100 Multiplexer Terminal | Access Node |
| | Element function | Fast Ethernet card | Router at customer's home | Access interface | Aggregation interface |
| **Bandwidth** | Unitary Bandwidth (Mbits / s) (Reception) | 100 | 3. 4 | 3. 4 | 3. 4 |
| | Unitary Bandwidth (Mbits / s) (Emission) | 100 | two | two | 12 |
| **Availability** | Availability | 100.0000% | 100.0000% | 100.0000% | 100.0000% |
| **Distance** | Distance (meters) | N / A | N / A | 5000 | N / A |
| **QoS** | QoS capability | N / A | TRUE | TRUE | TRUE |
| **Redundancy** | Redundancy (No. elements in parallel) | 1 | 1 | two | 1 |
| **THE** | LOS (Line of Sight Needed?) | N / A | N / A | N / A | N / A |
| **Frequency band** | Band (GHz) | N / A | N / A | N / A | N / A |
| **License** | Do you need a license? | FAKE | FAKE | FAKE | FAKE |

| Users | No. users: | 1 | 1 | N / A | N / A |
|---|---|---|---|---|---|
| Concurrence | Estimated average concurrency of users | N / A | 100.00% | N / A | N / A |
| Technology | Do you use wireless technology in any section? | DO NOT | DO NOT | DO NOT | DO NOT |
| Environment | Vector Environment (DENSE URBAN / URBAN / SUBURBAN / RURAL) | (1,1,1,0) | (1,1,1,0) | (1,1,1,0) | (1,1,1,0) |
| Attenuation by meteorology | Total decrease in Reception Bandwidth due to meteorological effects (Mbits / s) | 0 | 0 | 0 | 0 |
| | Total decrease in Broadcast Bandwidth due to meteorological effects (Mbits / s) | 0 | 0 | 0 | 0 |
| Ubiquity | Ubiquity at customer's address | YES | N / A | N / A | N / A |
| Health | Probability of provoking reluctance due to health risk (0 = NONE; 1 = LOW; 2 = MEDIUM; 3 = HIGH) | 0 | 0 | 0 | 0 |
| K (interest rate) | Type of interest | 1.00% | N / A | N / A | N / A |
| ARPU year 1 | Average revenue per user (Year 1) | € 14,000.00 | | | |
| ARPU year 2 | Average revenue per user (Year 1) | € 12,000.00 | | | |
| ARPU year 3 | Average revenue per user (Year 1) | € 12,000.00 | | | |
| CAPEX year 1 | Investments (Year 1) | € 150.00 | € 1,000.00 | € 1,000.00 | € 500.00 |
| CAPEX year 2 | Investments (Year 2) | | | | |
| CAPEX year 3 | Investments (Year 3) | | | | |
| OPEX year 1 | Operating Expenses (Year 1) | € 0.00000 | € 0.00000 | € 0.00000 | € 0.00000 |
| OPEX year 2 | Operating Expenses (Year 2) | € 0.00000 | € 0.00000 | € 0.00000 | € 0.00000 |
| OPEX year 3 | Operating Expenses (Year 3) | € 0.00000 | € 0.00000 | € 0.00000 | € 0.00000 |

*Table A.10: Scenario input parameters Dedicated point-to-point line 34 Mbits / s.*

**OUTPUT PARAMETERS $y_k$ and Figures of merit F1 and F2**

| | Output Parameters | $y_k$ | $a_k \cdot \overline{y}_k$ | $b_p \cdot y_k$ | Weighted Valuation F1 | F2 |
|---|---|---|---|---|---|---|
| **RECEPTION SPEED** | AVERAGE Bandwidth (Mbits / s per user) | 3. 4 | 0.0571 | 0.0000 | 107.9 6% | 40.74 % / K € |
| **EMISSION SPEED** | AVERAGE Bandwidth (Mbits / s per user) | 3. 4 | 4.4286 | 0.0000 | | |
| **AVAILABILITY** | Availability | 100.0000% | 0.1010 | 0.0000 | | |
| **DISTANCE** | Distance user to access point (meters) | 5000 | 0.1661 | 0.0000 | | |
| | Total distance user to access node (m) | 5,000 | 0.1661 | 0.0000 | | |
| **ARPU year 1** | Average revenue per user (Year 1) | € 14,000.00 | 0.0000 | 0.0000 | | |
| **ARPU year 2** | Average revenue per user (Year 1) | € 12,000.00 | 0.0000 | 0.0000 | | |
| **ARPU year 3** | Average revenue per user (Year 1) | € 12,000.00 | 0.0000 | 0.0000 | | |
| **CAPEX year 1** | Investments (Year 1) | € 2,650.00 | 0.0000 | 0.0000 | | |
| **CAPEX year 2** | Investments (Year 2) | € 0,000 | 0.0000 | 0.0000 | | |
| **CAPEX year 3** | Investments (Year 3) | € 0,000 | 0.0000 | 0.0000 | | |
| **OPEX year 1** | Operating Expenses (Year 1) | € 0,000 | 0.0000 | 0.0000 | | |
| **OPEX year 2** | Operating Expenses (Year 2) | € 0,000 | 0.0000 | 0.0000 | | |
| **OPEX year 3** | Operating Expenses (Year 3) | € 0,000 | 0.0000 | 0.0000 | | |
| **NPV** | Net Present Value at interest rate K | € 34,648.26 | 0.0000 | 0.0000 | | |
| **Net Cash Flow** | Net Cash Flow (interest rate K is not taken into account) | € 35,350.00 | 0.0000 | 0.0000 | | |
| **Payback Period (years)** | Amortization period | 1.00 | 0.0000 | 0.0000 | | |
| **COST** | CapEx + OpEx (year 1) | € 2,650.0000 | 0.0000 | € 2,650.0000 | | |
| **QoS** | QoS capability | TRUE | 1,0000 | 0.0000 | | |
| **THE** | LOS from user to access point (Line of Sight Necessary?) | N / A | 0.0000 | 0.0000 | | |
| | LOS from access point to access node required? | N / A | 0.0000 | 0.0000 | | |
| **LICENSE** | Do you need a license? | FAKE | 0.0000 | 0.0000 | | |
| **Ubiquity** | Ubiquity at customer's address | YES | 1,0000 | 0.0000 | | |
| **Health** | Probability of provoking reluctance due to health risk (0 = NONE; 1 = LOW; 2 = MEDIUM; 3 = HIGH) | 0 | -0.3333 | 0.0000 | | |

*Table A.11: Output parameters and F1 and F2 figures of merit on stage Dedicated line point to point 34 Mbits / s.$y_k$*

**OBTAINING THE MINIMUM NUMBER OF REDUNDANT ACCESSES R TO MEET CUSTOMER REQUIREMENTS**

| | Output Parameters | $y_k$ | Minimum rk value to meet customer requirements | |
|---|---|---|---|---|
| **RECEPTION SPEED** | AVERAGE Bandwidth per user in access (Mbits / s per user) | 3. 4 | 1 | |
| **EMISSION SPEED** | AVERAGE Bandwidth per user in access (Mbits / s per user) | 3. 4 | 1 | |
| **AVAILABILITY** | Availability | 100.0000% | 1 | |
| **DISTANCE** | Distance user to access point (meters) | 5,000 | COMPLIES | CONCLUSION |
| | Total distance user to access node (m) | 5,000 | COMPLIES | YES IT COMPLIES WITH: |
| **COST** | CapEx + OpEx (Year 1) | € 2,650.00 | COMPLIES | |
| **QoS** | QoS capability | TRUE | COMPLIES | R = 1 |
| **THE** | LOS from user to access point (Line of Sight Necessary?) | N / A | COMPLIES | |
| | LOS from access point to access node required? | N / A | COMPLIES | |
| **LICENSE** | Do you need a license? | FAKE | COMPLIES | |
| **Ubiquity** | Ubiquity at customer's address | YES | COMPLIES | |
| **Health** | Probability of provoking reluctance due to health risk (0 = NONE; 1 = LOW; 2 = MEDIUM; 3 = HIGH) | 0 | COMPLIES | |

*Table A.12: Minimum number of redundant accesses R for the technology Dedicated point-to-point 34 Mbits / s technology meets the established user requirements.*

## A.5 Scenario 7: 2 x ADSL redundant access

In this section we proceed to validate the model using a redundant 2 x ADSL access (ie: two ADSL accesses in parallel). The following tables show the input parameters of the model that are used as an example in this scenario, as well as the output parameters obtained by applying the UTEM model for this scenario. Any variation in input parameters, customer requirements, or user preferences will lead to different results.

**INPUT PARAMETERS** $x_{ij}$ of the UTEM MODEL

**SCENARIO NAME:** Redundant access 2 x ADSL

| | Input parameter | PC interface | Element 1 | Element 2 | Element 3 |
|---|---|---|---|---|---|
| **Element identification** | Element name | Wireless 802.11b / g adapter US ROBOTICS USR805420 | 3COM OfficeConnect 812 Router | DSLAM (Alcatel 7300) | Aggregation network |
| | Element function | Wi-Fi PC adapter | Router at customer's home | Access interface | Aggregation interface |
| **Bandwidth** | Unitary Bandwidth (Mbits / s) (Reception) | 100 | 10 | 10 | 10 |
| | Unitary Bandwidth (Mbits / s) (Emission) | 100 | 3 | 3 | 3 |
| **Availability** | Availability | 99.9962% | 99.9644% | 99.9990% | 100.0000% |
| **Distance** | Distance (meters) | N / A | N / A | 4500 | N / A |
| **QoS** | QoS capability | N / A | TRUE | TRUE | TRUE |
| **Redundancy** | Redundancy (No. elements in parallel) | 1 | two | two | two |
| **THE** | LOS (Line of Sight Needed?) | N / A | N / A | N / A | N / A |
| **Frequency band** | Band (GHz) | N / A | N / A | N / A | N / A |
| **License** | Do you need a license? | FAKE | FAKE | FAKE | FAKE |

| Users | No. users: | 1 | 1 | N / A | N / A |
|---|---|---|---|---|---|
| Concurrence | Estimated average concurrency of users | N / A | 100.00% | N / A | N / A |
| Technology | Do you use wireless technology in any section? | N / A | 100.00% | N / A | N / A |
| Environment | Vector Environment (DENSE URBAN / URBAN / SUBURBAN / RURAL) | N / A | 100.00% | N / A | N / A |
| Attenuation by meteorology | Total decrease in Reception Bandwidth due to meteorological effects (Mbits / s) | DO NOT | DO NOT | DO NOT | DO NOT |
| | Total decrease in Broadcast Bandwidth due to meteorological effects (Mbits / s) | (1,1,1,1) | (1,1,1,1) | (1,1,1,1) | (1,1,1,1) |
| Ubiquity | Ubiquity at customer's address | 0 | 0 | 0 | 0 |
| Health | Probability of provoking reluctance due to health risk (0 = NONE; 1 = LOW; 2 = MEDIUM; 3 = HIGH) | 0 | 0 | 0 | 0 |
| K (interest rate) | Type of interest | YES | N / A | N / A | N / A |
| ARPU year 1 | Average revenue per user (Year 1) | 1 | 0 | 0 | 0 |
| ARPU year 2 | Average revenue per user (Year 1) | 1.00% | | | |
| ARPU year 3 | Average revenue per user (Year 1) | € 833.08 | | | |
| CAPEX year 1 | Investments (Year 1) | € 727.20 | | | |
| CAPEX year 2 | Investments (Year 2) | € 727.20 | | | |
| CAPEX year 3 | Investments (Year 3) | € 15.00 | € 200.00 | € 200.00 | € 200.00 |
| OPEX year 1 | Operating Expenses (Year 1) | | | | |
| OPEX year 2 | Operating Expenses (Year 2) | | | | |
| OPEX year 3 | Operating Expenses (Year 3) | € 0.0006 | € 0.0711 | € 0.0020 | € 0.0000 |

*Table A.13: Scenario input parameters Redundant access 2 x ADSL.*

**OUTPUT PARAMETERS $y_k$ and Figures of merit F1 and F2**

| | Output Parameters | $y_k$ | $a_k \cdot \overline{y}_k$ | $b_P \cdot y_k$ | Weighted Valuation | |
|---|---|---|---|---|---|---|
| | | | | | F1 | F2 |
| RECEPTION SPEED | AVERAGE Bandwidth (Mbits / s per user) | twenty | -0.1429 | 0.0000 | 37.93 % | 61.67 % / K € |
| EMISSION SPEED | AVERAGE Bandwidth (Mbits / s per user) | 6 | 0.4286 | 0.0000 | | |
| AVAILABILITY | Availability | 99.9962% | 0.0626 | 0.0000 | | |
| DISTANCE | Distance user to access point (meters) | 4500 | 0.1494 | 0.0000 | | |
| | Total distance user to access node (m) | 4,500 | 0.1494 | 0.0000 | | |
| ARPU year 1 | Average revenue per user (Year 1) | € 833.39 | 0.0000 | 0.0000 | | |
| ARPU year 2 | Average revenue per user (Year 1) | € 727.20 | 0.0000 | 0.0000 | | |
| ARPU year 3 | Average revenue per user (Year 1) | € 727.20 | 0.0000 | 0.0000 | | |
| CAPEX year 1 | Investments (Year 1) | € 615.00 | 0.0000 | 0.0000 | | |
| CAPEX year 2 | Investments (Year 2) | € 0.00 | 0.0000 | 0.0000 | | |
| CAPEX year 3 | Investments (Year 3) | € 0.00 | 0.0000 | 0.0000 | | |
| OPEX year 1 | Operating Expenses (Year 1) | € 0.07 | 0.0000 | 0.0000 | | |
| OPEX year 2 | Operating Expenses (Year 2) | € 0.07 | 0.0000 | 0.0000 | | |
| OPEX year 3 | Operating Expenses (Year 3) | € 0.07 | 0.0000 | 0.0000 | | |
| NPV | Net Present Value at interest rate K | € 1,634.39 | 0.0000 | 0.0000 | | |
| Net Cash Flow | Net Cash Flow (interest rate K is not taken into account) | € 1,672.26 | 0.0000 | 0.0000 | | |
| Payback Period (years) | Amortization period | 1.00 | 0.0000 | 0.0000 | | |
| COST | CapEx + OpEx (year 1) | € 615.0737 | 0.0000 | € 615.0737 | | |
| QoS | QoS capability | TRUE | 1,0000 | 0.0000 | | |
| THE | LOS from user to access point (Line of Sight Necessary?) | N / A | 0.0000 | 0.0000 | | |
| | LOS from access point to access node required? | N / A | 0.0000 | 0.0000 | | |
| LICENSE | Do you need a license? | FAKE | 0.0000 | 0.0000 | | |
| Ubiquity | Ubiquity at customer's address | YES | 1,0000 | 0.0000 | | |
| Health | Probability of provoking reluctance due to health risk (0 = NONE; 1 = LOW; 2 = MEDIUM; 3 = HIGH) | 1 | -0.3333 | 0.0000 | | |

*Table A.14: Output parameters and F1 and F2 figures of merit in scenario Redundant access 2 x ADSL.$y_k$*

**OBTAINING THE MINIMUM NUMBER OF REDUNDANT ACCESSES R TO MEET CUSTOMER REQUIREMENTS**

| | Output Parameters | $y_k$ | Minimum rk value to meet customer requirements |
|---|---|---|---|
| RECEPTION SPEED | AVERAGE Bandwidth per user in access (Mbits / s per user) | twenty | two |
| EMISSION SPEED | AVERAGE Bandwidth per user in access (Mbits / s per user) | 6 | 1 |
| AVAILABILITY | Availability | 99.9962% | two |
| DISTANCE | Distance user to access point (meters) | 4500 | COMPLIES |
| | Total distance user to access node (m) | 4,500 | COMPLIES |
| COST | CapEx + OpEx (Year 1) | € 615.07 | COMPLIES |
| QoS | QoS capability | TRUE | COMPLIES |
| THE | LOS from user to access point (Line of Sight Necessary?) | N / A | COMPLIES |
| | LOS from access point to access node required? | N / A | COMPLIES |
| LICENSE | Do you need a license? | FAKE | COMPLIES |
| Ubiquity | Ubiquity at customer's address | YES | COMPLIES |
| Health | Probability of provoking reluctance due to health risk (0 = NONE; 1 = LOW; 2 = MEDIUM; 3 = HIGH) | 1 | COMPLIES |

CONCLUSION
YES IT COMPLIES WITH:

R = 2

*Table A.15: Minimum number of redundant accesses R for the Redundant Access 2 x ADSL technology to meet the established user requirements.*

## A.6 Scenario 8: ADSL in parallel with IEEE 802.11g Wi-Fi access point and WiMAX 802.16 backhaul

The following tables show the input parameters of the model that are used as an example in this scenario, as well as the output parameters obtained by applying the UTEM model for this scenario. Any variation in input parameters, customer requirements, or user preferences will lead to different results. In this case, the input and output parameters of the Parallel Submodel are shown, once the equivalent parameters of each individual access have been obtained through the Serial Submodel.

**INPUT PARAMETERS** $x_{ij}$ of the UTEM MODEL

**SCENARIO NAME:** ADSL in parallel with IEEE 802.11g WiFi access point and WiMAX 802.16 backhaul

|  | Input parameter | Access 1 | Access 2 |
|---|---|---|---|
| **Element identification** | Access Name | ADSL | 802.11g + WiMAX Backhaul |
|  | En Modo Respaldo ? (SI/NO) (RESPALDO = SI; AGREGADO = NO) | NO | NO |
| **Ancho de Banda** | Ancho de Banda Unitario (Mbits/s) (Recepción) | 10 | 0,121230769 |
|  | Ancho de Banda Unitario (Mbits/s) (Emisión) | 3 | 0,122461538 |
| **Disponibilidad** | Disponibilidad | 99,9597% | 99,9760% |
| **Distancia** | Distancia usuario a punto de acceso (metros) | 4.500 | 45 |
|  | Distancia total usuario a nodo de acceso (m) | 4.500 | 45.045 |
| **Coste Anual** | Coste Anual Servicio (€) | 315,04 € | 12,00 € |
| **QoS** | Capacidad para QoS | VERDADERO | VERDADERO |
| **Redundancia** | Redundancia (Nº elementos en paralelo) | 1 | 1 |
| **LOS** | LOS desde usuario a punto de acceso (Line of Sight Necesaria ?) | FALSO | VERDADERO |
|  | LOS desde punto de acceso a nodo de red de transporte necesaria? | FALSO | VERDADERO |
| **Banda de frecuencias** | Banda (GHz) | N/A | 2,4 |
| **Licencia** | Necesita Licencia ? | FALSO | VERDADERO |
| **Usuarios** | Nº usuarios: | 1 | 1 |

| Concurrencia | Concurrencia media estimada de usuarios | N/A | N/A |
|---|---|---|---|
| Tecnología | ¿Utiliza tecnología inalámbrica en algún tramo? | SI | SI |
| Entorno | Entorno (URBANO DENSO / URBANO / SUBURBANO / RURAL) | (1,1,1,1) | (1,1,1,1) |
| Atenuación por metorología | Disminución total de Ancho de Banda en Recepción por efectos meteorológicos (Mbits/s) | 0 | 0,03 |
| | Disminución total de Ancho de Banda en Emisión por efectos meteorológicos (Mbits/s) | 0 | 0,03 |
| Ubicuidad | Ubicuidad en domicilio de cliente | SI | SI |
| Salud | Probabilidad de suscitar reticencias por riesgo para la salud (0=NINGUNA; 1=BAJA; 2=MEDIA; 3=ALTA) | 1 | 2 |
| ARPU año 1 | Ingresos medios por usuario (Año 1) | 416,54 € | 174,00 € |
| ARPU año 2 | Ingresos medios por usuario (Año 1) | 363,60 € | 144,00 € |
| ARPU año 3 | Ingresos medios por usuario (Año 1) | 363,60 € | 144,00 € |
| CAPEX año 1 | Inversiones (Año 1) | 315,00 € | 66,67 € |
| CAPEX año 2 | Inversiones (Año 2) | 0,00 € | 0,00 € |
| CAPEX año 3 | Inversiones (Año 3) | 0,00 € | 0,00 € |
| OPEX año 1 | Gastos de Operación (Año 1) | 0,04 € | 0,00 € |
| OPEX año 2 | Gastos de Operación (Año 2) | 0,04 € | 0,00 € |
| OPEX año 3 | Gastos de Operación (Año 3) | 0,04 € | 0,00 € |

Tabla A.16: Parámetros de entrada en escenario ADSL en paralelo con punto de acceso WiFi IEEE 802.11g y backhaul WiMAX 802.16.

PARÁMETROS DE SALIDA $y_k$ y Figuras de mérito F1
y F2

| | Parámetros Salida | $y_k$ | $a_k \cdot \bar{y}_k$ | $b_P \cdot y_k$ | Valoración Ponderada F1 | Valoración Ponderada F2 |
|---|---|---|---|---|---|---|
| VELOCIDAD DE RECEPCIÓN | Ancho de Banda MEDIO (Mbits/s por usuario) | 10,09123077 | -0,28441 | 0,00000 | -22,19% | -58,12 %/K€ |
| VELOCIDAD DE EMISIÓN | Ancho de Banda MEDIO (Mbits/s por usuario) | 3,092461538 | 0,01321 | 0,00000 | | |
| DISPONIBILIDAD | Disponibilidad | 99,999990% | 0,10091 | 0,00000 | | |
| DISTANCIA | Distancia usuario a punto de acceso (metros) | 45 | 0,00083 | 0,00000 | | |
| | Distancia total usuario a nodo de acceso (m) | 4.500 | 0,14943 | 0,00000 | | |
| ARPU año 1 | Ingresos medios por usuario (Año 1) | 590,54 € | 0,00000 | 0,00000 | | |
| ARPU año 2 | Ingresos medios por usuario (Año 1) | 507,60 € | 0,00000 | 0,00000 | | |
| ARPU año 3 | Ingresos medios por usuario (Año 1) | 507,60 € | 0,00000 | 0,00000 | | |
| CAPEX año 1 | Inversiones (Año 1) | 381,67 € | 0,00000 | 0,00000 | | |
| CAPEX año 2 | Inversiones (Año 2) | 0,00 € | 0,00000 | 0,00000 | | |
| CAPEX año 3 | Inversiones (Año 3) | 0,00 € | 0,00000 | 0,00000 | | |
| OPEX año 1 | Gastos de Operación (Año 1) | 0,04 € | 0,00000 | 0,00000 | | |
| OPEX año 2 | Gastos de Operación (Año 2) | 0,04 € | 0,00000 | 0,00000 | | |
| OPEX año 3 | Gastos de Operación (Año 3) | 0,04 € | 0,00000 | 0,00000 | | |
| NPV | Valor actual neto (Net Present Value) a tipo de interés K | 1.196,96 € | 0,00000 | 0,00000 | | |
| Flujo de Caja Neto | Net Cash Flow (no se tiene en cuenta el tipo de interés K) | 1.223,96 € | 0,00000 | 0,00000 | | |
| Payback Period (años) | Período de Amortización | 1,00 € | 0,00000 | 0,00000 | | |
| COSTE | CapEx + OpEx (año 1) | 381,7048 € | 0,00000 | 381,7048 € | | |
| QoS | Capacidad para QoS | VERDADERO | 1,00000 | 0,00000 | | |
| LOS | LOS desde usuario a punto de acceso (Line of Sight Necesaria ?) | VERDADERO | -1,00000 | 0,00000 | | |
| | LOS desde punto de acceso a nodo de acceso necesaria? | VERDADERO | -1,00000 | 0,00000 | | |
| LICENCIA | Necesita Licencia ? | VERDADERO | -1,00000 | 0,00000 | | |
| Ubicuidad | Ubicuidad en domicilio de cliente | SI | 1,00000 | 0,00000 | | |
| Salud | Probabilidad de suscitar reticencias por riesgo para la salud (0=NINGUNA; 1=BAJA; 2=MEDIA; 3=ALTA) | 1 | -0,33333 | 0,00000 | | |

*Tabla A.17: Parámetros de salida y figuras de mérito F1 y F2 en escenario ADSL en paralelo con punto de acceso WiFi IEEE 802.11g y backhaul WiMAX 802.16.$y_k$*

**OBTENCIÓN DEL MÍNIMO NÚMERO DE ACCESOS REDUNDANTES R PARA CUMPLIR REQUISITOS DE CLIENTE**

| | Parámetros Salida | $y_k$ | Valor mínimo de rk para cumplir requisitos de cliente | |
|---|---|---|---|---|
| VELOCIDAD DE RECEPCIÓN | Ancho de Banda MEDIO por usuario en acceso (Mbits/s por usuario) | 10,09123077 | 3 | |
| VELOCIDAD DE EMISIÓN | Ancho de Banda MEDIO por usuario en acceso (Mbits/s por usuario) | 3,092461538 | 1 | |
| DISPONIBILIDAD | Disponibilidad | 99,999990% | 2 | |
| DISTANCIA | Distancia usuario a punto de acceso (metros) | 45 | CUMPLE | CONCLUSIÓN |
| | Distancia total usuario a nodo de acceso (m) | 4.500 | CUMPLE | SÍ CUMPLE CON: |
| COSTE | CapEx + OpEx (Año 1) | 381,70 € | CUMPLE | |
| QoS | Capacidad para QoS | VERDADERO | CUMPLE | R = 3 |
| LOS | LOS desde usuario a punto de acceso (Line of Sight Necesaria ?) | VERDADERO | CUMPLE | |
| | LOS desde punto de acceso a nodo de acceso necesaria? | VERDADERO | CUMPLE | |
| LICENCIA | Necesita Licencia ? | VERDADERO | CUMPLE | |
| Ubicuidad | Ubicuidad en domicilio de cliente | SI | CUMPLE | |
| Salud | Probabilidad de suscitar reticencias por riesgo para la salud (0=NINGUNA; 1=BAJA; 2=MEDIA; 3=ALTA) | 1 | CUMPLE | |

*Tabla A.18: Mínimo número de accesos redundantes R para que la tecnología ADSL en paralelo con punto de acceso WiFi IEEE 802.11g y backhaul WiMAX 802.16, cumpla los requisitos de cliente.*

## A.7 Escenario 9: VDSL

En las siguientes tablas, se muestran los parámetros de entrada del modelo que se utilizan como ejemplo en este escenario, así como los parámetros de salida obtenidos aplicando el modelo UTEM para este escenario. Cualquier variación en los parámetros de entrada, en los requisitos de cliente o en las preferencias de usuario, dará lugar a unos resultados diferentes.

**PARÁMETROS DE ENTRADA $x_{ij}$ del MODELO UTEM**

**NOMBRE ESCENARIO:** VDSL

| | Parámetro Entrada | Interfaz PC | Elemento 1 | Elemento 2 | Elemento 3 |
|---|---|---|---|---|---|
| **Identificación del elemento** | Nombre del elemento | Adaptador Wireless 802.11b/g U.S. ROBOTICS USR805420 | D-LINK DEV-311 VDSL Bridge Remote Unit | DSLAM (Alcatel 7300) | Red de agregación |
| | Función del elemento | Adaptador Wi-Fi PC | Router en domicilio de cliente | Interfaz Acceso | Interfaz Agregación |
| **Ancho de Banda** | Ancho de Banda Unitario (Mbits/s) (Recepción) | 100 | 50 | 50 | 50 |
| | Ancho de Banda Unitario (Mbits/s) (Emisión) | 100 | 5 | 5 | 5 |
| **Disponibilidad** | Disponibilidad | 99,9962% | 99,9644% | 99,9990% | 100,0000% |
| **Distancia** | Distancia (metros) | N/A | N/A | 600 | N/A |
| **QoS** | Capacidad para QoS | N/A | VERDADERO | VERDADERO | VERDADERO |
| **Redundancia** | Redundancia (Nº elementos en paralelo) | 1 | 1 | 1 | 1 |
| **LOS** | LOS (Line of Sight Necesaria ?) | N/A | N/A | N/A | N/A |

| Banda de frecuencias | Banda (GHz) | N/A | N/A | N/A | N/A |
|---|---|---|---|---|---|
| Licencia | Necesita Licencia ? | FALSO | FALSO | FALSO | FALSO |
| Usuarios | Nº usuarios: | 1 | 1 | N/A | N/A |
| Concurrencia | Concurrencia media estimada de usuarios | N/A | 100,00% | N/A | N/A |
| Tecnología | ¿Utiliza tecnología inalámbrica en algún tramo? | NO | NO | NO | NO |
| Entorno | VectorEntorno (URBANO DENSO / URBANO / SUBURBANO / RURAL) | (1,1,1,1) | (1,1,1,1) | (1,1,1,1) | (1,1,1,1) |
| Atenuación por metorología | Disminución total de Ancho de Banda en Recepción por efectos meteorológicos (Mbits/s) | 0 | 0 | 0 | 0 |
| | Disminución total de Ancho de Banda en Emisión por efectos meteorológicos (Mbits/s) | 0 | 0 | 0 | 0 |
| Ubicuidad | Ubicuidad en domicilio de cliente | SI | N/A | N/A | N/A |
| Salud | Probabilidad de suscitar reticencias por riesgo para la salud (0=NINGUNA; 1=BAJA; 2=MEDIA; 3=ALTA) | 1 | 0 | 0 | 0 |
| K (tipo de interés) | Tipo de interés | 1,00% | N/A | N/A | N/A |
| ARPU año 1 | Ingresos medios por usuario (Año 1) | 592,94 € | | | |
| ARPU año 2 | Ingresos medios por usuario (Año 1) | 540,00 € | | | |
| ARPU año 3 | Ingresos medios por usuario (Año 1) | 540,00 € | | | |
| CAPEX año 1 | Inversiones (Año 1) | 15,00 € | 125,00 € | 125,00 € | 100,00 € |
| CAPEX año 2 | Inversiones (Año 2) | | | | |
| CAPEX año 3 | Inversiones (Año 3) | | | | |
| OPEX año 1 | Gastos de Operación (Año 1) | 0,0006 € | 0,0444 € | 0,0013 € | 0,0000 € |
| OPEX año 2 | Gastos de Operación (Año 2) | 0,0006 € | 0,0444 € | 0,0013 € | 0,0000 € |
| OPEX año 3 | Gastos de Operación (Año 3) | 0,0006 € | 0,0444 € | 0,0013 € | 0,0000 € |

*Tabla A.19: Parámetros de entrada en escenario VDSL.*

**PARÁMETROS DE SALIDA** $y_k$ **y Figuras de mérito F1 y F2**

| | Parámetros Salida | $y_k$ | $a_k \cdot \overline{y}_k$ | $b_P \cdot y_k$ | Valoración Ponderada F1 | Valoración Ponderada F2 |
|---|---|---|---|---|---|---|
| **VELOCIDAD DE RECEPCIÓN** | Ancho de Banda MEDIO (Mbits/s por usuario) | 50 | 0,2857 | 0,0000 | 32,30 % | 88,48 %/K€ |
| **VELOCIDAD DE EMISIÓN** | Ancho de Banda MEDIO (Mbits/s por usuario) | 5 | 0,2857 | 0,0000 | | |
| **DISPONIBILIDAD** | Disponibilidad | 99,9597% | -0,3065 | 0,0000 | | |
| **DISTANCIA** | Distancia usuario a punto de acceso (metros) | 600 | 0,0193 | 0,0000 | | |
| | Distancia total usuario a nodo de acceso (m) | 600 | 0,0193 | 0,0000 | | |
| **ARPU año 1** | Ingresos medios por usuario (Año 1) | 592,94 € | 0,0000 | 0,0000 | | |
| **ARPU año 2** | Ingresos medios por usuario (Año 1) | 540,00 € | 0,0000 | 0,0000 | | |
| **ARPU año 3** | Ingresos medios por usuario (Año 1) | 540,00 € | 0,0000 | 0,0000 | | |
| **CAPEX año 1** | Inversiones (Año 1) | 365,00 € | 0,0000 | 0,0000 | | |
| **CAPEX año 2** | Inversiones (Año 2) | 0,00 € | 0,0000 | 0,0000 | | |
| **CAPEX año 3** | Inversiones (Año 3) | 0,00 € | 0,0000 | 0,0000 | | |
| **OPEX año 1** | Gastos de Operación (Año 1) | 0,05 € | 0,0000 | 0,0000 | | |
| **OPEX año 2** | Gastos de Operación (Año 2) | 0,05 € | 0,0000 | 0,0000 | | |
| **OPEX año 3** | Gastos de Operación (Año 3) | 0,05 € | 0,0000 | 0,0000 | | |
| **NPV** | Valor actual neto (Net Present Value) a tipo de interés K | 1.279,03 € | 0,0000 | 0,0000 | | |
| **Flujo de Caja Neto** | Net Cash Flow (no se tiene en cuenta el tipo de interés K) | 1.307,80 € | 0,0000 | 0,0000 | | |
| **Payback Period (años)** | Período de Amortización | 175000% | 0,0000 | 0,0000 | | |
| **COSTE** | CapEx + OpEx (año 1) | 1,00 | 0,0000 | 0,0000 | | |
| **QoS** | Capacidad para QoS | 365,0463 € | 0,0000 | 365,0463 € | | |
| **LOS** | LOS desde usuario a punto de acceso (Line of Sight Necesaria ?) | VERDADERO | 1,0000 | 0,0000 | | |
| | LOS desde punto de acceso a nodo de acceso necesaria? | N/A | 0,0000 | 0,0000 | | |
| **LICENCIA** | Necesita Licencia ? | N/A | 0,0000 | 0,0000 | | |
| **Ubicuidad** | Ubicuidad en domicilio de cliente | FALSO | 0,0000 | 0,0000 | | |
| **Salud** | Probabilidad de suscitar reticencias por riesgo para la salud (0=NINGUNA; 1=BAJA; 2=MEDIA; 3=ALTA) | SI | 1,0000 | 0,0000 | | |

*Tabla A.20: Parámetros de salida y figuras de mérito F1 y F2 en escenario VDSL.* $y_k$

**OBTENCIÓN DEL MÍNIMO NÚMERO DE ACCESOS REDUNDANTES R PARA CUMPLIR REQUISITOS DE CLIENTE**

| | Parámetros Salida | $y_k$ | Valor mínimo de rk para cumplir requisitos de cliente | |
|---|---|---|---|---|
| VELOCIDAD DE RECEPCIÓN | Ancho de Banda MEDIO por usuario en acceso (Mbits/s por usuario) | 50 | 1 | |
| VELOCIDAD DE EMISIÓN | Ancho de Banda MEDIO por usuario en acceso (Mbits/s por usuario) | 5 | 1 | |
| DISPONIBILIDAD | Disponibilidad | 99,9597% | 2 | |
| DISTANCIA | Distancia usuario a punto de acceso (metros) | 600 | CUMPLE | CONCLUSIÓN |
| | Distancia total usuario a nodo de acceso (m) | 600 | CUMPLE | SÍ CUMPLE CON: |
| COSTE | CapEx + OpEx (Año 1) | 365,0463 € | CUMPLE | |
| QoS | Capacidad para QoS | VERDADERO | CUMPLE | R = 2 |
| LOS | LOS desde usuario a punto de acceso (Line of Sight Necesaria ?) | N/A | CUMPLE | |
| | LOS desde punto de acceso a nodo de acceso necesaria? | N/A | CUMPLE | |
| LICENCIA | Necesita Licencia ? | FALSO | CUMPLE | |
| Ubicuidad | Ubicuidad en domicilio de cliente | SI | CUMPLE | |
| Salud | Probabilidad de suscitar reticencias por riesgo para la salud (0=NINGUNA; 1=BAJA; 2=MEDIA; 3=ALTA) | 1 | CUMPLE | |

*Tabla A.21: Mínimo número de accesos redundantes R para que la tecnología VDSL cumpla los requisitos de cliente establecidos.*